\documentclass[%
 reprint,
superscriptaddress,
nofootinbib,
 amsmath,amssymb,
 onecolumn,
prd,
]{revtex4-2}

\usepackage{graphicx}% Include figure files
\usepackage{dcolumn}% Align table columns on decimal point
\usepackage{bm}% bold math
\usepackage[colorlinks]{hyperref}% add hypertext capabilities
\usepackage{color}
\usepackage[caption=false]{subfig}
\def\beq{\begin{eqnarray}}
\def\eeq{\end{eqnarray}}

\usepackage{xspace}
\usepackage{amsmath}
\let\vec\mathbf

\usepackage{empheq}

\newcommand{\hMpc}{h\,\mathrm{Mpc}^{-1}}

\newcommand{\delD}[1]{(2\pi)^3\delta_\mathrm{D}\left({#1}\right)}

\newcommand{\fkp}{\mathrm{FKP}}

\newcommand{\av}[1]{\left\langle{#1}\right\rangle}

\newcommand{\vk}{\vec k}
\newcommand{\hk}{\hat{\vec k}}
\newcommand{\hn}{\hat{\vec n}}

\newcommand{\vp}{\vec p}
\newcommand{\vb}{\vec b}
\newcommand{\vd}{\vec d}

\newcommand{\vx}{\vec x}
\newcommand{\vy}{\vec y}
\renewcommand{\vr}{\vec r}
\newcommand{\hr}{\hat{\vec r}}
\newcommand{\vm}{\vec m}
\newcommand{\va}{\vec a}
\newcommand{\vn}{\vec n}
\newcommand{\Hi}{\mathsf{H}^{-1}}
\newcommand{\Ai}{\mathsf{A}^{-1}}
\newcommand{\A}{\mathsf{A}}
\newcommand{\C}{\mathsf{C}}
\newcommand{\B}{\mathsf{B}}

\newcommand{\Sig}{\mathsf{S}}
\newcommand{\N}{\mathsf{N}}
\newcommand{\Ci}{\mathsf{C}^{-1}}

\newcommand{\ft}[1]{\mathcal{F}\left[{#1}\right]}
\newcommand{\ift}[1]{\mathcal{F}^{-1}\left[{#1}\right]}
\newcommand{\fid}{\mathrm{fid}}
\newcommand{\Tr}[1]{\operatorname{Tr}\left[{#1}\right]}

\renewcommand{\H}{\mathsf{H}}
\newcommand{\fft}[1]{\mathrm{FFT}\left[{#1}\right]}
\newcommand{\ifft}[1]{\mathrm{FFT}^{-1}\left[{#1}\right]}

\definecolor{darkgreen}{RGB}{0,120,0}

\definecolor{brown}{RGB}{120,60,0}
\newcommand{\resub}[1]{#1}%\textcolor{darkgreen}{#1}}

\begin{document}

%\preprint{APS/123-QED}

\title{\Large Cosmology Without Window Functions: \\ \large II. Cubic Estimators for the Galaxy Bispectrum}% Force line breaks with \\

\author{Oliver H.\,E. Philcox}
\email{ohep2@cantab.ac.uk}
\affiliation{Department of Astrophysical Sciences, Princeton University,\\ Princeton, NJ 08540, USA}%
\affiliation{School of Natural Sciences, Institute for Advanced Study, 1 Einstein Drive,\\ Princeton, NJ 08540, USA}
%\author{\textit{et al.}}


\begin{abstract}
When analyzing the galaxy bispectrum measured from spectroscopic surveys, it is imperative to account for the effects of non-uniform survey geometry. Conventionally, this is done by convolving the theory model with the the window function; however, the computational expense of this prohibits full exploration of the bispectrum likelihood. In this work, we provide a new class of estimators for the \textit{unwindowed} bispectrum; a quantity that can be straightforwardly compared to theory. This builds upon the work of Philcox (2021) for the power spectrum, and comprises two parts (both obtained from an Edgeworth expansion): a cubic estimator applied to the data, and a Fisher matrix, which deconvolves the bispectrum components. In the limit of weak non-Gaussianity, the estimator is minimum-variance; furthermore, we give an alternate form based on FKP weights that is close-to-optimal and easy to compute. As a demonstration, we measure the binned bispectrum monopole of a suite of simulations both using conventional estimators and our unwindowed equivalents. Computation times are comparable, except that the unwindowed approach requires a Fisher matrix, computable in an additional $\mathcal{O}(100)$ CPU-hours. Our estimator may be straightforwardly extended to measure redshift-space distortions and the components of the bispectrum in arbitrary separable bases. The techniques of this work will allow the bispectrum to straightforwardly \resub{be} included in the cosmological analysis of current and upcoming survey data.
\end{abstract}

\maketitle

\section{Introduction}\label{sec: intro}

%\oliver{Is it worth rerunning with window-convolved n-bar? to check fkp vs ml?}
%\oliver{need to add correction factor into the code after its run + test whether it matters (e.g. for ML sims?)}
In the standard inflationary paradigm, the early Universe is well described by statistics that are Gaussian and close to scale-invariant \citep[e.g.,][]{1982PhLB..116..335L,1982PhRvL..48.1220A}, an assumption that is in exquisite agreement with cosmic microwave background (CMB) observations \citep[e.g.,][]{2020A&A...641A...6P}. In this limit, all cosmological information is encapsulated within the power spectrum of the observed field, or, equivalently, the two-point correlation function (2PCF), which can be easily measured, modeled, and fit. As the Universe evolves, non-linear structure growth shifts information from the primordial power spectrum into higher-order moments \citep[e.g.,][]{2015PhRvD..92l3522S}, the simplest of which is the bispectrum, or three-point correlation function (3PCF). At the epoch of recombination, the power spectrum still encodes almost all relevant information, thus we may safely neglect higher-order statistics when measuring $\Lambda$CDM parameters from the CMB. By the redshifts corresponding to current and future galaxy surveys, this is not the case; furthermore, higher-order statistics can carry signatures of \textit{primordial} non-Gaussianity (PNG), which is a key probe of inflationary physics.

The next decade will see the advent of vast large-scale structure (LSS) surveys such as DESI \citep{2016arXiv161100036D} and Euclid \citep{2011arXiv1110.3193L}. Unlike the CMB, current observations are far from being cosmic variance limited, thus the upcoming data releases will lead to significant enhancements in our constraining power on cosmological parameters. To extract maximal information from this treasure trove of information, we must combine two-point statistics with their higher-order variants (including the bispectrum), or apply some transformation of the density field \citep[e.g.,][]{2007ApJ...664..675E}. Such a synergistic approach has been oft-proposed in the literature, and is expected to give notable improvements in $\Lambda$CDM parameter constraints \citep{2017MNRAS.467..928G,2021JCAP...03..021A}, as well as non-standard parameters such as those underlying PNG \citep{2018MNRAS.478.1341K,2021JCAP...05..015M}, the neutrino sector \citep{2019JCAP...11..034C,2021JCAP...04..029H} and modified gravity \citep{2020arXiv201105771A}. 

Whilst there is a long history of estimating PNG parameters from the CMB using higher-order statistics \citep[e.g.,][]{1998MNRAS.299..805H,2000MNRAS.313..141V,2000PhRvD..62j3004G,2003MNRAS.341..623S,2005PhRvD..72d3003B,2006JCAP...05..004C,2009PhRvD..80d3510F,2014A&A...571A..24P}, comparatively few works have attempted to make use of the higher-point functions of the late Universe. There are notable exceptions however, in particular the historical bispectrum analyses of \citep{1982ApJ...259..474F,2001ApJ...546..652S,2005PhRvD..71f3001S}, as well as some more recent works \citep{2015MNRAS.451..539G,2017MNRAS.465.1757G,2017MNRAS.468.1070S,2018MNRAS.478.4500P,2020JCAP...05..005D}, which make use of SDSS data, albeit with several caveats. In configuration-space, the situation is similar. Early 3PCF estimates appear in the mid-1970s \citep[e.g.,][]{1975ApJ...196....1P,2001ASPC..252..201P,1998ApJ...503...37J,2004PASJ...56..415K,2006MNRAS.368.1507N}, and there has been a recent resurgence of interest, proving the 3PCF to be a useful statistic in its own right \citep{2011ApJ...737...97M,2015MNRAS.449L..95G,2017MNRAS.469.1738S,2018MNRAS.474.2109S}. Such approaches are now being extended to even higher-point functions, such as the 4PCF \citep{npcf_algo,4pcf_boss}, though we caution that configuration-space statistics are generically more difficult to model.

Performing robust inference using the galaxy bispectrum does not come without challenges. Firstly, the statistic is non-trivial to measure. Costs associated with its computation are significantly larger than those of the power spectrum, though a number of efficient estimators have been recently been developed \citep[e.g.,][]{2017MNRAS.472.2436W,2015PhRvD..92h3532S,2017JCAP...12..020R,2020MNRAS.492.1214P,philcox_fastfft}. Theoretical modeling is similarly difficult, despite being the subject of many works across a number of decades \citep[e.g.,][]{1998ApJ...496..586S,1997MNRAS.290..651M,1998MNRAS.300..747V,1998ApJ...496..586S,1999ApJ...517..531S,2001MNRAS.325.1312S,2000ApJ...544..597S,2015JCAP...05..007B,2015JCAP...10..039A}. In particular, whilst the matter bispectrum can be well-modelled perturbatively including its first-order corrections (one-loop) \citep{2015JCAP...10..039A}, we still lack a one-loop theory model for the \textit{galaxy} bispectrum, limiting any analysis to comparatively large scales. Many of the current tree-level models, whilst capable of modeling the bispectrum alone, are not self-consistent, and will introduce systematic errors if they are combined with the power spectrum in future high-precision analyses. An important goal of future work must be to develop and test bispectrum theory models, including via blind challenges analogous to \citep{2020PhRvD.102l3541N} for the power spectrum.

An additional difficulty in bispectrum analyses concerns dimensionality. With the power spectrum, the number of bins in the data-vector is usually far smaller than the number of available simulations, facilitating straightforward mock-based Gaussian analyses. In the bispectrum, this is rarely the case, which has led to some analyses opting to reduce the dimensionality by increasing the bin-width, potentially losing cosmological information \citep{2017MNRAS.465.1757G}. A number of low-dimensional analogs of the bispectrum have been proposed to counter this, including the skew-spectrum \citep{2001PhRvD..64d3516C,1998astro.ph.12271M,2015PhRvD..91d3530S}, line-correlation function \citep{2013ApJ...762..115O,2015MNRAS.453..797E}, integrated bispectrum \citep{2014JCAP...05..048C,2015JCAP...09..028C} and modal projections \citep{2012PhRvD..86f3511F,2012PhRvD..86l3524R,2013PhRvD..88f3512S,2021JCAP...03..105B}, though some lead to significant loss of information \citep{2021JCAP...03..105B}. An alternative route is to compress the sample and model bispectra, significantly decreasing the dimensionality whilst preserving the main information content \citep{2000ApJ...544..597S,2000MNRAS.317..965H,2021JCAP...03..105B,2018MNRAS.476.4045G,2019MNRAS.484.3713G,2019MNRAS.484L..29G,2021PhRvD.103d3508P,2018MNRAS.476L..60A}. Previous work has shown this to be a powerful option, capable of reducing the size of the bispectrum data-vector to $\sim$\,$30$ numbers for a DESI-like survey, without appreciable loss of constraining power \citep{2021PhRvD.103d3508P}.

Finally, the situation is complicated by the effects of non-uniform survey geometry. In practice, one does not have access to the galaxy overdensity field; rather, we have only the un-normalized density of galaxies and random particles. This leads to the following transformation in real- and Fourier-space:
\beq
    \delta(\vx) &\to& \delta_W(\vx) \equiv W(\vx)\delta(\vx), \quad 
    \delta(\vk) \to \delta_W(\vk)\equiv\int\frac{d\vp}{(2\pi)^3}W(\vk-\vp)\delta(\vp),
\eeq
where $\delta$ and $\delta_W$ are the true and windowed density fields, and $W(\vx)$ is the survey mask. In Fourier-space (where theory models are most naturally formulated), the action of the window is that of a convolution. This applies similarly to the bispectrum itself:
\beq\label{eq: bk-win}
    B(\vk_1,\vk_2,\vk_3) &\to& B_W(\vk_1,\vk_2,\vk_3)\equiv\int\frac{d\vp_1}{(2\pi)^3}\frac{d\vp_2}{(2\pi)^3}\frac{d\vp_3}{(2\pi)^3}W(\vk_1-\vp_1)W(\vk_2-\vp_2)W(\vk_3-\vp_3)B(\vp_1,\vp_2,\vp_3),
\eeq
where $B$ is the true bispectrum (predicted by theory), and $B_W$ its windowed equivalent, which is the quantity computed by most estimators. To compare data and theory, we have two choices: (a) convolve the theory model with the window via \eqref{eq: bk-win}, or (b) estimate the unwindowed bispectrum, $B$, directly. Traditionally, approach (a) is chosen, yet this is non-trivial. Unlike for the power spectrum, it is highly-expensive to perform the window convolution in full (though see \citep{2019MNRAS.484..364S} for an approach involving double Hankel transforms), especially considering that this must be done at each step in the eventual Markov chain Monte Carlo (MCMC) analysis. Simplifying assumptions are usually adopted; for example \citep{2015MNRAS.451..539G,2017MNRAS.465.1757G,2020JCAP...05..005D} applied the window only to the two power spectra appearing in the tree-level bispectrum model. This assumption is uncontrolled and unwarranted; indeed, \citep{2020JCAP...05..005D} discarded any bispectrum modes with $k_\mathrm{min}\leq 0.04\hMpc$ for this reason. Whilst the current size of survey error-bars permit such approximations on all but the largest scales, this will soon change.

\vskip 4pt 

In this work, we construct estimators for the \textit{unwindowed} bispectrum. Measuring such a quantity allows data and theory to be directly compared without the need for window-convolution, significantly simplifying the eventual MCMC analysis. Our approach is analogous to \citep{2021PhRvD.103j3504P}, which constructed unwindowed estimators for the galaxy power spectrum (see also \citep{2021arXiv210606324B} and the historical approaches of \citep{1997PhRvD..55.5895T,1998ApJ...499..555T,1997ApJ...480...22T,1998PhRvD..57.2117B,1999ApJ...510..551O,2002MNRAS.335..887T,2002ApJ...571..191T,2005astro.ph..3603H,2005astro.ph..3604H,2004ApJ...606..702T}). By optimizing the (non-Gaussian) likelihood for the pixelized galaxy field itself, we obtain a bispectrum estimator that is unbiased and (under certain assumptions) minimum-variance, forward-modeling the effects of survey geometry. This is similar to the CMB bispectrum estimators of \citep{2011MNRAS.417....2S}. Below, we derive the estimators in full and consider their practical implementation, both for the binned bispectrum, and an arbitrary (separable) basis decomposition. Contrary to that claimed in \citep{2021PhRvD.103j3504P}, the estimators do not have dependence on external simulations; rather, they can be formulated in a manner requiring only the data and knowledge of the survey-geometry. This approach will enable a future measurement of the bispectrum of SDSS galaxies (following \citep{2020JCAP...05..042I} for the power spectrum), including a robust treatment of the window function and a perturbative theory model encapsulating all effects relevant on quasi-linear scales. Although we focus on the rotation-averaged bispectrum monopole in this work, the higher-order anisotropic moments are known to be a useful source of cosmological information \citep{2019MNRAS.484..364S,2020MNRAS.497.1684S,2020JCAP...06..041G,2021arXiv210403976G}. Our approach may be similarly applied to this scenario, and further, can be used to extract the information from the field directly, without the need for binning or multipole decompositions, if a suitable basis can be found.  

\vskip 4pt 

The remainder of this paper is structured as follows. In \S\ref{sec: previous-work} we give an overview of our power spectrum estimator (building upon \citep{2021PhRvD.103j3504P}), laying the necessary groundwork for \S\ref{sec: estimator-derivation}, wherein our bispectrum estimator is derived. In \S\ref{sec: spectro-specialization}, we discuss specialization to spectroscopic surveys, before considering practical implementation of the algorithm in \S\ref{sec: implementation}. Bispectra of realistic mock surveys are presented in \S\ref{sec: application}, before we conclude in \S\ref{sec: summary}. Appendices \ref{appen: pixel-cov}\,\&\,\ref{appen: limits} provide details of our modelling of pixelation effects and limiting forms of the bispectrum estimators respectively. A \textsc{Python} implementation of our algorithm can be found on GitHub.\footnote{\href{https://github.com/oliverphilcox/BOSS-Without-Windows}{github.com/oliverphilcox/BOSS-Without-Windows}}.

\section{Overview of Previous Work}\label{sec: previous-work}
We begin with a brief summary of the optimal power spectrum estimators discussed in \citep{2021PhRvD.103j3504P} (itself building on \cite{1998ApJ...499..555T,1998PhRvD..57.2117B,1999ApJ...510..551O,2004ApJ...606..702T,2005astro.ph..3603H,2005astro.ph..3604H}), serving as an introduction to our bispectrum estimators. The treatment below is an updated version of the previous work, and removes the need for a suite of simulations.

\subsection{Quadratic Estimators}
Consider a vector $\vd$ of observations, for example the unnormalized galaxy overdensity measured by some survey in a set of pixels. Usually, this can be modelled as a sum of two components: a theory model $\vm$ (\textit{i.e.}\ the underlying galaxy density field) and a stochastic noise contribution $\vn$ (\textit{i.e.}\ Poisson noise), with covariances $\Sig \equiv \av{\vm\vm^T}$ and $\N \equiv \av{\vn\vn^T}$ respectively, assuming $\av{\vm}=\av{\vn} = 0$.\footnote{In this work, angle brackets represent an average over realizations of the data-set and the underlying random fields.} Since both $\vm$ and $\vn$ are random fields, the dependence on the underlying physical parameters, $\vp$, enters only through the moments of the data-set, thus it is useful to first write down the likelihood of the data, $L[\vd]$. Under Gaussian assumptions, the negative log-likelihood is given by
\beq\label{eq: neg-log-like-Gaussian}
    \ell_G[\vd](\vp) = -\log L_G[\vd](\vp) = \frac{1}{2}\vd^T\Ci(\vp)\vd + \frac{1}{2}\mathrm{Tr}\log\C(\vp) + \text{const.},
\eeq
where $\C(\vp)$ is the pixel covariance matrix, such that $\C(\vp^\mathrm{true})=\av{\vd\vd^T}\equiv \C_D$ where $\vp^\mathrm{true}$ is the parameter set that generates $\vd$. In this section, $\vp$ is the set of \textit{power spectrum amplitudes}, according to some binning. The assumption of Gaussianity holds if the number of modes is large and the underlying density field is linear. Violation of these constraints is the subject of \S\ref{sec: estimator-derivation}.

To obtain an estimate for the parameter vector $\vp$, we need simply extremize the likelihood \eqref{eq: neg-log-like-Gaussian}. In general, this is non-trivial, but can be performed straightforwardly if one first expands $\ell_G(\vp)$ around some fiducial power spectrum $\vp^\fid$:
\beq
    \ell_G[\vd](\vp^\fid+\delta\vp) \approx \ell_G(\vp^\fid) + \delta \vp^T\nabla_{\vp}\ell_G + \frac{1}{2}\delta\vp^T\left(\nabla_{\vp}\nabla_{\vp'}\ell_G\right)\delta\vp',
\eeq
writing $\delta\vp = \vp-\vp^\fid$ and assuming all gradients to be evaluated at $\vp^\fid$. If one assumes $\vp$ to be close to $\vp^\fid$ (such that $\C(\vp^\fid)\approx \C_D$), this leads to the following Newton-Raphson estimate:
\beq
    \hat\vp \approx \vp^\fid -\left[\nabla_{\vp}\nabla_{\vp'}\right]^{-1}\nabla_{\vp'}\ell_G.
\eeq
Inserting the gradients, we obtain the maximum-likelihood (ML) power spectrum estimator 
\begin{empheq}[box=\fbox]{align}\label{eq: ml-pk-estimator}
    \hat{p}^\mathrm{ML}_\alpha &= p_\alpha^\fid + \frac{1}{2}\sum_\beta F_{\alpha\beta}^{-1,\rm ML}\Tr{\Ci\C_{,\beta}\Ci\left(\vd\vd^T-\C\right)}\\\nonumber
    &=\frac{1}{2}\sum_\beta F_{\alpha\beta}^{-1, \rm ML}\Tr{\Ci\C_{,\beta}\Ci\left(\vd\vd^T-\N\right)},
\end{empheq}
where $\vp \equiv \{p_\alpha\}$, $\C_{,\alpha} \equiv \partial \C/\partial p_\alpha$, and we define the (realization-averaged) Fisher matrix
\beq
    F_{\alpha\beta}^{\rm ML} &=& \frac{1}{2}\Tr{\Ci\C_{,\alpha}\Ci\C_{,\beta}}.
\eeq
In the above, we assume all quantities to be evaluated at the fiducial cosmology, \textit{i.e.}\ $\C \equiv \C(\vp^\fid)$. The estimator is quadratic in the data, and involves applying a filter $\Ci\C_{,\alpha}\Ci$ to two copies of $\vd$ then removing additive and multiplicative bias terms. As shown in \citep{2021PhRvD.103j3504P}, it is unbiased (assuming $\C(\vp^\fid)$ can be robustly computed), and minimum-variance if (a) the likelihood is Gaussian and (b) $\C(\vp^\fid) = \C(\vp^\mathrm{true})$.\footnote{Note that the degree of non-optimality is quadratic in $\C(\vp^\fid)-\C_D$, and thus expected to be small.} Extensions to include non-Gaussian noise were also considered in the former work, resulting in a cubic correction term.

Whilst \eqref{eq: ml-pk-estimator} is the optimal estimator (subject to the above caveats), it is just a special case of a more general quadratic estimator, given by
\beq\label{eq: general-qe}
    \hat{p}_\alpha &=& p_\alpha^\fid + \frac{1}{2}\sum_\beta F^{-1}_{\alpha\beta}\Tr{\Hi\C_{,\beta}\Hi\left(\vd\vd^T-\C\right)}\\\nonumber
    &=& \frac{1}{2}\sum_\beta F^{-1}_{\alpha\beta}\Tr{\Hi\C_{,\beta}\Hi\left(\vd\vd^T-\N\right)}\\\nonumber
    F_{\alpha\beta} &=& \frac{1}{2}\Tr{\Hi\C_{,\alpha}\Hi\C_{,\beta}},
\eeq
where $\H$ is a positive-definite weighting matrix, and $\C_{,\alpha}$ is again the derivative of the full covariance with respect to the parameters of interest. The ML estimator simply requires the data to be inverse-covariance weighted, \textit{i.e.}\ $\Hi = \Ci$. Given that the full $\Ci$ matrix is often expensive to compute, \eqref{eq: general-qe} can lead to significant expedition (though a slight loss of optimality) if some straightforwardly invertible weighting matrix $\H$ can be found that is close to $\C_D$.

Two forms of the power spectrum estimators are given in \eqref{eq: ml-pk-estimator}\,\&\,\eqref{eq: general-qe}. Previous work \citep{2021PhRvD.103j3504P} focused on the first choice, which explicitly requires a fiducial model, $\vp^{\rm fid}$, for the power spectrum parameters. Both this and the pixel \resub{covariance} $\C$ can be defined using a suite of simulations, thus the equations take the form of difference estimators, which are unbiased by construction if the data and simulations share the same parameters. However, the reliance on a suite of simulations is a practical restriction on the estimator's use. For this reason, we adopt the second form of the estimators in this work; these do not require mocks and significantly reduce the dependence on some fiducial model (with the general quadratic estimator obviating it entirely, assuming $\C(\vp)$ to be linear in $\vp$).

\subsection{Power Spectrum Estimation}\label{subsec: previous-pk}

The principal utility of the above estimators is to measure the galaxy power spectrum multipoles binned in wavenumber (hereafter the `band-powers'). This is convenient since, neglecting non-Gaussianity of the noise, the band-powers enter only in the two-point covariance of the signal. To implement the estimators, we must first specify the covariance matrix and its dependence on $\vp$. Following \citep{2021PhRvD.103j3504P}, we assume our data-set to be the measured overdensity of galaxy survey, \textit{i.e.}\ $d(\vr) = \hat n_g(\vr)-\hat n_r(\vr)$, where $\hat n_g(\vr)$ and $\hat n_r(\vr)$ are discretely-sampled galaxy and random density fields at position $\vr$. Ignoring weights and pixelation for simplicity, these have the pairwise expectations
\beq\label{eq: random-field-expectations}
    \av{\hat n_g(\vr)\hat n_g(\vr')} &=& n(\vr)n(\vr')[1+\xi(\vr,\vr')]+n(\vr)\delta_\mathrm{D}(\vr-\vr')\\\nonumber
    \av{\hat n_r(\vr)\hat n_r(\vr')} &=& n(\vr)n(\vr')+n(\vr)\delta_\mathrm{D}(\vr-\vr')\\\nonumber
    \av{\hat n_g(\vr)\hat n_r(\vr')} &=& n(\vr)n(\vr'),
\eeq
where $n(\vr)$ is the background number density of the survey and $\xi(\vr,\vr') = \av{\delta(\vr)\delta(\vr')}$ is the two-point correlation function of the galaxy overdensity field. The Dirac functions, $\delta_\mathrm{D}$, arise due to the discrete nature of the density fields and source the Poisson noise term. From this, we obtain the signal and noise covariances between two points, $\vr$ and $\vr'$:
\beq\label{eq: simple-cov}
    \C(\vr,\vr') &\equiv& \Sig(\vr,\vr') + \N(\vr,\vr')\\\nonumber
    \Sig(\vr,\vr') &\approx& n(\vr)n(\vr')\int_{\vk}e^{i\vk\cdot(\vr-\vr')}P(\vk)\\\nonumber
    \N(\vr,\vr') &\approx& n(\vr)\delta_\mathrm{D}(\vr-\vr'),
\eeq
where $P(\vk)$ is the galaxy power spectrum and $\int_{\vk}\equiv \int\frac{d\vk}{(2\pi)^3}$. If we wish to measure the monopole power spectrum in some bin $\alpha$, the relevant covariance derivative becomes
\beq
    \C_{,\alpha}(\vr,\vr') &\approx& n(\vr)n(\vr')\int_{\vk}e^{i\vk\cdot(\vr-\vr')}\Theta^\alpha(k),
\eeq
where we have written $P(\vk) \approx \sum_\alpha p_\alpha \Theta^\alpha(k)$, introducing the binning function $\Theta^\alpha(k)$, which is unity in bin $\alpha$ and zero else. In reality, the situation is complicated by the effects of pixelation, particle weights, and redshift-space distortions. These can be straightforwardly included, and the full forms for $\C$ are presented in Appendix \ref{appen: pixelized-2pt}, including a more advanced treatment of pixelation than \citep{2021PhRvD.103j3504P}.

Given the above covariances, one may compute estimates for the band-powers $\{p_\alpha\}$ using either the ML estimator \eqref{eq: ml-pk-estimator} or the general form \eqref{eq: general-qe}. This requires application of both $\C_{,\alpha}$ and $\Ci$ or $\Hi$ to the data $\vd$. Since the pixelized density fields are usually of high-dimension, it is impractical to store any of the covariance matrices in full, and infeasible to invert them; we can avoid this by considering only the matrices' action on a pixelized field (which can be straightforwardly computed using Fast Fourier Transforms), and using conjugate-gradient descent methods to invert $\C$ \citep{1999ApJ...510..551O}. On small scales, a useful approximation to $\Ci$ is provided by
\beq\label{eq: fkp-weights}
    \Ci(\vr,\vr') \approx \Hi_\mathrm{FKP} &\equiv& \frac{\delta_\mathrm{D}(\vr-\vr')}{n(\vr)\left[1+n(\vr)P_\fkp\right]},
\eeq
with $P_\fkp\sim 10^4h^{-3}\mathrm{Mpc}^{-3}$. This is analogous to the well-known FKP weighting scheme of \citep{1994ApJ...426...23F}, and found to be a useful approximation in \citep{2021PhRvD.103j3504P}, since it does not require numerical matrix inversion. An analogous form for this including pixelation effects and particle weights is given in \eqref{eq: fkp-with-pix}.% Appendix \ref{appen: pixelized-2pt}. 

Finally, the quadratic estimators require the traces
\beq
    \Tr{\Hi\C_{,\alpha}\Hi\N}, \quad \Tr{\Hi\C_{,\alpha}\Hi\C_{,\beta}}
\eeq
(using the form of \eqref{eq: general-qe} which does not involve a fiducial spectrum). Given that the matrices are too large to be directly computed, these may seem difficult to obtain. However, as in \citep{1999ApJ...510..551O,2011MNRAS.417....2S}, they may be computed via Monte Carlo methods, first writing
\beq
    \Tr{\Hi\C_{,\alpha}\Hi\C} = \av{\vec{a}^T\Hi\C_{,\alpha}\Hi\C\mathsf{A}^{-1}\vec{a}}, \quad \Tr{\Hi\C_{,\alpha}\Hi\C_{,\beta}} = \av{\vec{a}^T\Hi\C_{,\alpha}\Hi\C_{,\beta}\mathsf{A}^{-1}\vec{a}},
\eeq
where $\{\vec{a}\}$ are a set of simulated maps that satisfy $\av{\vec{a}\vec{a}^T} = \mathsf{A}$, for some covariance matrix $\mathsf{A}$. Whilst one could set $\mathsf{A} = \mathsf{C}$ and use mock catalogs for this purpose (as proposed in \citep{2021PhRvD.103j3504P}), this requires perfect knowledge of $\mathsf{C}$ in the mock cosmology, which may be difficult to obtain. In this work, we will use uniformly distributed particles for this purpose, which have a simple, and invertible, form for $\mathsf{A}$ given in \eqref{eq: unif-covariance-A}\,\&\,\eqref{eq: unif-inv-covariance-Ainv}. Computing the traces in this way is efficient, and incurs a Monte Carlo error scaling as $\sqrt{1+1/N_\mathrm{mc}}$ when using $N_\mathrm{mc}$ maps. %In fact it is \textit{not} required to have simulations with $\av{\vm\vm^T} = \C$; we can in fact use any set of random maps with a known and calculable covariance. This is discussed in \S\ref{sec: implementation}

\subsection{Properties}
Before continuing, we note a number of important properties of the optimal power spectrum estimator. 
\begin{itemize}
    \item \textbf{Window Function}: Unlike the standard power spectrum estimator of \citep{1994ApJ...426...23F} (and later \citep{2006PASJ...58...93Y,2017JCAP...07..002H}), the quadratic estimators measure the \textit{unwindowed} power spectrum, \textit{i.e.}\ the output is not convolved with the survey window function. This occurs since the effects of survey-geometry are forward-modelled through the covariance matrix, and allows us to compare measured and theoretical power spectra directly. When using the FKP weights of \eqref{eq: fkp-weights}, our approach is equivalent to that of \citep{2021arXiv210606324B}.
    \item \textbf{Optimality}: Assuming Gaussianity, the quadratic estimator with $\H = \C$ achieves the tightest possible constraints on the band-powers, in the sense that it saturates the Cram\'er-Rao limit. This will be particularly notable for large-scale analyses, such as those constraining primordial non-Gaussianity \citep[e.g.,][]{2019JCAP...09..010C}.\footnote{There is a important caveat to this statement, as noted in \citep{2021arXiv210606324B}. If one uses the same maximum wavenumber, $k_\mathrm{max}$, for both unwindowed and windowed power spectrum estimates, the signal-to-noise of the windowed estimates will be generically slightly larger. This occurs since the window function mixes in modes of larger $k$, which have smaller variance. If one restricts to the same \textit{pre-convolved} $k$-modes in both cases, the quadratic estimator will be at least as constraining as the windowed FKP approach.} 
    \item \textbf{Compression}: As shown in \citep{2021PhRvD.103j3504P}, quadratic estimators can be used to directly measure the coefficients of the power spectrum under some linear compression scheme. This significantly reduces dimensionality, and obviates the need for $k$-space binning.
    \item \textbf{Gridding and Shot-Noise}: The estimators may be formulated as the difference between a quantity measured in simulations and data (as in the first line of \ref{eq: ml-pk-estimator}\,\&\,\ref{eq: general-qe}). This removes the leading-order effects of unmeasured modes, discretization and non-Poissonian shot-noise, facilitating the use of coarser pixelation grids. Alternatively, the estimators can be constructed without this (as in the second line of \eqref{eq: ml-pk-estimator}\,\&\,\eqref{eq: general-qe}, which will be generally assumed here), which removes the need for a fiducial cosmology, and thus a suite of simulations.
    \item \textbf{Integral Constraint}: When analyzing the output from conventional (windowed) power spectrum estimators, we must account for the integral constraint, \textit{i.e.}\ the fact that the overall survey density is not known. In the quadratic estimator framework, such effects are shunted into the fundamental $k$-mode thus may be ignored if this mode is not analyzed.
\end{itemize}
We refer the reader to \citep{2021PhRvD.103j3504P} for a more in-depth discussion of these effects, as well as the estimator's application to the measurement of power spectrum multipoles from a suite of simulations.

% \begin{itemize}
%     \item Power spectrum likelihood, and its optimization
%     \item Rough form of the pixel covariance matrix
%     \item Quadratic estimator and assumptions
%     \item Form without difference estimator and with uniform random simulations.
%     \item Conclusions from \citep{2021PhRvD.103j3504P}.
%     \item Note lack of integral constraints and $k_\mathrm{max}$ vs FKP form.
% \end{itemize}

\section{The Cubic Bispectrum Estimator}\label{sec: estimator-derivation}
A maximum-likelihood estimator for the bispectrum may be derived in an analogous manner to the above. Before doing so, we outline a number of reasons why this is of use.
\begin{itemize}
    \item Whilst the window function can be straightforwardly included in the theory model for the power spectrum via 1D Hankel transforms, this is considerably more difficult for the bispectrum, since it requires a six-dimensional convolution integral. Whilst some procedures do exist \citep[e.g.,][]{2019MNRAS.484..364S}, they are computationally expensive, prohibiting full parameter exploration.
    \item The maximum-likelihood estimator derived below is optimal (in the Cram\'er-Rao sense) in the limit of weak non-Gaussianity.  This allows us to extract maximal information from the bispectrum, effectively giving an increase in survey volume.
    \item We do not require a fiducial model for the bispectrum to use the cubic estimator, just a fiducial power spectrum model (cf.\,\S\ref{sec: previous-work}), if $\Ci$ weights are assumed. This is useful since obtaining an accurate fiducial bispectrum model, and simulations which reproduce it, is difficult. Furthermore, our estimator is standalone; \textit{i.e.}\ it does not require a suite of realistic simulations.
    \item Bispectra are generally high-dimensional, often involving measurements from $\gtrsim 10^3$ triangles. This requires a large number of mocks to compute a sample covariance, which has limited some previous analyses \citep{2017MNRAS.465.1757G}. Using analogous techniques to those developed for the power spectrum in \citep{2021PhRvD.103j3504P}, we may avoid this by directly estimating a set of basis coefficients rather than the full bispectrum, in a similar vein to \citep{2021JCAP...03..105B}.
    \item To compare theoretical and observed bispectra, the theory model should properly be integrated over the finite $k$-bins. This is expensive in practice, and can be avoided by directly estimating basis coefficients, which are defined from the unbinned bispectra.
\end{itemize}

Below, we consider the derivation of the general bispectrum estimator, before specializing to the case of spectroscopic surveys in \S\ref{sec: spectro-specialization}. Part of the below parallels a similar derivation for the CMB bispectrum in \cite{2005PhRvD..72d3003B}, and we will adopt several of the associated tricks used in \citep{2011MNRAS.417....2S}. 

\subsection{Non-Gaussian Likelihood}
To derive constraints on bispectrum parameters, $\vb$, (which could be the binned bispectrum estimates or some other summary statistic), we first require a likelihood for the data which contains them. Since such parameters are necessarily absent in the covariance, we require the non-Gaussian likelihood, which, in the limit of mild non-Gaussianity, is given by the Edgeworth expansion:
\beq\label{eq: edgeworth-expansion}
    L[\vd](\vb) &=& L_G[\vd]\left[1 + \frac{1}{3!}\mathsf{B}^{ijk}\mathcal{H}_{ijk} + \frac{1}{4!}\mathsf{T}^{ijkl}\mathcal{H}_{ijkl} + \frac{1}{6!}\left(\mathsf{B}^{ijk}\mathsf{B}^{lmn}+\mathrm{9\,perms.}\right)\mathcal{H}_{ijklmn}+...\right]
\eeq
\citep[e.g.,][]{2005PhRvD..72d3003B,2017arXiv170903452S}. Here, we have written the Gaussian likelihood of \eqref{eq: neg-log-like-Gaussian} as $L_G$, which is independent of $\vb$, and denoted the connected three- and four-point expectations of the density field by
\beq
    \B^{ijk} = \av{d^id^jd^k}, \quad \mathsf{T}^{ijkl} = \av{d^id^jd^kd^l} - \left[\av{d^id^j}\av{d^kd^l} + \text{2 perms.}\right].
\eeq
We use Latin indices to denote pixels in the dataset, such that $d^i\equiv d(\vr_i)$, and assume Einstein summation over repeated indices. The Hermite tensors, $\mathcal{H}$, used in \eqref{eq: edgeworth-expansion} may be defined in terms of $h_i \equiv \left[\Ci \vd\right]_i$ by
\beq\label{eq: hermite-def}
    \mathcal{H}_{ijk} &=& h_ih_jh_k - \left[h_i\Ci_{jk} + \mathrm{2\,perms.}\right]\\\nonumber
    \mathcal{H}_{ijkl} &=& h_ih_jh_kh_l - \left[h_ih_j\Ci_{kl} + \mathrm{5\,perms.}\right] + \left[\Ci_{ij}\Ci_{kl} + \mathrm{2\,perms.}\right]\\\nonumber
    \mathcal{H}_{ijklmn} &=& h_ih_jh_kh_lh_mh_n - \left[h_ih_jh_kh_l\Ci_{mn} + \mathrm{14\,perms.}\right]\\\nonumber
    &&\,\quad + \left[h_ih_j\Ci_{kl}\Ci_{mn} + \mathrm{44\,perms.}\right] -\left[\Ci_{ij}\Ci_{kl}\Ci_{mn} + \mathrm{14\,perms.}\right]
\eeq
\citep{2017arXiv170903452S}, where $\Ci_{ij}\equiv \Ci(\vr_i,\vr_j)$. Each tensor is symmetric under any permutation of its arguments, and all involve the data weighted by the inverse covariance, just as in \S\ref{sec: previous-work}. The likelihood \eqref{eq: edgeworth-expansion} is an expansion in non-Gaussianity, and thus valid if its effects are small. In practice, this can be ensured by restricting the analysis to relatively large scales. 

From the above, we may construct a negative log-likelihood for the data as
\beq\label{eq: edgeworth-log-expansion}
    \ell[\vd](\vb) &=& \ell_G[\vd](\vb) - \frac{1}{3!}\mathsf{B}^{ijk}\mathcal{H}_{ijk} - \frac{1}{4!}\mathsf{T}^{ijkl}\mathcal{H}_{ijkl} + \frac{1}{72}\mathsf{B}^{ijk}\mathsf{B}^{lmn}\left[\mathcal{H}_{ijk}\mathcal{H}_{lmn}-\mathcal{H}_{ijklmn}\right]+ \mathcal{O}(\B^3),
\eeq
absorbing the 10 permutations of $\B^{ijk}\B^{lmn}$ into the totally symmetric tensor $\mathcal{H}_{ijklmn}$.

\subsection{Cubic Estimators}
To obtain an estimator for the bispectrum parameters $\vb$, we need simply extremize \eqref{eq: edgeworth-log-expansion}. Here, we assume $\vb$ be encapsulated solely in the three-point expectation $\B^{ijk}$, \textit{i.e.}\ we ignore contributions from the higher-order correlators (arising due to Poisson corrections). As for the power spectrum, we first expand the negative log-likelihood in a Taylor series about some fiducial parameter set, here chosen as $\vb^\fid = \vec 0$;
\beq\label{eq: bk-logl}
    \ell[\vd](\vb) = \ell [\vd](\vec 0) + \vb^T\nabla_{\vb}\ell[\vd](\vec 0) + \frac{1}{2}\vb^T\left(\nabla_{\vb}\nabla_{\vb'}\ell[\vd](\vec 0)\right)\vec p' + \ldots.
\eeq
This is simply an expansion in non-Gaussianity, \textit{i.e.}\ the $n$-th order term contains a product of $n$ three-point correlators. Whilst one could instead expand around $\vb^\fid \neq \vec 0$ (and thus obtain a closer-to-optimal estimator if $\vb^\fid$ is well-chosen), this will introduce dependence on a fiducial bispectrum model, and is thus ignored. Maximization of \eqref{eq: bk-logl} leads to the estimator
\beq
    \hat{\vb} = \av{-\left[\nabla_{\vb}\nabla_{\vb'}\ell[\vd](\vec 0)\right]}^{-1}\nabla_{\vb'}\ell[\vd](\vec 0) + \ldots,
\eeq
where we have additionally taken the expectation of the inverse term, as for the power spectrum case. In the limit of vanishing three- and higher-point correlators in the data (\textit{i.e.}\ equivalence of true and fiducial parameters), this is the maximum-likelihood (ML) solution, as we show below. The relevant derivatives are easily obtained from \eqref{eq: edgeworth-log-expansion};
\beq
    \partial_\alpha\ell[\vd](\vec 0) &=& -\frac{1}{6}\B_{,\alpha}^{ijk}\mathcal{H}_{ijk}+...\\\nonumber
    \partial_\alpha\partial_\beta\ell[\vd](\vec 0) &=& -\frac{1}{36}\B_{,\alpha}^{ijk}\B_{,\beta}^{lmn}\left[\mathcal{H}_{ijklmn}-\mathcal{H}_{ijk}\mathcal{H}_{lmn}\right] + \ldots,
\eeq
noting that the Gaussian part, $\ell_G$, and the four-point correlator, $\mathsf{T}^{ijkl}$, are independent of the bispectrum coefficients. We have additionally used that the Hermite tensors are totally symmetric and asserted that the three-point correlator $\B^{ijk}$ is linear in $\vb$ (which is true both for the binned bispectrum and the basis components in some linear compression scheme). The ML bispectrum estimator becomes
\beq\label{eq: optimal-bispectrum-estimator}
    \boxed{\hat{b}^\mathrm{ML}_\alpha = \sum_{\beta} F^{-1,\mathrm{ML}}_{\alpha\beta}\hat{q}^\mathrm{ML}_\beta,}
\eeq
subject to the definitions
\beq
    \hat{q}^\mathrm{ML}_\alpha \equiv \frac{1}{6}\B^{ijk}_{,\alpha}\mathcal{H}_{ijk},\quad&& F^\mathrm{ML}_{\alpha\beta} \equiv \frac{1}{36}\B_{,\alpha}^{ijk}\B_{,\beta}^{lmn}\left[\av{\mathcal{H}_{ijk}\mathcal{H}_{lmn}}-\av{\mathcal{H}_{ijklmn}}\right].
\eeq
This requires the expectation of a Hermite 6-tensor and a pair of 3-tensors (defined in \ref{eq: hermite-def}); following a tedious, but elementary, calculation, these give
\beq\label{eq: Hermite-expectations}
    \av{\mathcal{H}_{ijk}\mathcal{H}_{lmn}} &=& \left(\av{h_ih_jh_k}\av{h_lh_mh_n} + \text{9\,perms.}\right) + \left(\av{h_ih_l}\av{h_jh_m}\av{h_kh_n}+\text{5\,perms.}\right)\\\nonumber
    &=& \left(\Ci_{ii'}\Ci_{jj'}\Ci_{kk'}\mathsf{B}^{i'j'k'}\Ci_{ll'}\Ci_{mm'}\Ci_{nn'}\mathsf{B}^{l'm'n'} + \mathrm{9\,perms.}\right) + \left(\Ci_{il}\Ci_{jm}\Ci_{kn} + \text{5\,perms.}\right)\\\nonumber
    \av{\mathcal{H}_{ijklmn}} &=& \av{h_ih_jh_k}\av{h_lh_mh_n} + \mathrm{9\,perms.}\\\nonumber
    &=& \Ci_{ii'}\Ci_{jj'}\Ci_{kk'}\mathsf{B}^{i'j'k'}\Ci_{ll'}\Ci_{mm'}\Ci_{nn'}\mathsf{B}^{l'm'n'} + \mathrm{9\,perms.},
\eeq
dropping correlators above second-order in $\B$, and noting that the two-point piece of $\av{\mathcal{H}_{ijklmn}}$ vanishes by orthonomality (Hermite tensors are Gaussian-orthonormal with the zeroth order tensor being unity). This leads to the simplified forms:
\beq\label{eq: q-alpha-def}
    \boxed{\hat{q}^\mathrm{ML}_\alpha = \frac{1}{6}\B_{,\alpha}^{ijk}\left[\Ci\vd\right]_i\left(\left[\Ci\vd\right]_j\left[\Ci\vd\right]_k - 3\Ci_{jk}\right)}
\eeq
\beq\label{eq: F-ab-def}
    \boxed{F_{\alpha\beta}^\mathrm{ML} = \frac{1}{36}\B_{,\alpha}^{ijk}\B_{,\beta}^{lmn}\left[\Ci_{il}\Ci_{jm}\Ci_{kn} + \mathrm{5\,perms.}\right] = \frac{1}{6}\B_{,\alpha}^{ijk}\B_{,\beta}^{lmn}\Ci_{il}\Ci_{jm}\Ci_{kn},}
\eeq
assuming $\B^{ijk}$ to be symmetric under $\{i,j,k\}$ permutations and working to lowest non-trivial order in $\B$. This matches that found in, for example, \citep{2011MNRAS.417....2S,2012PhRvD..86f3511F}.

Much like the power spectrum estimator of \eqref{eq: ml-pk-estimator} involved a filter $\C_{,\alpha}$ applied to two copies of the inverse-covariance-weighted data, $\Ci\vd$, the bispectrum estimator of \eqref{eq: optimal-bispectrum-estimator} involves a filter $\B_{,\alpha}$ applied to three copies of $\Ci\vd$. In both cases, the Fisher matrix is a trace over two copies of the filter, weighted by some number of inverse covariance matrices. However, the bispectrum Fisher matrix does \textit{not} depend on any bispectrum amplitudes, unlike for the optimal power spectrum estimator. This occurs since we have assumed the non-Gaussianity to be weak, fixing the fiducial bispectrum to zero. This choice does not lead to a bias in the estimator, though it will give a slight loss of optimality if the true bispectrum is large. We further note the subtraction of a term involving $[\Ci\vd]_i \Ci_{jk}$ in \eqref{eq: q-alpha-def}; this averages to zero and is not included in the simple estimators of \citep{2005PhRvD..71f3001S,2015MNRAS.451..539G,2017MNRAS.465.1757G}, though \citep{2012PhRvD..86f3511F} note it to be important. In the limit of uniform survey density, this contributes only to the $\vk = \vec 0$ mode (Appendix \ref{appen: limits}).

As for the power spectrum case, the form of \eqref{eq: optimal-bispectrum-estimator} motivates a more general cubic bispectrum estimator:
\beq\label{eq: general-bispecturm-estimator}
    \hat{b}_\alpha &=& \sum_\beta F_{\alpha\beta}^{-1}\hat{q}_\beta\\\nonumber
    \hat{q}_\alpha &=& \frac{1}{6}\B_{,\alpha}^{ijk}\left[\Hi\vd\right]_i\left(\left[\Hi\vd\right]_j\left[\Hi\vd\right]_k - 3\Hi_{jk}\right), \qquad     F_{\alpha\beta} = \frac{1}{6}\B_{,\alpha}^{ijk}\B_{,\beta}^{lmn}\Hi_{il}\Hi_{jm}\Hi_{kn}
\eeq
for invertible weight matrix $\H$. This approaches the ML solution in the limit $\H\to\C_D$, where $\C$ is the covariance of the data.

\subsection{Estimator Properties}\label{subsec: cubic-estimator-properties}
\paragraph{Bias}
Taking the expectation of $\hat{q}_\alpha$, the bias of the general cubic estimator \eqref{eq: general-bispecturm-estimator} is easily considered:
\beq\label{eq: bk-bias}
    \mathbb{E}[\hat{q}_\alpha] &=& \frac{1}{6}\B_{,\alpha}^{ijk}\Hi_{ii'}\Hi_{jj'}\Hi_{kk'}\left(\av{d_{i'}d_{j'}d_{k'}}-3\Hi_{j'k'}\av{d_{i'}}\right)\\\nonumber
    &=& \frac{1}{6}\Hi_{ii'}\Hi_{jj'}\Hi_{kk'}\B_{,\alpha}^{ijk}\sum_\beta b^\mathrm{true}_\beta \B_{,\beta}^{i'j'k'} = \sum_{\beta}b^\mathrm{true}_\beta F_{\alpha\beta}\\\nonumber
    \Rightarrow \mathbb{E}[\hat{b}_\alpha] &=& b^\mathrm{true}_\alpha,
\eeq
where we have written $[\Hi\vd]_i\equiv \Hi_{ii'}d_{i'}$ and assumed that the bispectrum parameters $\vb$ fully determine $\B$, \textit{i.e.}\ that $\B= \sum_{\alpha}b^\mathrm{true}_\alpha\B_{,\alpha}$. This holds true also for the ML estimator of \eqref{eq: optimal-bispectrum-estimator}, setting $\H = \C$. Note that we ignore any biases from higher-order terms in the likelihood, \textit{i.e.}\ those second order in $\B$. These contribute if $\vb\neq\vec 0$, and could be removed at leading order by including an additional `bias' term in the estimator.

When considering binned bispectrum estimates, relation \eqref{eq: bk-bias} strictly only holds if we measure all possible wavenumber bins, thus a more appropriate expansion is $\B = \sum_{\alpha}b^\mathrm{true}_\alpha \B_{,\alpha} + \Delta \B$, where $\Delta\B$ contains the bispectrum contributions outside the region of interest. This leads to a bias
\beq
    \Delta b_\alpha \equiv \mathbb{E}[\hat b_\alpha]-b^\mathrm{true}_\alpha = \sum_\beta F_{\alpha\beta}^{-1}\,\B_{,\beta}^{ijk}\Hi_{ii'}\Hi_{jj'}\Hi_{kk'}\Delta\B_{i'j'k'}.
\eeq
For optimal weights and a uniform survey of infinite volume, $\Ci_{ii'}\propto \delta^\mathrm{K}_{ii'}$, thus the bias arises from the term $\B_{,\beta}^{ijk}\Delta\B_{ijk}$. This is vanishing for all $\beta$, since $\Delta\B$ contains contributions only from modes not in $\{b_\beta\}$. In more realistic scenarios, this mode-coupling bias is expected to be small, assuming the survey window to be relatively compact. \citep{2021PhRvD.103j3504P} suggested that one should estimate a slightly larger range of wavenumber bins than used in an eventual analysis to ameliorate such a bias; we apply this principle in \S\ref{sec: implementation}.

Special care is required if some bispectrum decomposition scheme is adopted in which there is an additional linear term $\bar{\B}$ not included in the parameter set $\vb$, \textit{i.e.}\ if $\B = \sum_{\alpha}b_\alpha \B_{,\alpha}+\bar{\B}$. An example of this is the singular value decomposition scheme proposed in \citep{2021PhRvD.103d3508P}, which expands the statistic around some fiducial mean. In this case, we must modify the general estimator to ensure it remains unbiased:
\beq
    \bar{b}_\alpha &\rightarrow& \sum_\beta F^{-1}_{\alpha\beta}\left[\hat{q}_\beta - \bar{q}_\beta\right], \quad \bar{q}_\alpha =\frac{1}{6}\Hi_{ii'}\Hi_{jj'}\Hi_{kk'}\B_{,\alpha}^{ijk}\bar{\B}^{i'j'k'}
\eeq
An alternative approach, which is somewhat easier to estimate, is to expand the log-likelihood around the mean $\bar{\B}$ instead of zero; in this instance, we estimate the \textit{difference} between $b_\alpha$ and $\bar b_\alpha$, and may drop the $\bar{q}_\alpha$ term. Assuming $\bar{\B}$ to be a reasonable bispectrum estimate, this will not lead to a significant loss of optimality. 

\paragraph{Covariance}
Assuming the optimal weighting $\H=\C$, the covariance of the bispectrum estimator \eqref{eq: optimal-bispectrum-estimator} can be derived via Wick's theorem;
\beq
    \operatorname{cov}\left(\hat{b}^\mathrm{ML}_\alpha,\hat{b}^\mathrm{ML}_\beta\right) = \sum_{\gamma\delta}F^{-1}_{\alpha\gamma}F^{-1}_{\beta\delta}\operatorname{cov}\left(\hat{q}^\mathrm{ML}_\gamma,\hat{q}^\mathrm{ML}_\delta\right),
\eeq
with 
\beq\label{eq: covariance-bispectrum}
    \operatorname{cov}\left(\hat{q}^\mathrm{ML}_\alpha,\hat{q}^\mathrm{ML}_\beta\right) &\equiv& \mathbb{E}[\hat{q}^\mathrm{ML}_\alpha\hat{q}^\mathrm{ML}_\beta]-\mathbb{E}[\hat{q}^\mathrm{ML}_\alpha]\mathbb{E}[\hat{q}^\mathrm{ML}_\beta] = \frac{1}{36}\B^{ijk}_{,\alpha}\B^{lmn}_{,\beta}\left[\av{\mathcal{H}_{ijk}\mathcal{H}_{lmn}}-\av{\mathcal{H}_{ijk}}\av{\mathcal{H}_{lmn}}\right]\\\nonumber
    &=& \frac{1}{6}\B^{ijk}_{,\alpha}\B^{lmn}_{,\beta}\Ci_{il}\Ci_{jm}\Ci_{kn} + \frac{5}{18}\Ci_{ii'}\Ci_{jj'}\Ci_{kk'}\Ci_{ll'}\Ci_{mm'}\Ci_{nn'}\B^{i'j'k'}\B^{l'm'n'}\B^{ijk}_{,\alpha}\B^{lmn}_{,\beta}.
\eeq
This uses the expectation of two Hermite 3-tensors given in \eqref{eq: Hermite-expectations} and is valid at second order in $\B$. In the limit of small bispectrum, this is equal to the inverse Fisher matrix, such that
\beq\label{eq: optimal-cov}
     \operatorname{cov}\left(\hat{b}^\mathrm{ML}_\alpha,\hat{b}^\mathrm{ML}_\beta\right) = F^{-1,\mathrm{ML}}_{\alpha\beta}+\mathcal{O}(\B^2).
\eeq
The calculation for general weighting $\H$ is analogous but lengthy, and is thus omitted from this publication.

\paragraph{Optimality}
According to the Cram\'er-Rao theorem, for the estimator $\hat{b}^\mathrm{ML}_\alpha$ to be optimal, it must satisfy
\beq\label{eq: Cramer-Rao}
    \left.\operatorname{cov}\left(\hat{b}^\mathrm{ML}_\alpha,\hat{b}^\mathrm{ML}_\beta\right)\right|_\mathrm{CR\,bound} = \av{\frac{\partial^2\ell[\vd](\vb)}{\partial b_\alpha\partial b_\beta}}^{-1}\equiv \mathcal{I}_{\alpha\beta}^{-1},
\eeq
where the right-hand-side is the inverse Fisher information for negative log-likelihood $\ell[\vd](\vb)$ depending on parameters $\vb$ and data $\vd$ \citep[e.g.][]{2009arXiv0906.0664H}. Since our Fisher matrix $F_{\alpha\beta}^\mathrm{ML}$ is defined as the (realization-averaged) second derivative of the log-likelihood, $\mathcal{I}_{\alpha\beta} = F^\mathrm{ML}_{\alpha\beta}$, and \eqref{eq: optimal-cov} demonstrates that the ML estimator satisfies its Cram\'er-Rao bound in the limit $\B\to0$. We thus obtain optimality in the limit of weak non-Gaussianity, assuming inverse covariance weighting and neglecting non-Gaussian contributions from stochastic effects. A corollary of this is that any estimator with $\H\neq \C$ must give larger error-bars, since the covariance of $\hat{b}_\alpha$ will necessarily be different from \eqref{eq: optimal-cov}, and no unbiased estimator can exceed its Cram\'er-Rao bound.

\section{Specialization to Spectroscopic Surveys}\label{sec: spectro-specialization}
We now consider the form of the bispectrum estimators (\ref{eq: optimal-bispectrum-estimator}\,\&\,\ref{eq: general-bispecturm-estimator}) when applied to spectroscopic surveys. We will specialize to the binned isotropic \resub{(\textit{i.e.}\ rotationally-averaged)} bispectrum in \S\ref{subsec: spec-binned}, before commenting on alternative bases in \S\ref{subsec: spec-bases}.

\subsection{General Formalism}\label{subsec: spec-general}
As in \S\ref{sec: previous-work}, we assume the data vector to be a (pixelized) field of data-minus-randoms, such that $d(\vr) \equiv \hat n_g(\vr)-\hat n_r(\vr)$. For clarity, we ignore the effects of pixelation and particle weights in this section; the full estimators including such phenomena are presented in Appendix \ref{appen: pixelized-3pt}.

Assuming Poisson statistics for the two discrete density fields, the two- and three-point correlators are given by
\beq
    \C^{ij} &\equiv& \av{d^id^j} = n(\vr_i)n(\vr_j)\xi(\vr_i, \vr_j) + n(\vr_i)\delta_\mathrm{D}(\vr_i-\vr_j)\\\nonumber
    \B^{ijk} &\equiv& \av{d^id^jd^k} = n(\vr_i)n(\vr_j)n(\vr_k)\zeta(\vr_i,\vr_j,\vr_k) + \left[n(\vr_i)n(\vr_j)\delta_\mathrm{D}(\vr_i-\vr_k)\xi(\vr_i-\vr_j) + \text{2 perms.}\right]\\\nonumber
    &&\,+\,n(\vr_i)\delta_\mathrm{D}(\vr_i-\vr_j)\delta_\mathrm{D}(\vr_j-\vr_k),
\eeq
applying the results of \eqref{eq: random-field-expectations}. This uses the background number density field $n(\vr)$, the 2PCF $\xi(\vr_1,\vr_2) = \av{\delta(\vr_1)\delta(\vr_2)}$ and the 3PCF $\zeta(\vr_1,\vr_2,\vr_3) = \av{\delta(\vr_1,\vr_2,\vr_3)}$, retaining dependence on all position vectors for generality. The terms involving Dirac deltas arise from Poisson contractions of the density field,; these take a slightly different form after incorporating pixelation effects (Appendix \ref{appen: pixel-cov}). Writing the 2PCF and 3PCF in terms of their Fourier-space counterparts, $P(\vk)$ and $B(\vk_1,\vk_2,\vk_3)$, we obtain
\beq\label{eq: Bijk-def-Fourier}
    \C^{ij} &=& n(\vr_i)n(\vr_j)\int_{\vk}P(\vk)e^{i\vk\cdot(\vr_i-\vr_j)} + n(\vr_i)\delta_\mathrm{D}(\vr_i-\vr_j)\\\nonumber
    \B^{ijk} &=& n(\vr_i)n(\vr_j)n(\vr_k)\int_{\vk_1\vk_2\vk_3}B(\vk_1,\vk_2,\vk_3)e^{i\vk_1\cdot\vr_i+i\vk_2\cdot\vr_j+i\vk_3\cdot\vr_k}\delD{\vk_1+\vk_2+\vk_3}\\\nonumber
    &&\,+\,\left[n(\vr_i)n(\vr_j)\delta_\mathrm{D}(\vr_i-\vr_k)\int_{\vk}P(\vk)e^{i\vk\cdot(\vr_i-\vr_j)} + \text{2 perms.}\right] + n(\vr_i)\delta_\mathrm{D}(\vr_i-\vr_j)\delta_\mathrm{D}(\vr_j-\vr_k),
\eeq
where the Dirac delta function $\delD{\vk_1+\vk_2+\vk_3}$ arises from translation invariance. Additionally, we have made the flat-sky approximation, assuming the power spectrum and bispectrum to be independent of the position vectors $\vr$. This is done only for clarity; we can retain dependence of the statistics on the \textit{local} line-of-sight (e.g., in the Yamamoto approximation \citep{2006PASJ...58...93Y}), as done for the power spectrum multipoles in \citep{2021PhRvD.103j3504P}. Full treatment of the three-pixel correlator allowing for this and other effects can be found in \eqref{eq: Bijk-pixel}.

In the estimators of \S\ref{sec: estimator-derivation}, $\B^{ijk}$ does not enter directly; instead we require its derivatives with respect to a set of bispectrum parameters. Here, we assume the gravitational bispectrum $B(\vk_1,\vk_2,\vk_3)$ to be a linear sum of templates, such that
\beq\label{eq: Bk-decomposition}
    B(\vk_1,\vk_2,\vk_3) = \sum_\alpha b_\alpha \omega^\alpha(\vk_1,\vk_2,\vk_3),
\eeq
where $\vb\equiv \{b_\alpha\}$ are the parameters of interest, and $\omega^\alpha(\vk_1,\vk_2,\vk_3)$ are some templates, which are of relevance only if $\vk_1+\vk_2+\vk_3=\vec 0$.\footnote{If necessary, we can introduce an offset such that $B(\vk_1,\vk_2,\vk_3) = \sum b_\alpha \omega^\alpha(\vk_1,\vk_2,\vk_3) + \bar{B}(\vk_1,\vk_2,\vk_3)$. This is of relevance when considering SVD decompositions, and modifies the bispectrum estimator slightly, as discussed in \S\ref{subsec: cubic-estimator-properties}.} This decomposition is fully applicable to the simple case of binned bispectrum estimates as well as more nuanced schemes; the former is discussed in \S\ref{subsec: spec-binned}.

Using \eqref{eq: Bk-decomposition}, the derivatives of \eqref{eq: Bijk-def-Fourier} are straightforward,
\beq\label{eq: cumulant-deriv-applied}
    \B^{ijk}_{,\alpha} &=& n(\vr_i)n(\vr_j)n(\vr_k)\int_{\vk_1\vk_2\vk_3}\omega^\alpha(\vk_1,\vk_2,\vk_3)e^{i\vk_1\cdot\vr_i+i\vk_2\cdot\vr_j+i\vk_3\cdot\vr_k}\delD{\vk_1+\vk_2+\vk_3},
\eeq
leading to the following $\hat{q}_\alpha$ estimator (from \ref{eq: general-bispecturm-estimator});
\beq\label{eq: q-alpha-tmp}
    \hat{q}_\alpha &=& \frac{1}{6}\B_{,\alpha}^{ijk}[\Hi\vd]_i\left([\Hi\vd]_j[\Hi\vd]_k-3\Hi_{jk}\right)\\\nonumber
    &=& \frac{1}{6}\int_{\vk_1\vk_2\vk_3}\omega^\alpha(\vk_1,\vk_2,\vk_3)\ft{n\Hi\vd}(\vk_1)\ft{n\Hi\vd}(\vk_2)\ft{n\Hi\vd}(\vk_3)\delD{\vk_1+\vk_2+\vk_3}\\\nonumber
    &&\,-\frac{1}{2}\int_{\vk_1\vk_2\vk_3}\omega^\alpha(\vk_1,\vk_2,\vk_3)\ft{n\Hi\vd}(\vk_1)\av{\ft{n\Hi\va}(\vk_2)\ft{n\Ai\va}(\vk_3)}\delD{\vk_1+\vk_2+\vk_3}.
\eeq
In the above, we have performed the spatial integrals (which are simply Fourier transforms, denoted by $\mathcal{F}$), and, in the second line, written $\Hi_{jk} \equiv \Hi_{jj'}\A^{j'k'}\Ai_{k'k} \equiv \av{\Hi_{jj'}a^{j'}\Ai_{kk'}a^{k'}}$, where $\{\vec a\}$ are a set of simulated maps with known covariance $\A$, \textit{i.e.}\ $\av{\vec a\,\vec a^T} = \A$. This is similar to \cite{2011MNRAS.417....2S} and is the same trick used to evaluate the ML power spectrum bias term in \S\ref{sec: previous-work}. Unlike in previous work, we do not enforce $\A = \H$; instead we use uniformly distributed particles for this purpose (as in \S\ref{sec: previous-work}), removing the dependence on $N$-body simulations. The corresponding form for $\Ai$ (including pixelation effects) is given in \eqref{eq: unif-inv-covariance-Ainv}.
%. When adopting the ML weights, $\H = \C$, we can use mock simulations to compute this term, as in \citep{2012PhRvD..86f3511F}.

In practice, the above expression is difficult to compute due to the momentum-conserving delta-function. A useful assumption is that the filter functions, $\omega^\alpha$, are separable, such that
\beq\label{eq: bk-separability}
    \omega^\alpha(\vk_1,\vk_2,\vk_3) = \prod_{i=1}^3\omega^{\alpha,i}(\vk_i).
\eeq
This is the case for the binned bispectrum estimates considered below. Writing the Dirac function as the integral of a complex exponential, this yields a straightforward form for the $\hat{q}_\alpha$ estimator:
\beq\label{eq: q-alpha-applied-separable}
    \hat{q}_\alpha &=& \frac{1}{6}\int d\vr\,g^{\alpha,1}[\vd](\vr)g^{\alpha,2}[\vd](\vr)g^{\alpha,3}[\vd](\vr)\\\nonumber
    &&\,-\frac{1}{6}\int d\vr\,\left[g^{\alpha,1}[\vd](\vr)\av{g^{\alpha,2}[\va](\vr)\tilde{g}^{\alpha,3}[\va](\vr)}+\text{2\,perms.}\right],
\eeq
where we have defined the functions
\beq\label{eq: g-alpha-init}
    g^{\alpha,i}[\vy](\vr) &=& \int_{\vk}e^{-i\vk\cdot\vr}\omega^{\alpha,i}(\vk)\int d\vr'\,e^{i\vk\cdot\vr'}n(\vr')[\Hi\vy](\vr')\\\nonumber
    \tilde{g}^{\alpha,i}[\vy](\vr) &=& \int_{\vk}e^{-i\vk\cdot\vr}\omega^{\alpha,i}(\vk)\int d\vr'\,e^{i\vk\cdot\vr'}n(\vr')[\Ai\vy](\vr').
\eeq
These may be efficiently computed using Fast Fourier Transforms (FFTs).

In this manner, the $\hat{q}_\alpha$ estimator may be evaluated as a set of FFTs and, for the $\vr$-integrals, real-space summations. Note that we have allowed the templates to depend explicitly on the direction $\vk_i$; whilst no such dependence is required \resub{when measuring the rotationally-averaged bispectrum}, it will arise in \resub{generalizations to higher-order moments}.% redshift-space generalizations. %Computation of the bias term $\bar{q}_\alpha$ may be more difficult if $\bar{B}$ is not separable; however, this is independent of the data and thus only needs to be done once, thus slower direct integration techniques will likely suffice.

To form the full bispectrum estimator, we additionally require an explicit form for the Fisher matrix \eqref{eq: F-ab-def}. When using ML weighting schemes, one option is to compute $F_{\alpha\beta}^\mathrm{ML}$ as the (inverse) sample covariance of a set of $N_{\rm mc}$ Monte-Carlo realizations of $\hat{q}_\alpha^\mathrm{ML}$, using \eqref{eq: covariance-bispectrum}. However, this requires a suite of realistic simulations, works only for ML weights and is slow to converge, since the Fisher matrix must be invertible. As in \citep{2011MNRAS.417....2S,2021PhRvD.103j3504P}, we instead compute the full matrix via Monte Carlo averages. This proceeds similarly to that of $\hat{q}_\alpha$; we first rewrite
\beq\label{eq: HHH-separation}
\frac{1}{2}\Hi_{il}\left(\Hi_{jm}\Hi_{kn}+\Hi_{jn}\Hi_{km}\right) &\equiv& \frac{1}{2}\Hi_{il}\Hi_{jj'}\Hi_{kk'}\mathsf{A}^{-1}_{mm'}\mathsf{A}^{-1}_{nn'}\left(\mathsf{A}^{j'm'}\mathsf{A}^{k'n'}+\mathsf{A}^{j'n'}\mathsf{A}^{k'm'}\right)
\eeq
(assuming interchange symmetry under $\{j,m\}\leftrightarrow\{k,n\}$), then replace the two-point cumulants with Monte-Carlo averages, using
\beq
    \A^{j'm'}\A^{k'n'}+\A^{j'n'}\A^{k'm'} = \av{a^{j'}a^{k'}a^{m'}a^{n'}}-\av{a^{j'}a^{k'}}\av{a^{m'}a^{n'}},
\eeq
for Monte Carlo simulations $\va$ satisfying $\av{\va\,\va^T} = \A$, assuming the contributions from higher-point correlators to be small.\footnote{When using a set of $N$ uniformly distributed particles to define $\va$, the two- and four-point correlators are $\mathsf{A}_{ij} = \delta^\mathrm{K}_{ij}\,a(\vr_i)$ and $\mathsf{A}_{ijkl} = \delta^\mathrm{K}_{ij}\delta^\mathrm{K}_{jk}\delta^\mathrm{K}_{kl}\,a(\vr_i)$, indexing galaxies by $i,j,k,l$ and ignoring pixelation effects for simplicity. Taking the trace, $\sum_i \mathsf{A}_{iiii} = N$, $\sum_i \mathsf{A}_{ii}\mathsf{A}_{ii} = N^2$, thus the four-point correlator is negligible, for $N\gg 1$.}
 This gives
\beq\label{eq: F-general-tmp}
    F_{\alpha\beta} &=& \frac{1}{12}\av{\left(\B_{,\alpha}^{ijk}\Hi_{jj'}\Hi_{kk'}a^{j'}a^{k'}\right)\Hi_{il}\left(\B_{,\beta}^{lmn}\Ai_{mm'}\Ai_{nn'}a^{m'}a^{n'}\right)}\\\nonumber
    && -\frac{1}{12}\av{\B_{,\alpha}^{ijk}\Hi_{jj'}\Hi_{kk'}a^{j'}a^{k'}}\Hi_{il}\av{\B_{,\beta}^{lmn}\Ai_{mm'}\Ai_{nn'}a^{m'}a^{n'}}.
\eeq

Whilst \eqref{eq: F-general-tmp} seems complicated, it can be expressed as a vector product. To see this, first define
\beq\label{eq: phi-applied}
    \phi_\alpha^i[\va] &=& \B_{,\alpha}^{ijk}\Hi_{jj'}\Hi_{kk'}a^{j'}a^{k'}\\\nonumber
    &=& \int_{\vk_1\vk_2\vk_3}e^{i\vk_1\cdot\vr_i}\omega^\alpha(\vk_1,\vk_2,\vk_3)\delD{\vk_1+\vk_2+\vk_3}n(\vr_i)[n\Hi\va](\vk_2)[n\Hi\va](\vk_3)\\\nonumber
    &=& \frac{1}{3}\int_{\vk_1} d\vr\,e^{i\vk_1\cdot\left(\vr_i-\vr\right)}\left(\omega^{\alpha,1}(\vk_1)n(\vr_i)g^{\alpha,2}[\va](\vr)g^{\alpha,3}[\va](\vr)+\text{2\,perms.}\right)\\\nonumber
    &=&\frac{1}{3}n(\vr_i)\left(\ift{\omega^{\alpha,1}(\vk)\ft{g^{\alpha,2}[\va]g^{\alpha,3}[\va]}(\vk)}(\vr_i)+\text{2\,perms.}\right),
\eeq
and analogously $\tilde\phi_\alpha^i[\va]$ with $\Hi\to\Ai$, and thus $g\to\tilde{g}$. Each of these can be evaluated using FFTs, assuming separability of the bispectrum templates. The Fisher matrix is thus
\beq\label{eq: fish-applied}
    F_{\alpha\beta} = \frac{1}{12}\left(\av{\phi_{\alpha}^i\Hi_{il}\tilde\phi_{\beta}^l}-\av{\phi_{\alpha}^i}\Hi_{il}\av{\tilde\phi_\beta^l}\right),
\eeq
which can be evaluated using a matrix inversion and a real-space sum. In practice, the second term is small except on the largest scales, as discussed in Appendix \ref{appen: limits}. When considering the ML estimators, we must perform one matrix inversion per element of $\vb$ (as well as one to define $\Ci\va$), thus the computation time scales as $(N_\mathrm{bins}+1)$, indicating the benefits of an efficient data compression scheme. As before, the error on $F_{\alpha\beta}$ scales as $\sqrt{1+1/N_\mathrm{mc}}$ for $N_\mathrm{mc}$ simulations, implying that $N_\mathrm{mc}$\,$\sim$\,$50$ is requires to compute the statistic to percent-level accuracy. This applies also to the second term in $\hat{q}_\alpha$.

\subsection{Application to the Binned Bispectrum}\label{subsec: spec-binned}
We now specialize to the measurement of the bispectrum monopole in a set of $k$-bins. This is the quantity obtained from most standard estimators \citep[e.g.,][]{2015MNRAS.451..539G,2017MNRAS.465.1757G,2005PhRvD..71f3001S}. Here, we will assume the parameter $b_\alpha$ to be the measured bispectrum amplitude in an ordered triplet of $k$-bins, indexed by $\alpha\equiv\{a,b,c\}$ with $a\leq b\leq c$. In this case, the bispectrum can be decomposed separably as
\beq\label{eq: Bk-decomposition-binned}
    B(\vk_1,\vk_2,\vk_3) \approx \sum_\alpha \frac{1}{\Delta_\alpha} b_\alpha\left[\Theta^a(k_1)\Theta^b(k_2)\Theta^c(k_3)+\text{5 perms.}\right],
\eeq
where the binning function $\Theta^\alpha$ is unity in bin $a$ and zero else. We additionally require the symmetry factor $\Delta_\alpha$, defined as
\beq\label{eq: Delta-alpha-def}
    \Delta_\alpha = \begin{cases} 6 & \text{ if } a = b = c\\ 2 & \text{ if } a = b \text{ or } a = c \text{ or } b = c\\ 1 & \text{else,}\end{cases}
\eeq
to account for the permutations of \eqref{eq: Bk-decomposition-binned}. This may be familiar from the Gaussian covariance of the bispectrum \citep[e.g.,][]{1998ApJ...496..586S}. The momentum conservation condition, $\vk_1+\vk_2+\vk_3 = \vec 0$, places strong constraints on the triangle bins, in particular $|k_a-k_b|<k_c<k_a+k_b$ where $k_i$ is the center of bin $i$, if we ignore the effects of finite bin widths. Note that our decomposition ignores any directional information about the triangle, \textit{i.e.}\ we consider only the \textit{\resub{rotationally-averaged}} bispectrum. \resub{In the absence of redshift-space distortions, this is expected to capture all possible bispectrum information; in the alternate case, it will encode only the dominant component. Generalizations to the anisotropic moments are considered in \S\ref{subsec: spec-bases}.}

% (though see \S\ref{subsec: spec-bases} for a generalization). \resub{If there are no redshift-space distortions, the rotationally-averaged estimator will measure all possible bispectrum information; in the presence of anisotropy, it captures only part of the information.}

%\footnote{When comparing model to data, we must bin the theory model also to avoid significant errors due to the finite sizes of the bins and the $\vk_1+\vk_2+\vk_3=\vec 0$ condition.} 

The cumulant derivative $\B_{,\alpha}^{ijk}$, \eqref{eq: cumulant-deriv-applied}, becomes
\beq
    \B_{,\alpha}^{ijk} = \frac{n(\vr_i)n(\vr_j)n(\vr_k)}{\Delta_\alpha}\int_{\vk_1\vk_2\vk_3}\left[\Theta^a(k_1)\Theta^b(k_2)\Theta^c(k_3)+\text{5 perms.}\right]e^{i\vk_1\cdot\vr_i+i\vk_2\cdot\vr_j+i\vk_3\cdot\vr_k}\delD{\vk_1+\vk_2+\vk_3},
\eeq
leading to the following form for $\hat{q}_\alpha$ (from \ref{eq: q-alpha-applied-separable}):
\beq\label{eq: q-alpha-binned}
    \hat{q}_\alpha = \frac{1}{\Delta_\alpha}\int d\vr\,\left\{g^{a}[\vd](\vr)g^{b}[\vd](\vr)g^c[\vd](\vr)-\left(g^a[\vd](\vr)\av{g^b[\va](\vr)\tilde g^c[\va](\vr)}+\text{2 perms.}\right)\right\},
\eeq
incorporating the permutation symmetries. This uses the definitions
\beq\label{eq: g-alpha-binned}
    g^a[\vy](\vr) = \int_{\vk}e^{-i\vk\cdot\vr}\Theta^a(k)\int d\vr' e^{i\vk\cdot\vr'}n(\vr')[\Hi\vy](\vr') &\equiv& \ift{\Theta^a(k)\ft{n\Hi\vy}(\vk)}(\vr)\\\nonumber
    \tilde g^a[\vy](\vr) = \int_{\vk}e^{-i\vk\cdot\vr}\Theta^a(k)\int d\vr' e^{i\vk\cdot\vr'}n(\vr')[\Ai\vy](\vr') &\equiv& \ift{\Theta^a(k)\ft{n\Ai\vy}(\vk)}(\vr),
\eeq
analogous to \eqref{eq: g-alpha-init}. The Fisher matrix is given by 
\beq\label{eq: F-ab-binned}
    F_{\alpha\beta} = \frac{1}{12}\left(\av{\phi_{\alpha}^i\Hi_{il}\tilde\phi_{\beta}^l}-\av{\phi_{\alpha}^i}\Hi_{il}\av{\tilde\phi_\beta^l}\right),
\eeq
as in \eqref{eq: fish-applied}, where the $\phi$ coefficients of \eqref{eq: phi-applied} simplify to
\beq\label{eq: phi-alpha-binned}
    \phi_\alpha[\va](\vr) &=&\frac{2n(\vr)}{\Delta_\alpha}\ift{\Theta^{a}(k)\ft{g^{b}[\va]g^{c}[\va]}(\vk)}(\vr)+\text{2 perms.}\\\nonumber
    \tilde\phi_\alpha[\va](\vr) &=&\frac{2n(\vr)}{\Delta_\alpha}\ift{\Theta^{a}(k)\ft{\tilde g^{b}[\va]\tilde g^{c}[\va]}(\vk)}(\vr)+\text{2 perms.}
\eeq
Analogous definitions including the effects of pixelation and particle weights can be found in \eqref{eq: g-alpha-pixel}\,\&\,\eqref{eq: phi-alpha-pixel}.

The approach described above has several differences from standard approach: (1) we subtract a (zero-mean) bias term in \eqref{eq: q-alpha-binned}, (2) the data \resub{are} weighted by $n\Hi$ in \eqref{eq: g-alpha-binned}, (3) we include a symmetry factor $\Delta_\alpha$, and (4) we normalize by a geometry-dependent factor $F_{\alpha\beta}$, rather than the bin volumes. The first three are included for the sake of optimality, and arise naturally from the ML solution, whilst (4) allows measurement of the unwindowed statistic. In Appendix \ref{appen: limits}, we demonstrate that the above procedure reduces to conventional bispectrum estimators \citep[e.g.,][]{2017MNRAS.472.2436W} in the limit of uniform, and large, $\bar{n}$ \eqref{eq: p-alpha-uniform-limit}, \resub{additionally noting that we have not subtracted a Poissonian shot-noise term}.

\subsection{Alternative Bases}\label{subsec: spec-bases}

In the above discussion, we have assumed that the bispectrum can be fully parametrized by the lengths of three triangle sides, $\{k_1,k_2,k_3\}$. In reality, redshift-space distortions (RSD) make the bispectrum anisotropic, giving additional dependence on the relative orientation of $\{\vk_1,\vk_2,\vk_3\}$ triplet and the (local) line-of-sight $\hn$. Thanks to azimuthal symmetry about $\hn$, this is specified by only two additional variables. A number of parametrizations of the bispectrum anisotropy are possible \citep[e.g.,][]{2015PhRvD..92h3532S,2019MNRAS.484..364S,2020JCAP...06..041G}; here, we will consider that of \citep{2015PhRvD..92h3532S}:
\beq\label{eq: bk-multipole-binned}
    B(\vk_1,\vk_2,\vk_3;\hn) \approx \sum_\alpha \frac{1}{\Delta_\alpha}b_\alpha\left[\Theta^a(k_1)\Theta^b(k_2)\Theta^c(k_3)\mathcal{L}_\ell(\hk_3\cdot\hn)+\text{5 perms.}\right],
\eeq
where $\mathcal{L}_\ell$ is a Legendre polynomial, $\hk_3\cdot\hn$ is the angle between $\vk_3$ and the line-of-sight, $\hn$, and we have neglected dependence on the second angular coordinate for simplicity. In this case, the binning is now specified by four numbers; $\alpha = \{a,b,c,\ell\}$, with $\ell=0$ reproducing the bispectrum monopole of \S\ref{subsec: spec-bases}. 

Computation of the bispectrum multipoles, $b_\alpha$, appearing in \eqref{eq: bk-multipole-binned} is possible via similar methods to before. In particular, the cumulant derivative becomes
\beq
    \B_{,\alpha}^{ijk} &=& \frac{n(\vr_i)n(\vr_j)n(\vr_k)}{\Delta_\alpha}\int_{\vk_1\vk_2\vk_3}\left[\Theta^a(k_1)\Theta^b(k_2)\Theta^c(k_3)\mathcal{L}_\ell(\hk_3\cdot\hr_k)+\text{5 perms.}\right]e^{i\vk_1\cdot\vr_i+i\vk_2\cdot\vr_j+i\vk_3\cdot\vr_k}\\\nonumber
    &&\,\times\,\delD{\vk_1+\vk_2+\vk_3},
\eeq
setting the local line-of-sight to $\hn = \hr_k$ (as in the Yamamoto approximation of \citep{2006PASJ...58...93Y}). Only the $\vk_3$ part is more difficult to compute: this leads to the form
\beq
    \hat{q}_\alpha = \frac{1}{\Delta_\alpha}\int d\vr\,\left\{g^a[\vd](\vr)g^b[\vd](\vr)g^{c,\ell}[\vd](\vr) - \left(g^a[\vd](\vr)\av{\tilde g^b[\va](\vr)g^{c,\ell}[\va](\vr)}+\text{2 perms.}\right)\right\},
\eeq
with
\beq
    g^{a,\ell}[\vy](\vr) &=& \int_{\vk}e^{-i\vk\cdot\vr}\Theta^a(k)\int d\vr'\,n(\vr')\left[\Hi\vy\right](\vr')\mathcal{L}_\ell(\hk\cdot\hr')\\\nonumber
    &=& \frac{4\pi}{2\ell+1}\sum_m\ift{\Theta^a(k)Y^*_{\ell m}(\hk)\ft{n\Hi\vy Y_{\ell m}}(\vk)},
\eeq
expanding the Legendre polynomial in terms of spherical harmonics. The Fisher matrix can be computed similarly. This estimator gives a practical manner with which to estimate the bispectrum multipoles, and we note that its computation time is not significantly greater than that of the isotropic estimator, except due to the larger number of bins.

The estimators of \S\ref{subsec: spec-general} may also be applied to the scenario in which the bispectrum is represented as a sum of templates (e.g.,\,\ref{eq: Bk-decomposition}) rather than a large number of binned estimates \citep[cf.\,][]{2021PhRvD.103j3504P}. Whilst a suitably chosen decomposition scheme can significantly reduce the number of bispectrum elements (and thus the computation time) \citep{2021PhRvD.103d3508P}, this is non-trivial, since we require the templates to follow the separable form of \eqref{eq: bk-separability}. A simple approach would be to measure only the galaxy bias parameters from the bispectrum, assuming tree-level perturbation theory to be valid. In this case, the bispectrum \textit{is} separable into a sum of $\sim 6$ distinct components \citep{2015PhRvD..91d3530S,2021JCAP...03..105B}. However, this decomposition becomes more difficult when higher-loop effects, redshift-space distortions and the Alcock-Paczynski effect are included, or when cosmological parameters must also be constrained. It seems likely that approximate methods such modal decompositions \citep{2012PhRvD..86f3511F} may assist with this; we leave a comprehensive discussion of this and other subtleties to future work.

\section{Practical Implementation}\label{sec: implementation}
Whilst \S\ref{sec: spectro-specialization} gives the explicit formulae required to apply the bispectrum estimators to a galaxy survey, we are left with a number of practical questions, including how to estimate the underlying field $n(\vr)$ and how to efficiently implement the $\hat{q}_\alpha$ and $F_{\alpha\beta}$ estimators. We discuss this below, as well as details pertaining to our simulations and computation.

\subsection{Computation Strategy}

To compute the first term in $\hat{q}_\alpha$ \eqref{eq: q-alpha-binned}, we require the following intermediate quantities:
\begin{itemize}
    \item $\Hi\vd$: Assuming FKP weights, this can be computed via \eqref{eq: fkp-with-pix}, and requires four FFTs, or none, if pixelation effects are ignored \eqref{eq: fkp-weights}. For ML weights with $\H = \C$, the application of the inverse is performed using conjugate gradient descent (CGD), as in \citep{2021PhRvD.103j3504P}. This generally converges in a few tens of steps, and requires repeated computation of $\C\vx$ for some map $\vx$. This is achieved using chained FFTs, as discussed in Appendix \ref{appen: pixelized-2pt}.
    \item $g^a[\vd](\vr)$: Using the pixelized definition given in \eqref{eq: g-alpha-pixel}, we require $N_k+3$ FFTs to compute all relevant $g^a$ maps given $\Hi\vd$ (assuming $N_k$ bins per dimension). This reduces to $N_k+1$ FFTs if pixelation effects are not included. Assuming each map contains $N_\mathrm{pix}$ pixels and is stored as an array of floats, the set of all $g^a$ maps requires $4N_kN_\mathrm{pix}$ bytes of memory, which may be large for high-resolution maps or fine binning. As an example, using 512 grid-points per dimension with 30 $k$-bins leads to a total memory requirement of $\approx 2$\,GB.
    \item Contribution to $\hat{q}_\alpha$: As in \eqref{eq: q-alpha-binned}, this can be computed by performing (inexpensive) real-space sums over products of three $g^a$ functions, subject to the triangle conditions on $\{a,b,c\}$. In total, this gives $N_\mathrm{bins}\sim N_k^3$ scalar coefficients. Depending on the machine requirements, one may need to store the $g^a$ functions to disk then load each sequentially to compute $\hat{q}_\alpha$. 
\end{itemize}

The second term in $\hat{q}_\alpha$ requires the average of pairs of $g^a[\va]$ and $\tilde g^a[\va]$ functions. We use uniformly distributed particles to define $\{\va\}$, created by Poisson sampling a spatially invariant background number density (here set to $\bar n = 10^{-3}h^{3}\mathrm{Mpc}^{-3}$), then assigning the sampled particles to a grid, with the normalization chosen such that $\av{\va} = \vec 0$. Given $\va$, we estimate both $g^a[\vec a]$ and $\tilde g^a[\vec a]$ filters via \eqref{eq: g-alpha-binned}, using the exact form of $\mathsf{A}^{-1}$ given in \eqref{eq: unif-inv-covariance-Ainv}, whose implementation requires two FFTs. The pairwise product of these filters are saved to disk, and can then be used in combination with the data $g^a[\vec d]$ maps to compute $\hat{q}_\alpha$ via \eqref{eq: q-alpha-binned}. In practice, we use $N_\mathrm{mc} = 50$ realizations to define the average $\av{g^b[\vec a]\tilde g^c[\vec a]}$; since the error scales as $\sqrt{1+1/N_\mathrm{mc}}$ this is sufficient to give errors well below the statistical threshold.

The Fisher matrix is obtained in a similar fashion using the same set of $\va$ maps. Given the previously computed $g^a[\vec a]$ and $\tilde g^a[\vec a]$ maps, we further compute:
\begin{itemize}
    \item $\phi_\alpha[\vec a]$: By \eqref{eq: phi-alpha-binned}, this is simply computed using two Fourier transforms per $\{a,b,c\}$ permutation, or four when the pixelation effects are included \eqref{eq: phi-alpha-pixel}. Since the Fisher matrix requires $\av{\phi_\alpha[\vec a]}$, a simple approach would be to save each map to disk. We caution that this is \textit{not} practically achievable in many contexts, since it requires $4N_\mathrm{bins}N_\mathrm{pix}N_\mathrm{mc}$ of (temporary) storage if all bins are computed at once. In practice, one only needs to save the partial sum of all computed maps, greatly reducing the storage requirements.
    \item $\tilde\phi_\alpha[\vec a]$: This is computed analogously to $\phi_\alpha[\vec a]$, but using the $\tilde g^a[\vec a]$ filters.
    \item $\Hi\phi_\alpha[\vec a]$: The action of $\Hi$ on the $\phi_\alpha$ maps can be computed using FFTs, and, if ML weights are applied, CGD inversion. Since $N_\mathrm{bins}(N_\mathrm{mc}+1)$ such maps must be computed, this is the rate-limiting step of the ML algorithm, as each requires $\mathcal{O}(100)$ FFTs. 
    \item $F_{\alpha\beta}$: The Fisher matrix is constructed using \eqref{eq: F-ab-binned}, requiring $\av{\phi_\alpha\Hi\tilde\phi_\beta}$ and $\av{\phi_\alpha}\Hi\av{\tilde\phi_\beta}$. The MC averages can be accumulated from each $\va$ simulation, and stored as a matrix of size $N_\mathrm{bins}\times N_\mathrm{bins}$.
\end{itemize}

It is clear that computation of the bispectrum using cubic estimators is a relatively intensive procedure. The principal work lies in computing $g^a$ and $\Hi\phi_\alpha$ maps, requiring $\mathcal{O}(N_k)$ and $\mathcal{O}(N_\mathrm{bins})$ FFTs respectively. The latter dominates in practice, particularly when using ML weights, since we must apply an inverse matrix $\Hi$ to each of $N_\mathrm{mc}$ random simulations in $N_\mathrm{bins}$ bins. However, computation of the Fisher matrix (and the $\av{g^b[\vec a]\tilde g^c[\vec a]}$ bias term) is \textit{independent} of both data and any $N$-body simulations, with its estimation requiring only the background number density $n(\vr)$ and the fiducial band-powers (if using ML weights). Usually, one computes summary statistics on both the data and a large number of mock catalogs; each simulation requires only computation of $\hat{q}_\alpha$, and is thus fast. %Comparing our (unwindowed) cubic estimators to conventional approaches \citep[e.g.,][]{2005PhRvD..71f3001S,2015MNRAS.451..539G,2017MNRAS.465.1757G}, we see that the computation time per simulation is very similar for both, particularly when FKP weights are used. Our approach simply requires an additional data-independent Fisher matrix calculation, which is required to remove the window function.

\subsection{Mock Catalogs}

Foreshadowing the eventual application of the above estimators to the BOSS DR12 galaxy sample \citep{2017MNRAS.470.2617A}, we first consider their use on a set of mock galaxy catalogs taken from the \textsc{MultiDark-Patchy} (hereafter `Patchy') suite \cite{2016MNRAS.460.1173R,2016MNRAS.456.4156K}.\footnote{These are publicly available at \href{https://data.sdss.org/sas/dr12/boss/lss/}{data.sdss.org/sas/dr12/boss/lss}.} These share the BOSS geometry and selection functions, and are calibrated to have similar power spectrum and bispectrum to the observational data. As in \citep{2021PhRvD.103j3504P}, we use only the north Galactic cap with the `z1' redshift cut $0.2<z<0.5$ \citep[e.g.,][]{2017MNRAS.466.2242B,2020JCAP...05..042I}. This has mean redshift $ z = 0.38$ and a total volume $V=1.46h^{-3}\mathrm{Gpc}^3$, containing $\sim$\,$5\times 10^5$ galaxies. We additionally use a Patchy random catalog which has the same selection functions as the data but $50\times$ larger particle density. The observed redshifts and angular positions of both galaxies and randoms are converted to Cartesian co-ordinates using the fiducial matter density $\Omega_m = 0.31$ \citep{2017MNRAS.466.2242B}, before they are painted to a uniform grid using a triangular-shaped cloud mass assignment scheme, implemented using \textsc{nbodykit} \citep{2018AJ....156..160H}. The discretization grid uses pixels of twice the width as those in the official BOSS release; this is more than sufficient for our purposes (cf.\,\S\ref{sec: application}) and gives a Nyquist frequency $k_{\rm Nyq} = 0.3\hMpc$. The data vector $\vd$ is then computed as the difference between galaxies and (rescaled) random particles, including the incompleteness weights, $w_c$, of \citep{2017MNRAS.466.2242B}. Note that we do not add FKP weights at this point.

To define the background density field, $n(\vr)$, we make use of the publicly available survey mask files as well as the redshift distribution, $\bar n(z)$, from the Patchy random catalogs (themselves calibrated from the BOSS data). In practice, we first histogram the randoms in redshift bins (weighted by $w_c$), before transforming these to volumetric bins via the Jacobian $dV/dz = 4\pi r^2(z) dr/dz$, where $r(z)$ is the comoving distance to redshift $z$. We then evaluate $\bar n(z(\vr))$ at the center of each pixel, before multiplying by the angular \textsc{mangle} mask $\phi(\hat{\vr})$ and normalizing such that $\int d\vr\,n^3(\vr)$ matches the value obtained from the random catalog (including completeness weights). This differs from the approach of \citep{2021PhRvD.103j3504P}, which assumed $n(\vr)$ to be equal to the pixelized field of randoms, and is preferred since (a) it does not include pixelation effects, and (b) it gives a smooth field, \textit{i.e.}\ there are no regions within the survey area where $n(\vr)=0$, which can cause instabilities upon inversion.%the field is not blurred by the mass assignment windows. The latter is important for proper treatment of pixelation effects.

To apply the ML estimators, we additionally require a fiducial power spectrum for the signal covariance, as in \eqref{eq: sig-cov-pix}. Since the bispectrum estimator is unbiased for all choices of invertible $\H$, the precise power spectrum is not critical, though the procedure is suboptimal if it is far from the true power spectrum.\footnote{This is unlike the approach of \citep{2021PhRvD.103j3504P}, whereupon the fiducial spectrum was a crucial component of the power spectrum estimator. This was a consequence of using the first form of \eqref{eq: ml-pk-estimator}\,\&\,\eqref{eq: general-qe} which contains explicit dependence on a fiducial set of band-powers. In this work, we focus on the second form, which does not require such considerations.} Here, we follow \citep{2021PhRvD.103j3504P}, and model the unwindowed power spectra using a fit to 1000 Patchy simulations using \textsc{class-pt} \citep{2020PhRvD.102f3533C}. No fiducial spectrum is required when using FKP weights, \textit{i.e.}\ $\H = \H_\fkp$.

\subsection{Unwindowed Bispectrum Computation}\label{subsec: bk-computation}

Given the above, we may apply the bispectrum estimators of \S\ref{sec: estimator-derivation}\,\&\,\ref{sec: spectro-specialization} to the Patchy simulations. For this demonstration, we adopt a $k$-space binning with width $\Delta k = 0.01\hMpc$ from $k_\mathrm{min} = 0\hMpc$ to $k_\mathrm{max} = 0.16\hMpc$, giving $N_k = 16$. Allowing for the effects of finite bin widths, this gives $N_\mathrm{bins} = 508$ triplets obeying the triangle conditions, \textit{i.e.}\ containing wavevectors $\{\vk_1,\vk_2,\vk_3\}$ satisfying $\vk_1+\vk_2+\vk_3=\vec 0$.\footnote{When comparing data and theory, it is common to evaluate the theory model in the center of each $k$-bin, \textit{i.e.}\ at $\{\bar k_1,\bar k_2,\bar k_3\}$. We caution that some of our allowed bins violate the triangle conditions on $\bar k_i$, \textit{i.e.}\ they do not satisfy $|\bar k_1-\bar k_2|\leq \bar k_3\leq \bar k_1 + \bar k_2$, and should not be included in the fit. If the theory model is integrated over a finer grid of wavenumbers, these bins are important to include however.} After computation, we discard any bins with $k<0.01\hMpc$ or $k>0.15\hMpc$, since these are not fully window-corrected (cf.\,\citep{2021PhRvD.103j3504P}); this leaves $N_{\rm bins} = 399$ elements. A finer binning would lead to a much larger dimensionality; this does not pose problems in the analysis if a compression scheme is applied before the likelihood is computed \citep[e.g.,][]{2000ApJ...544..597S,2021PhRvD.103d3508P}, though it will increase the computation time.

Here, we compute the bispectrum of $999$ Patchy simulations, using $N_\mathrm{mc}=50$ uniform random simulations to define the linear $\hat{q}_\alpha$ term and the Fisher matrix. For our fiducial run, we assume an FKP weighting matrix \eqref{eq: fkp-weights}, and do not include the forward-modelling of pixelation effects described in Appendix \ref{appen: pixel-cov}.\footnote{We do, however, include a factor $\tilde\psi^{-1}(\vk_1)\tilde\psi^{-1}(\vk_2)\tilde\psi^{-1}(\vk_3)$ in the $\hat{q}_\alpha$ estimator (where $\tilde\psi$ is the Fourier-space pixelation window) to ensure that our bispectra are not biased. This is further discussed in Appendix \ref{appen: pixelized-3pt}.} To test the various hyperparameters, we consider a number of alternative runs, each with $100$ simulations: (1) using a coarser pixel grid with $k_{\rm Nyq} = 0.2\hMpc$, (2) using $N_{\rm mc} = 100$ MC simulations, (3) including the full pixelation effects of Appendix \ref{appen: pixel-cov}, (4) omitting the second, variance reducing, term in \eqref{eq: q-alpha-binned}, and (5) using ML weights. The impact of these assumptions will be assessed in \S\ref{subsec: results-hyperparam}.

All computations are performed in \textsc{python}, making extensive use of the \textsc{pyfftw} package to perform FFTs.\footnote{Our code is publicly available at \href{https://github.com/oliverphilcox/BOSS-Without-Windows}{github.com/oliverphilcox/BOSS-Without-Windows}.} When using FKP (ML) weights, we require $100$ ($140$) CPU-hours to compute the data bispectrum contributions from all 999 mocks (matching the computation time of any standard bispectrum estimator), plus an additional $150$ ($850$) CPU-hours in total for the uniform random contributions. Whilst non-negligible, this runtime is not unreasonable, especially considering that the bulk of the time is spent computing the Fisher matrix which only needs to be done once. This is further reduced if the bin-width is increased or when using a coarser pixel grid.

\subsection{Windowed Bispectrum Estimators}\label{subsec: win-estimators}

To assess the impact of the survey window function on the bispectrum, we additionally compute the statistic using conventional estimators. In this case, the output bispectrum is binned, FKP-weighted (before gridding), and convolved with the survey window function. Following \citep{2017MNRAS.465.1757G}, the windowed bispectrum of data $d$ can be written
\beq\label{eq: win-estimator}
    \hat b_\alpha^{\rm win} = \frac{1}{I_3}\frac{1}{V_\alpha}\int_{\vk_1\vk_2\vk_3}\delD{\vk_1+\vk_2+\vk_3}d(\vk_1)d(\vk_2)d(\vk_3)\Theta^a(k_1)\Theta^b(k_2)\Theta^c(k_3),
\eeq
where the Dirac delta function enforces momentum conservation and the bin volume is given by
\beq\label{eq: win-volume}
    V_\alpha = \int_{\vk_1\vk_2\vk_3}\delD{\vk_1+\vk_2+\vk_3}\Theta^a(k_1)\Theta^b(k_2)\Theta^c(k_3).
\eeq
\eqref{eq: win-estimator} also includes the normalization factor $I_3 = \int d\vr\,n^3(\vr)w_\fkp^3(\vr)$, where $w_\fkp(\vr)=\left[1+n(\vr)P_\fkp\right]^{-1}$ is the FKP weight. This can be computed from the random catalog (requiring the inclusion of completeness weights) as in \citep{2017MNRAS.465.1757G}. Rewriting the Dirac functions in \eqref{eq: win-estimator}\,\&\,\eqref{eq: win-volume} as complex exponentials as before, gives a more practical form for the estimator:
\beq\label{eq: win-estimator-applied}
    \hat b_\alpha^{\rm win} = \frac{1}{I_3}\frac{\int d\vr\,\mathcal{D}_a(\vr)\,\mathcal{D}_b\,(\vr)\,\mathcal{D}_c(\vr)}{\int\,d\vr\,\mathcal{U}_a(\vr)\,\mathcal{U}_b(\vr)\,\mathcal{U}_c(\vr)},
\eeq
defining 
\beq
    \mathcal{D}_a(\vr) = \int_{\vk}e^{-i\vk\cdot\vr}d(\vk)\Theta_a(\vk), \qquad \mathcal{U}_a(\vr) = \int_{\vk}e^{-i\vk\cdot\vr}\Theta_a(\vk),
\eeq
the set of which can be computed using $(2N_k+1)$ FFTs. Here, we apply the estimator to 999 Patchy simulations, which requires $\sim$ 100 CPU-hours in total.

\section{Results}\label{sec: application}

% \begin{itemize}
%     \item Results with FKP and ML weights
%     \item Results with $g=2$
%     \item Results with MAS
%     \item Results with 100 randoms
%     \item Results without q-bar
% \end{itemize}

\subsection{Bispectrum Estimates}\label{subsec: results-bk}

\begin{figure}
    \centering
    \includegraphics[width=\textwidth]{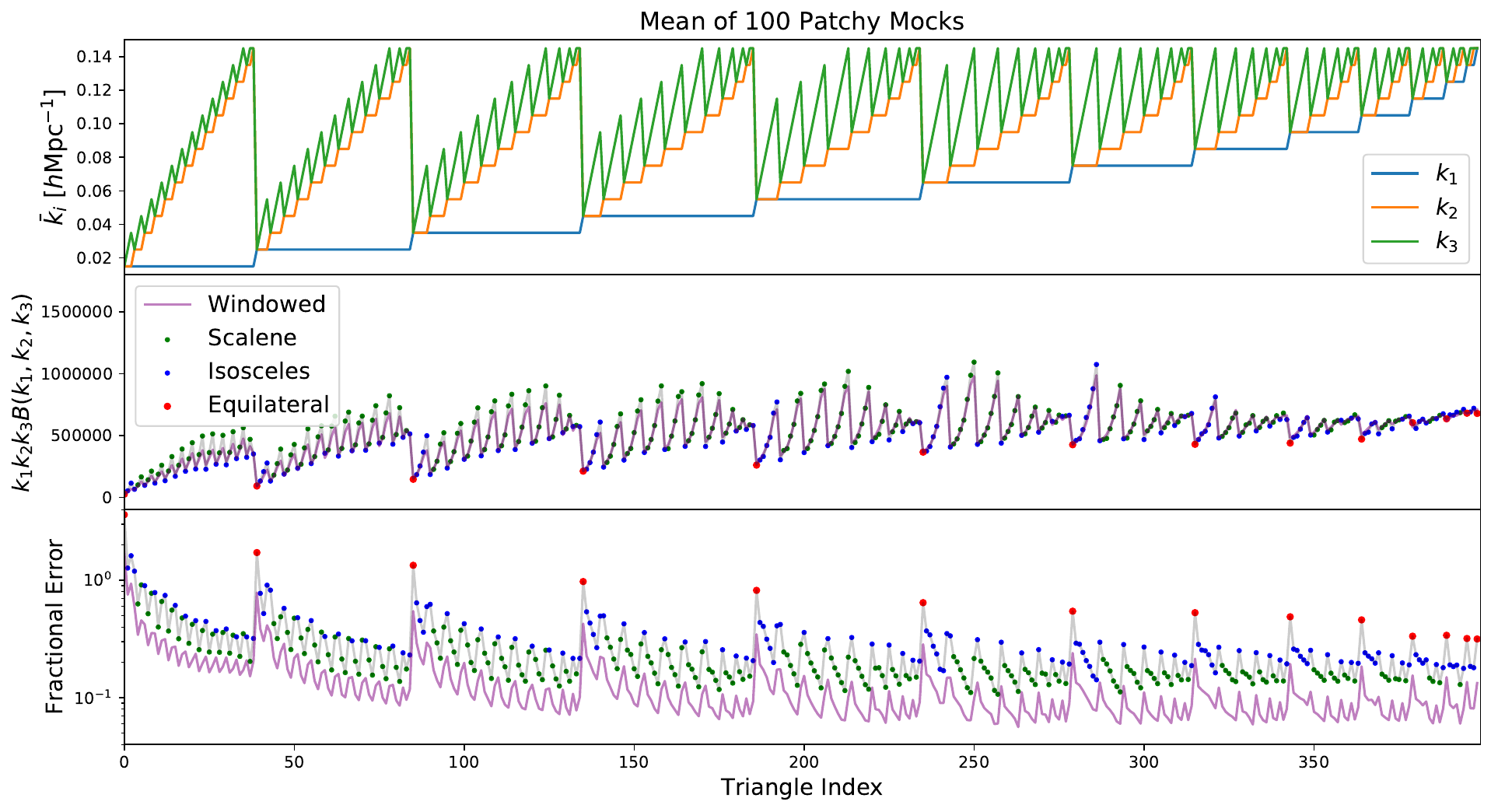}
    \caption{Unwindowed bispectrum measurements from 999 Patchy simulations at $z = 0.57$, obtained using the cubic estimators of this work. In the first panel, we show the bin-averaged triangle sides $\{k_1,k_2,k_3\}$ corresponding to each one-dimension wavenumber bin, with the second panel giving the associated binned bispectrum estimates (normalized by $k_1k_2k_3$), averaged across all mocks. The third panel displays the fractional error in the bispectrum relevant for a single mock dataset. We show triangle configurations corresponding to scalene ($k_1\neq k_2\neq k_3$), isosceles ($k_1=k_2<k_3$ or $k_1<k_2=k_3$) and equilateral ($k_1=k_2=k_3$) triangles in green, blue and red respectively; the error-bars for isosceles (equilateral) triangles are inflated by a factor $\sim$\,$2$ ($\sim$\,$6$) as expected. To compute these bispectra, we assume an FKP weighting \eqref{eq: fkp-weights}, a Nyquist frequency of $k_{\rm Nyq} = 0.3\hMpc$, $N_{\rm mc} = 50$ Monte Carlo simulations, and do not forward model pixelation effects (cf.\,Appendix \ref{appen: pixel-cov}). We additionally show results from the \textit{windowed} bispectrum estimator in purple; whilst these have smaller fractional errors, this is primarily due to increased bin-to-bin covariances, as demonstrated in Fig.\,\ref{fig: correlation-patchy}. The figure shows results only for the bispectrum monopole (averaged over triangle rotations); the estimators of this work could be simply extended to include higher multipoles sourced by redshift-space distortions.}
    \label{fig: bk-patchy}
\end{figure}

In Fig.\,\ref{fig: bk-patchy}, we display the main results of this work; a set of binned bispectrum measurements from 999 Patchy mocks using FKP weights \eqref{eq: fkp-weights}, which are not convolved with the survey window function. The measurements exhibit a distinctive `sawtooth' pattern (analogous to those of \citep{2017MNRAS.465.1757G}), which is a result of projecting a 3D data-set into one-dimension. In full, this arises since $k_1k_2k_3B(k_1,k_2,k_3)$ is an increasing function of $k_i$. Considering the fractional error, it is clear that we have detected a non-zero bispectrum at high significance, with most bins having a signal-to-noise above unity (ignoring correlations). We note a clear distinction between the three types of triangles shown in Fig.\,\ref{fig: bk-patchy}; scalene triangles exhibit significantly smaller error-bars than those of isosceles configurations, themselves smaller than the equilateral triangles error-bars. This arises due to the number of ways in which one can obtain a triangle of a given configuration (\textit{i.e.}\ $\Delta_\alpha$, \ref{eq: Delta-alpha-def}), and matches that found in previous works \citep[e.g.,][]{2019MNRAS.484.3713G}.

Also shown in Fig.\,\ref{fig: bk-patchy} are the \textit{windowed} bispectrum measurements from Patchy, computed using the estimators of \S\ref{subsec: win-estimators}. On short scales (high-$k$, corresponding to the rightmost data-points of Fig.\,\ref{fig: bk-patchy}), the mean bispectrum measurements from the unwindowed and windowed estimators appear highly consistent, implying that the window function's impact is minimal. Moving to larger scales (low-$k$), the window function distortions become significant, particularly for squeezed triangles (containing two long sides and a short side). Given that the BOSS window function primarily contains power on large scales \citep{2021PhRvD.103j3504P}, this matches our expectations. Some of the most important bispectrum parameters (such as $f_{\rm NL}$) have signatures concentrated on large scales, thus these results highlight the importance of a proper window function treatment.

\begin{figure}
    \centering
    \includegraphics[width=0.5\textwidth]{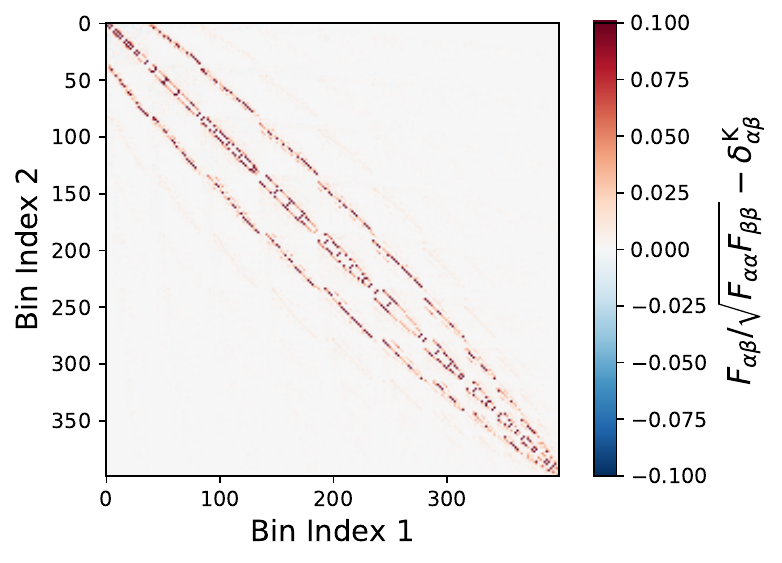}
    \caption{Fisher matrix for the unwindowed bispectrum estimates shown in Fig.\,\ref{fig: bk-patchy}. As discussed in the main text, this removes effects of the survey window function, practically acting as a deconvolution matrix. Here, we plot the matrix using the same binning scheme as Fig.\,\ref{fig: bk-patchy}, normalizing by the diagonal elements for clarity. The largest correlations are between bins $k_1'=k_1\pm\Delta k$, $k_2'=k_2$, $k_3'=k_3$ (or permutations); other correlations are found to be small.}
    \label{fig: fisher-patchy}
\end{figure}

To further assess the effects of window function convolution, it is useful to look at the Fisher matrix, $F_{\alpha\beta}$ \eqref{eq: F-ab-binned}. In the limit of an ideal survey geometry (Appendix \ref{appen: limits}), the matrix is diagonal and simply encodes the bin volumes and normalization; outside this limit, it has a more complex structure, including off-diagonal contributions, which act to deconvolve the bispectrum (cf.\,\citep{2021arXiv210606324B}). In Fig.\,\ref{fig: fisher-patchy}, we plot the Fisher matrix for the Patchy simulations, again assuming $\H = \H_\fkp$. Here, we find off-diagonal correlations up to $\sim$\,$10\%$ in neighboring bins (\textit{i.e.}\ those in which a single element of $\alpha=\{a,b,c\}$ and $\beta=\{a',b',c'\}$ differs by one), but small corrections elsewhere. This indicates that the primary effect of the window function is to convolve neighbouring bins (due to the compact nature of the window function), and motivates our choice of scale-cuts (\S\ref{subsec: bk-computation}).

\begin{figure}
    \centering
    \includegraphics[width=0.8\textwidth]{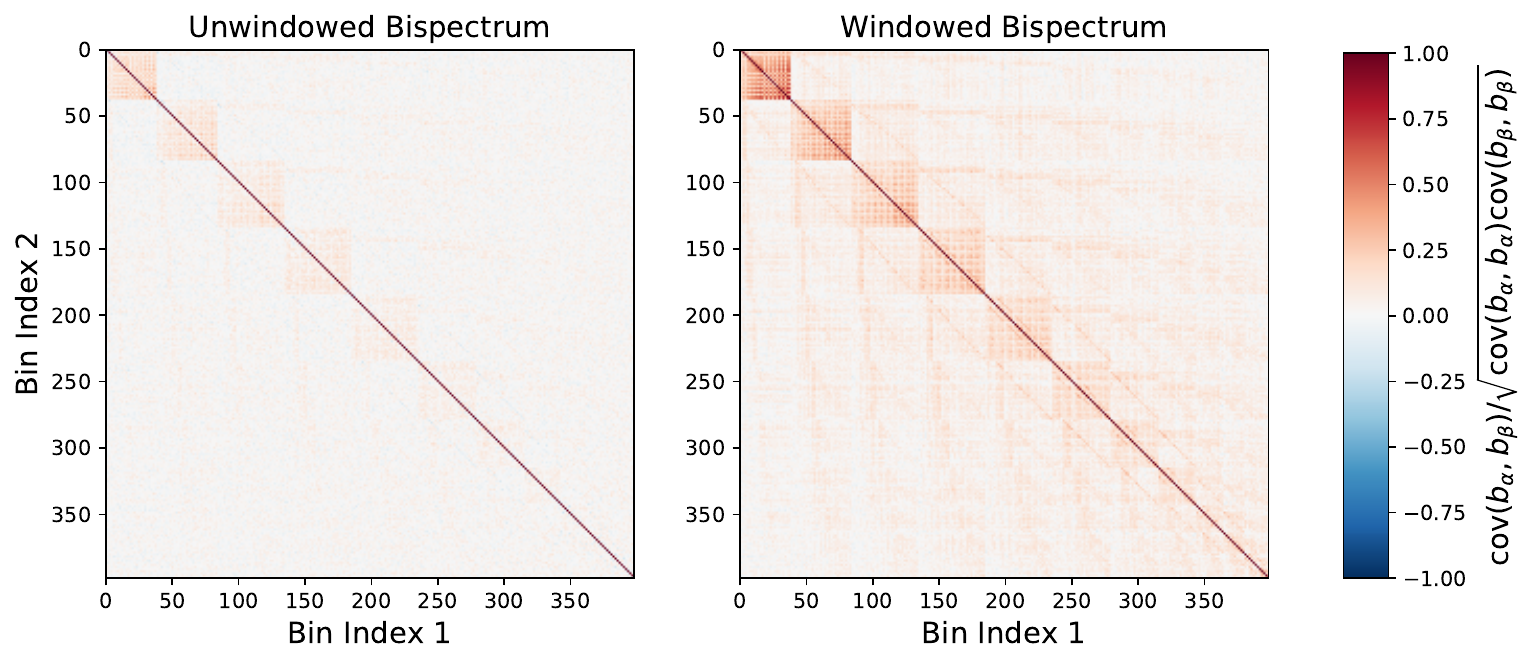}
    \caption{Correlation matrices for the unwindowed and windowed bispectrum measurements shown in Fig.\,\ref{fig: bk-patchy}. These are computed from the sample covariance of 999 individual Patchy bispectra and follow the same bin-ordering as Fig.\,\ref{fig: bk-patchy}. Whilst the unwindowed spectra have a close-to-diagonal covariance, this is not true for the windowed measurements, and highlights the smoothing effect of the window function.}
    \label{fig: correlation-patchy}
\end{figure}

From the fractional errors shown in Fig.\,\ref{fig: bk-patchy}, it may seem that the windowed bispectrum estimator achieves significantly higher precision than the unwindowed estimator. In fact, this is not the case, since the individual windowed data-points are far more correlated than their unwindowed equivalents; an effect sourced by the non-uniform geometry. As a demonstration, we plot the bispectrum correlation matrices (equal to the covariance matrices normalized by their diagonals) in Fig.\,\ref{fig: correlation-patchy}. In the unwindowed case, the matrix is close to diagonal, though there remain some off-diagonal correlations in closely separated bins.\footnote{For ML weights, and in the limit of zero bispectrum, the covariance matrix is equal to the inverse of the Fisher matrix (cf.\,\ref{eq: optimal-cov}).} In contrast, we observe a much increased correlation structure in the windowed estimates, with correlation coefficients approaching unity at low-$k$ (where the effects of the window function are largest). This highlights the utility of the windowed estimator, and matches the conclusions of \citep{2021PhRvD.103j3504P} for the power spectrum.\footnote{Given that the bispectra obtained from the estimators of \S\ref{sec: spectro-specialization} are not convolved with the window function, it may be tempting to model their covariance perturbatively, ignoring window effects. This is not true however, since the impact of the window on the covariance is different to that on the data.}

\subsection{Dependence on Hyperparameters}\label{subsec: results-hyperparam}

\begin{figure}
    \centering
    \includegraphics[width=0.9\textwidth]{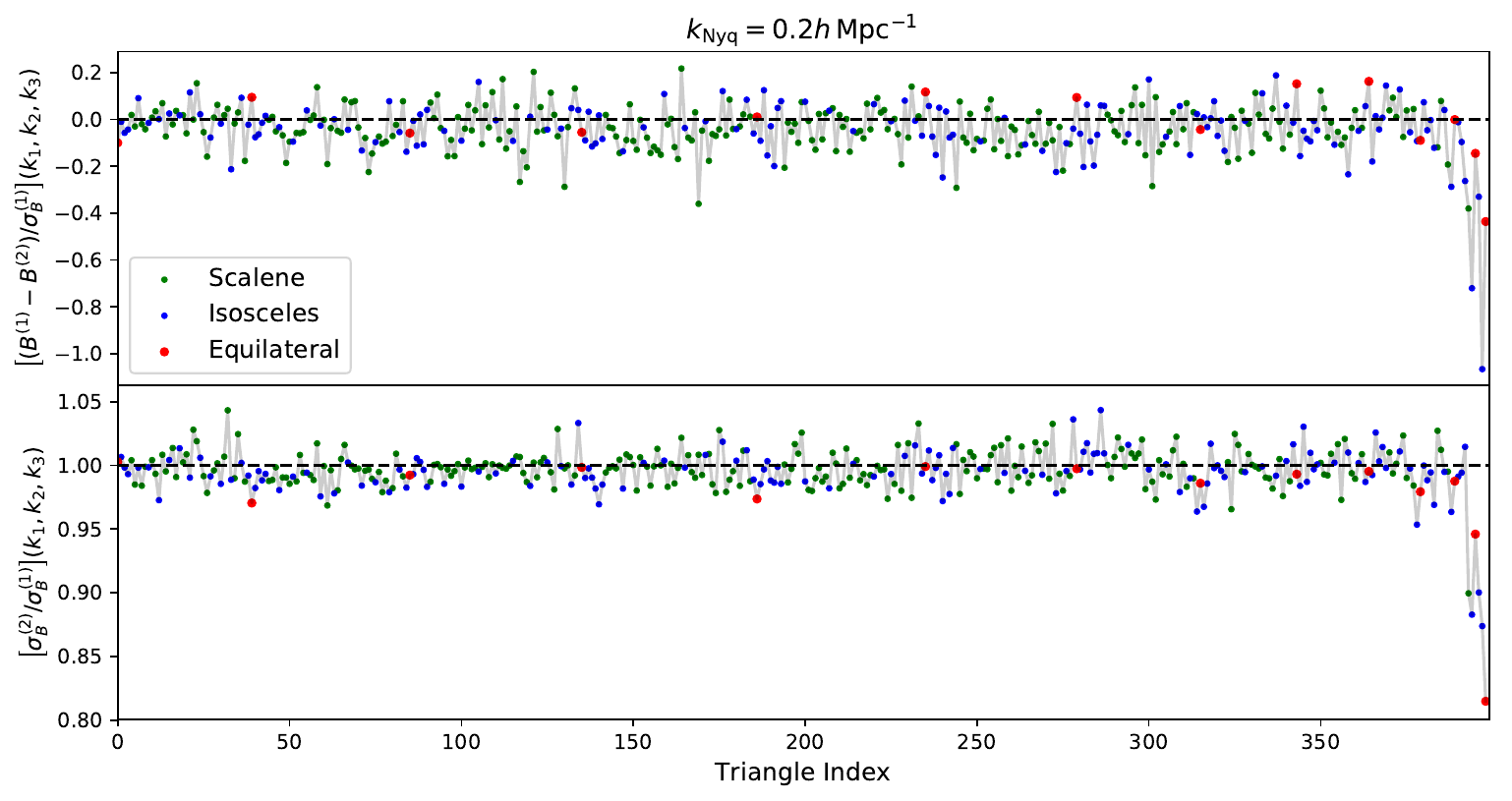}
    \caption{Dependence of the unwindowed bispectra on the Fourier-space grid-size. This compares two sets of bispectra; those computed with a Nyquist frequency $k_{\rm Nyq} = 0.3\hMpc$ (shown in Fig.\,\ref{fig: bk-patchy}) and those with $k_{\rm Nyq} = 0.2\hMpc$. The top panel shows the difference between the two estimates as a fraction of the statistical error of the fiducial estimates, whilst, in the bottom panel, we give the ratio of statistical errors. The bins are ordered in the same fashion as Fig.\,\ref{fig: bk-patchy}, and both datasets include 100 Patchy simulations. Here, increasing the number of grid-points is found to have an insignificant effect, except on the smallest scales.}
    \label{fig: comparison-g3}
\end{figure}

To test the dependence of the output spectra on the algorithm hyperparameters, we perform a number of additional bispectrum analyses, as noted in \S\ref{subsec: bk-computation}. Firstly, Fig.\,\ref{fig: comparison-g3} considers the effects of using a coarser Fourier-space grid, reducing the Nyquist frequency from $k_{\rm Nyq} = 0.3\hMpc$ to $0.2\hMpc$ (which can be compared to $k_{\rm max} = 0.15\hMpc$). Though the noise properties differ somewhat, we find consistent results between the two data-sets across all triangle bins, except those in which all three triangle sides have large wavenumbers. This matches our expectations, and implies that the fiducial value of $k_{\rm Nyq} = 0.3\hMpc$ is more than sufficient for our choice of $k_{\rm max}$.

\begin{figure}
    \centering
    \includegraphics[width=0.9\textwidth]{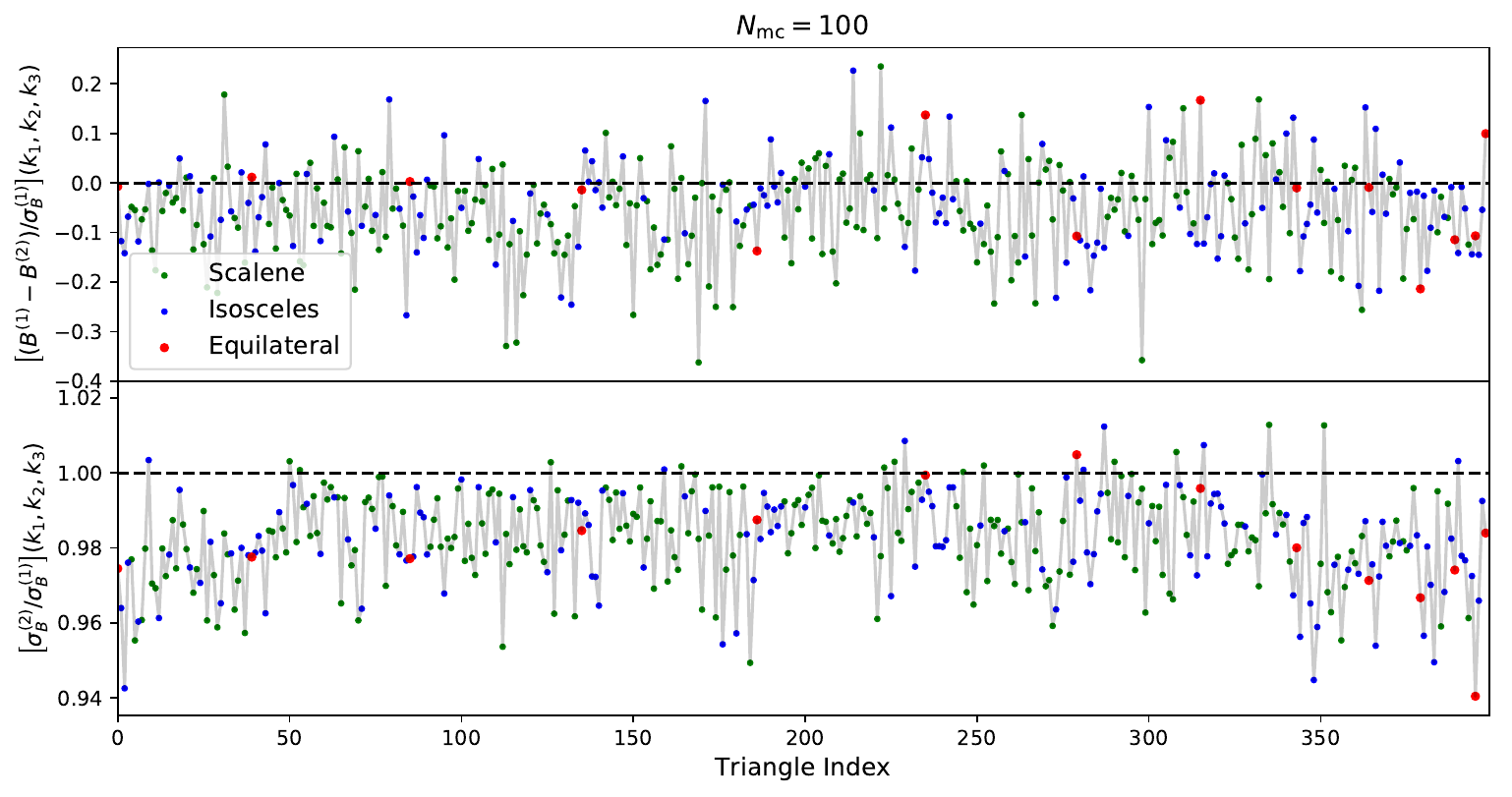}
    \caption{As Fig.\,\ref{fig: comparison-g3}, but assessing the impact of the number of Monte Carlo simulations used to define the Fisher matrix. Here we compare the case of $N_{\rm mc}=100$ to the fiducial results with $N_{\rm mc} = 50$. The error is expected to scale as $\sqrt{1+1/N_{\rm mc}}$; this matches that found in the figure.}
    \label{fig: comparison-Nmc}
\end{figure}

Fig.\,\ref{fig: comparison-Nmc} assesses the impact of Monte Carlo noise. As noted in \S\ref{sec: spectro-specialization}, we use a set of $N_{\rm mc} = 50$ uniform random realizations to compute the Fisher matrix (and the second term of $\hat{q}_\alpha$), in order to sidestep a computationally prohibitive matrix multiplication. When $N_{\rm mc}$ is increased to $100$, the results are statistically consistent, yet we find a $\sim$\,$2\%$ reduction in the error-bar. Given that the expected error is $\sqrt{1+1/N_{\rm mc}}$, this matches expectations. In future runs, it may be desired to use a larger number of simulations than the fiducial $N_{\rm mc} = 50$, though we note that this is still a small (though multiplicative) contribution to the overall error budget.

\begin{figure}
    \centering
    \includegraphics[width=0.9\textwidth]{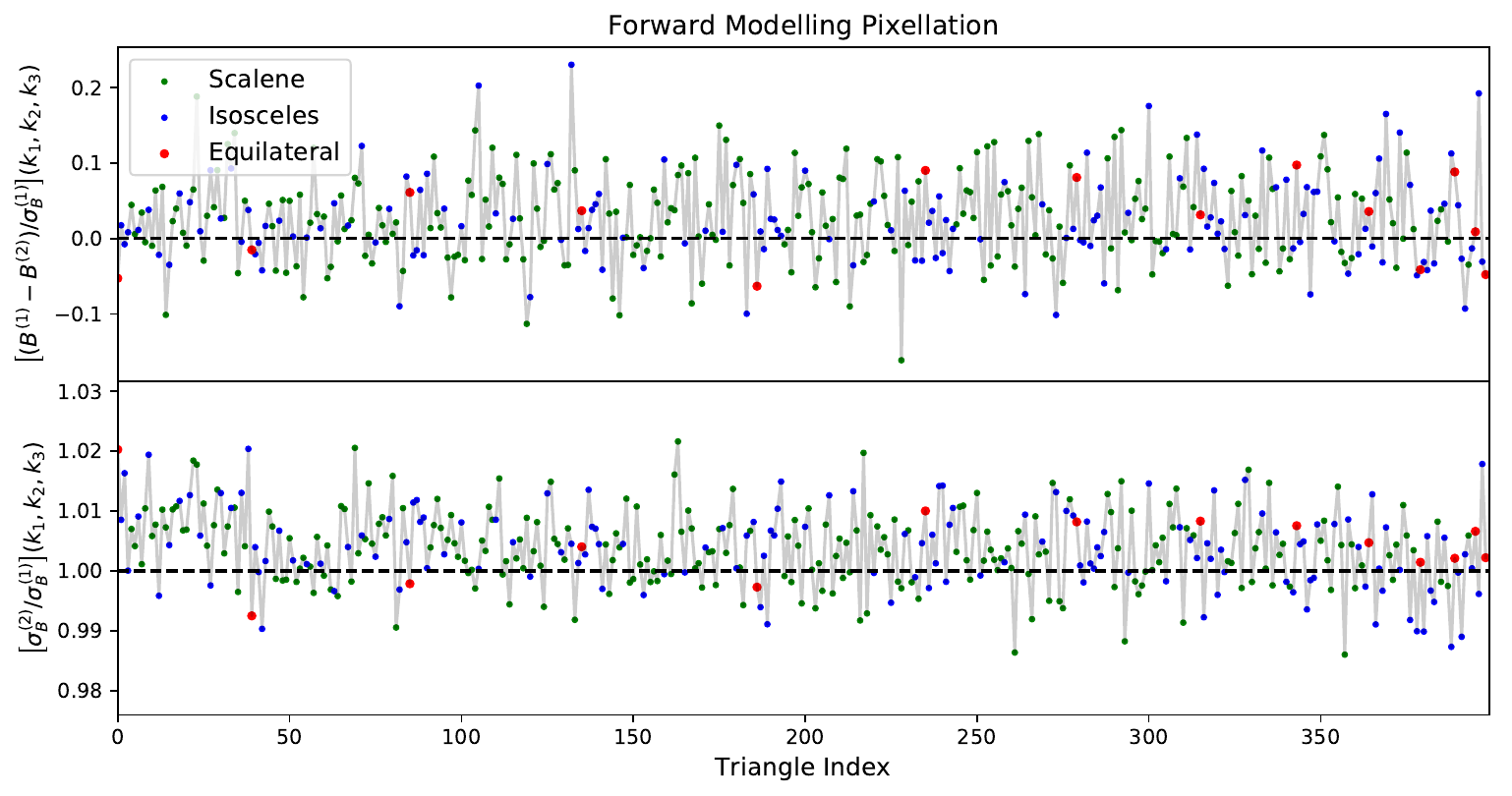}
    \caption{As Fig.\,\ref{fig: comparison-g3}, but testing the impact of forward modelling the pixelation effects, as described in Appendix \ref{appen: pixel-cov}. This is found to have a very small effect in practice, since the survey size is much larger than the pixel width.}
    \label{fig: comparison-mas}
\end{figure}

In Appendix \ref{appen: pixel-cov}, we present a full treatment of the two- and three-point cumulants of the \textit{gridded} density field. This allows one to forward-model the pixelation effects, instead of simply dividing by the pixelation mask in the final step of the bispectrum estimator computation. Bispectra computed using this approach are compared to the fiducial set in Fig.\,\ref{fig: comparison-mas}. In this case, we do not find any marked improvement with this approach; rather there appears to be a slight increase in the variances (though we caution that this may be sourced by a decrease in bin-to-bin correlations). Whilst this approach may be of use for analyses with more complex window functions, it is not found to be important here, and is disfavored since it requires more FFT operations.

\begin{figure}
    \centering
    \includegraphics[width=0.9\textwidth]{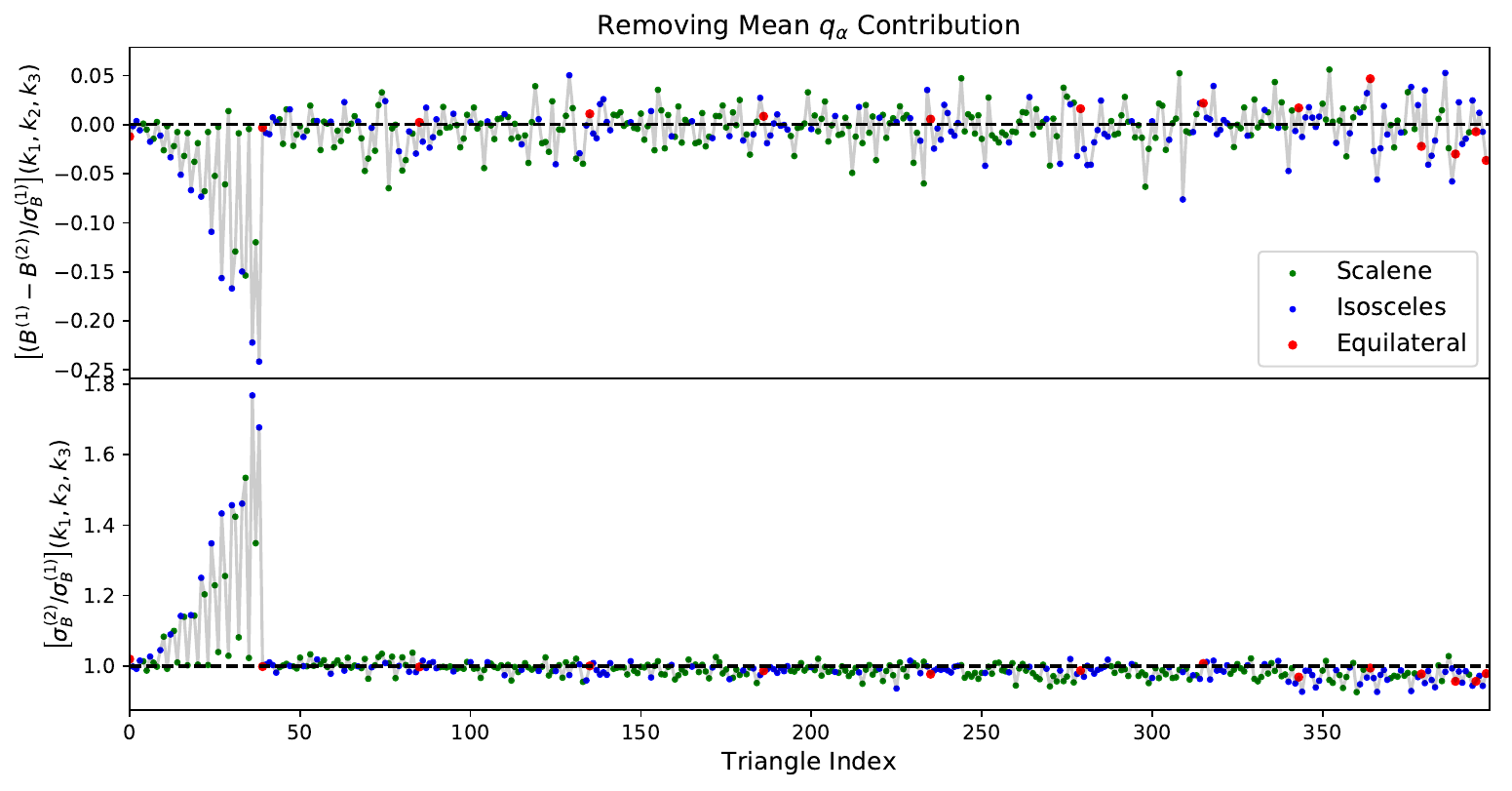}
    \caption{As Fig.\,\ref{fig: comparison-g3}, but assessing the importance of the second term in $\hat{q}_\alpha$ \eqref{eq: q-alpha-binned}, which is ignored in conventional bispectrum estimators. Removal of this term is found to significantly increase the error on large scales; this is consistent with the expected behavior found in Appendix \ref{appen: limits}, and highlights the importance of this contribution in $f_{\rm NL}$-based studies.}
    \label{fig: comparison-qbar}
\end{figure}

An additional test concerns the second term appearing in the $\hat{q}_\alpha$ quantity, \textit{i.e.}\ that involving $\Ci_{ij}[\Ci \vd]_k$ (and permutations) in \eqref{eq: q-alpha-def}. On average, this is expected to be zero, and furthermore, it contributes only to the $\vk = \vec 0$ mode for a uniform survey geometry (Appendix \ref{appen: limits}). Fig.\,\ref{fig: comparison-qbar} shows the effects of removing it from the bispectrum estimator; we find a large increase in the bispectrum variances in the lowest $k$-bin (with one leg in the range $[0.01,0.02]\hMpc$), but negligible impact to other bins.\footnote{There is also a slight bias, though we again caution that error-bars are most strongly correlated on these scales.} We conclude that this term should certainly be included when the measuring large-scale bispectrum modes, but is of little importance if these are not included in the analysis.

\begin{figure}
    \centering
    \includegraphics[width=0.9\textwidth]{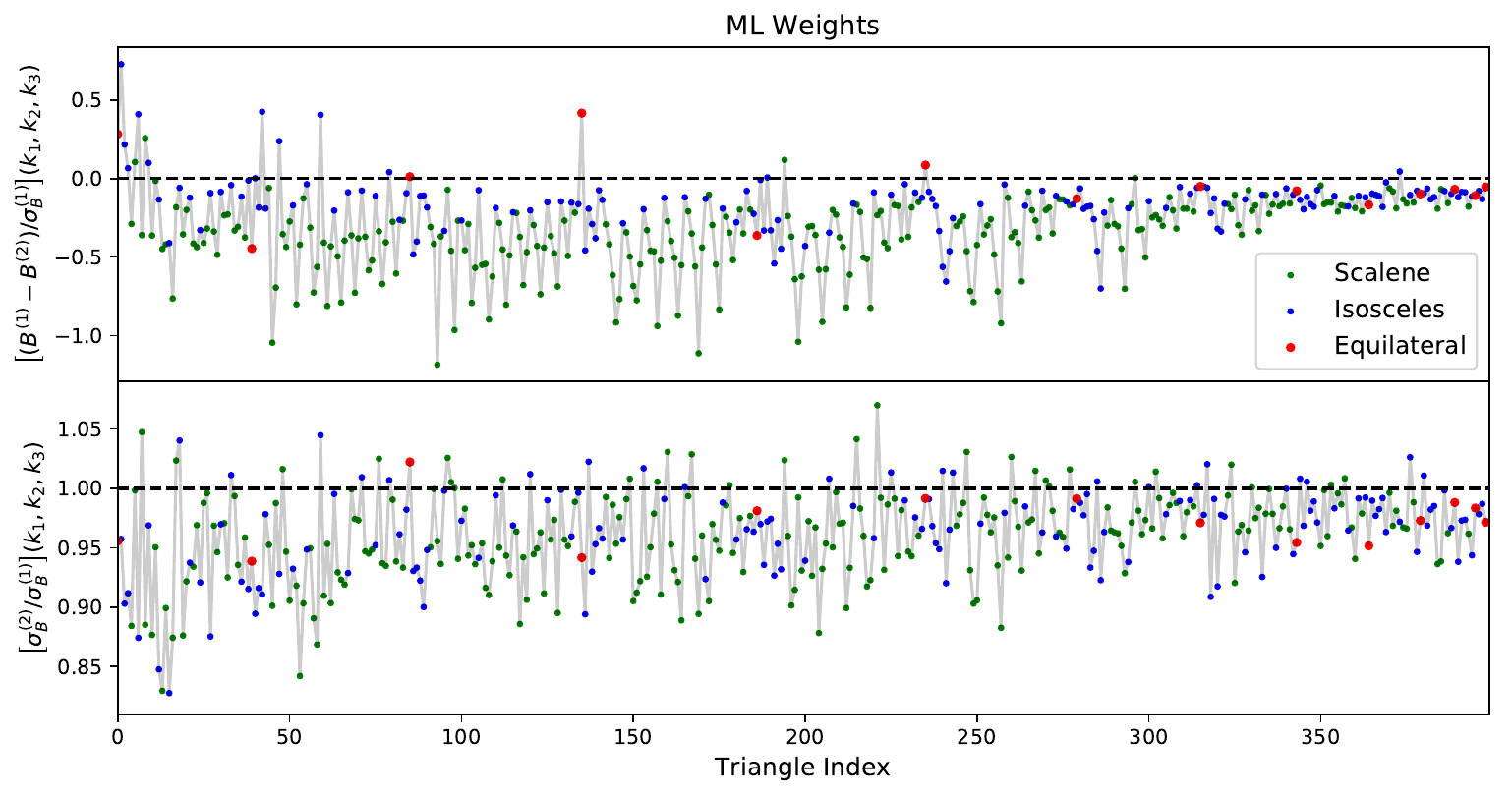}
    \caption{As Fig.\,\ref{fig: comparison-g3}, but adopting maximum-likelihood weights, $\Ci$ \eqref{eq: pixel-like-config}, rather than the short-scale FKP approximation \eqref{eq: fkp-weights}. As in \citep{2021PhRvD.103j3504P}, this is not expected to significantly improve constraints from a BOSS-like survey, since the window function contains mostly large-scale power and the shot-noise is large. Here, we find a $\sim$5\% decrease in the error-bars; however, this is associated with a non-trivial underprediction of the bispectrum. This may arise from the difficulty of inverting the survey mask (since it contains zeros, which do not quite match those of the data-set due to pixelation effects). In practice, we expect the difference to be absorbed in bispectrum bias parameters in any cosmological analysis, thus this is unlikely to be a cause for concern.}
    \label{fig: comparison-ml}
\end{figure}

Finally, we assess the impact of using ML weights in the bispectrum estimator (rather than the FKP scheme of \ref{eq: fkp-weights}). A similar comparison was performed for the power spectrum in \citep{2021PhRvD.103j3504P}; in that case, the ML weights were found to have negligible impact on both the output power spectra and the corresponding cosmological parameter constraints. This was attributed to the compact nature of the window function (\textit{i.e.}\ with power concentrated on large scales), the relatively high shot-noise, and the narrow $k$-bins adopted. The analogous results for the bispectrum are shown in Fig.\,\ref{fig: comparison-ml}. Unlike for the power spectrum, we find a significant ($\sim$\,$5\%$) reduction in the bispectrum variances, somewhat increasing on large scales. More troublingly, we find also a shift in the spectrum compared to that of FKP weights, which is not expected (since the bispectrum estimators should be unbiased for any choice of weighting matrix $\H$). We hypothesize that this is caused by zeros in the survey mask and the effects of gridding; since the data \resub{are} smoothed by the pixelation window but the underlying density $n(\vr)$ is not, there will be pixels in which $n(\vr)$ is zero but the data \resub{are} non-zero. These can cause instabilities upon inversion, which will affect the ML and FKP weights differently, particularly impacting the ML weights due to their more complex character. We defer further consideration of this effect to future work; however, we note that the principle effect is that of a global renormalization, which may be degenerate with the galaxy bias parameters.% (even when combined with the power spectrum), thus this is unlikely to be a major issue in practice.

\section{Summary}\label{sec: summary}

In the effort to harvest maximal information from current and future spectroscopic surveys, the bispectrum plays a key r\^ole. In combination with the galaxy power spectrum, it has been shown to significantly enhance constraining power on $\Lambda$CDM parameters \citep{2017MNRAS.467..928G,2021JCAP...03..021A}, as well as strengthening the bounds on additional phenomena such as primordial non-Gaussianity, neutrino masses and modified gravity \citep{2018MNRAS.478.1341K,2021JCAP...05..015M,2019JCAP...11..034C,2021JCAP...04..029H,2020arXiv201105771A}. However, several challenges must be overcome before its full potential can be realized, in particular regarding its measurement, modeling and dimensionality. In this work, we have introduced new estimators for the bispectrum, which, unlike previous approaches, output spectra which are not convolved with the survey window. These are analogous to the `quadratic estimators' used in power spectrum analyses of old (and \citep{2021PhRvD.103j3504P}), and are constructed by maximizing a pixel-likelihood based on the Edgeworth expansion. We provide general forms for the estimator using both optimal and FKP-like \citep{1994ApJ...426...23F} weighting schemes, with the latter providing a simple-to-implement approximation to the maximum-likelihood (ML) solution. The general estimator consists of two parts: a cubic estimator applied to the data, and a Fisher matrix computed from the survey mask, which acts as a deconvolution matrix (cf.\,\citep{2021arXiv210606324B}).

When considering the binned bispectrum monopole, the bispectrum estimators may be straightforwardly implemented using FFTs, and, if optimal weights are adopted, conjugate gradient descent methods. Their computation requires only the data, a random catalog, and knowledge of the survey mask. The estimators were applied to a suite of Patchy simulations and shown to be efficient, with the window function corrections taking $\sim$\,$150$ ($850)$ CPU-hours to compute with FKP (ML) weights, irrespective of the number of mocks analyzed. Comparison of the windowed and unwindowed bispectra showed highly consistent results on small-scales, but significant differences at low-$k$, due to the non-uniform survey geometry. Furthermore, the unwindowed bispectrum bins were found to be substantially less correlated than the windowed bins, simplifying their interpretation.

Measuring bispectra in this manner has a number of benefits over standard approaches. First and foremost, it avoids the need to window-convolve the theory model in an MCMC analysis. This is a computationally expensive procedure (involving double Hankel transforms \citep{2019MNRAS.484..364S}), thus its removal will significantly expedite parameter inference studies, and obviate the need for survey-specific analysis pipelines. Secondly, the estimators approach the minimum-variance solution (assuming weak non-Gaussianity). Whilst this is unlikely to be of importance for relatively uniform surveys such as BOSS \citep{2021PhRvD.103j3504P} (with the ML weights in fact causing a slight bias due to pixelation effects, and FKP weights being close to optimal), the unwindowed bispectrum estimator does feature an additional linear term which is shown to significantly reduce the error-bars on the large-scale bispectrum. An additional use concerns the global integral constraint, arising from the unknown mean survey density. Since our measurements are, by definition, unwindowed, these effects should be restricted to the first $k$-bin, thus may be ignored if this is not included in the analysis.

There are a number of ways in which the ideas of this work can be extended. In particular, one may wish to construct analogous estimators for the \textit{anisotropic} bispectrum multipoles \citep{2015PhRvD..92h3532S,2020JCAP...06..041G,2021arXiv210403976G,2019MNRAS.484..364S}. This follows the procedure noted in \S\ref{subsec: spec-bases}, and will allow all the information present in the three-point statistic to be captured. More abstract bases for the bispectrum would also be of use; essentially, one decomposes the bispectrum into a set of separable shapes, whose amplitudes can be directly estimated. For the simple case of constraining galaxy biases from a tree-level model, such a formalism exists \citep{2015PhRvD..91d3530S,2021JCAP...03..020S}, though we caution that finding such a decomposition in a more general case may be difficult. An additional application would be to compute windowless \textit{skew-spectra} using cubic estimators; such statistics have been shown to capture equivalent information to the galaxy bispectra \citep{2020JCAP...04..011M,2021JCAP...03..020S} in a friendlier form. Finally, an important application of the windowless bispectra presented herein is to the BOSS galaxy sample. Combining the above measurements with a bispectrum theory model in a full-shape framework will tighten parameter constraints from power-spectrum based analyses \citep{2020JCAP...05..042I,2021PhRvD.103b3538P}, further pinning down the Universe's composition and evolution.

% \begin{itemize}
%     \item Why cubic estimators are important? (windows, optimality, integral-constraints, ease for MCMC)
%     \item Can derive + implement efficiently
%     \item Discussion of Patchy results and versus windowed forms.
%     \item Future work I: more efficient separable basis, including RSD
%     \item Future work II: cosmology from the BOSS Bispectrum and Power Spectrum. (Link to bispectrum theory model).
% \end{itemize}

\begin{acknowledgments}
\footnotesize
It is a pleasure to thank Giovanni Cabass, Emanuele Castorina, Mikhail Ivanov, Azadeh Moradinezhad Dizgah, Marko Simonovi\'c, David Spergel and Matias Zaldarriaga for insightful conversations without which this work would not have been possible. \resub{We are additionally grateful for insightful comments from the referee.} OHEP acknowledges funding from the WFIRST program through NNG26PJ30C and NNN12AA01C, \resub{and thanks the Simons Foundation for additional support}.

The authors are pleased to acknowledge that the work reported in this paper was substantially performed using the Princeton Research Computing resources at Princeton University, which is a consortium of groups led by the Princeton Institute for Computational Science and Engineering (PICSciE) and the Office of Information Technology's Research Computing Division. Additional computations were performed on the Helios cluster at the Institute for Advanced Study.

\end{acknowledgments}

\appendix 

\section{Pixel Correlation Functions}\label{appen: pixel-cov}

Below, we present a full discussion of the pixel correlators used in this work, including the effects of discretization, particle weights, and redshift space distortions. Our covariance modeling represents an improvement to that of \citep{2021PhRvD.103j3504P}, particularly with regards to the treatment of pixelation, which was previously ignored.\footnote{This does not affect the conclusions of the earlier work, since their estimators were formulated as differences from a set of simulations, and thus unbiased by construction.}

Firstly, we define the data and random density fields as weighted sums over Dirac delta functions;
\beq
    \hat{n}'_g(\vr) = \sum_a w_d^a \delta_\mathrm{D}(\vr-\vr_a) \qquad \hat{n}'_r(\vr) = \alpha\,\sum_b w_r^b\delta_\mathrm{D}(\vr-\vr_b),
\eeq
where $a$ and $b$ represent data and random particle indices, and $\alpha \equiv \sum_a w_d^a/\sum_b w_r^b$. We add apostrophes to indicate weighted fields. Under a Poisson average, $\av{\hat{n}_g'(\vr)} = n(\vr)[1+\delta(\vr)]$ and $\av{\alpha\,\hat{n}'_r(\vr)} = n(\vr)$, where $n(\vr)$ is the background number density (assuming that the weights correct for sampling errors), \textit{i.e.}\ $n(\vr) = \lim_{\alpha\rightarrow0}\alpha\,\hat{n}_r(\vr)$.

After painting the density fields to a grid, the overdensity in pixel $i$ is given by
\beq\label{eq: pixelized-di-field}
    d^i = \int d\vr\,[\hat{n}_g'(\vr)-\alpha\,\hat{n}_r'(\vr)]\psi(\vr-\vr_i) = \sum_a w_d^a \psi(\vr_i-\vr_a) - \alpha\,\sum_b w_r^b \psi(\vr_i-\vr_b),
\eeq
where $\psi(\vr)$ is some mass-assignment (compensation) function with compact support. A typical choice for this is the `triangle-shaped-cloud' interpolation scheme, whose functional form can be found in \citep{2005ApJ...620..559J}.

\subsection{Two-Point Statistics}\label{appen: pixelized-2pt}

Given the pixelized density field \eqref{eq: pixelized-di-field}, we may construct the two-point covariance $\C^{ij} \equiv\av{d^id^j}$. To do so, we first define the pairwise expectations of the weighted fields, analogously to \eqref{eq: random-field-expectations}:
\beq
    \av{\hat{n}'_g(\vr)\hat{n}'_g(\vr')} &=& n(\vr)n(\vr')[1+\xi(\vr-\vr')]+\delta_D(\vr-\vr')\av{\sum_a (w_d^a)^2\delta_D(\vr-\vr_a)}\\\nonumber
    \av{\alpha\,\hat{n}'_r(\vr)\hat{n}'_g(\vr')} &=& n(\vr)n(\vr')\\\nonumber
    \av{\alpha^2\hat{n}'_r(\vr)\hat{n}'_r(\vr')} &=& n(\vr)n(\vr')+\alpha^2\delta_D(\vr-\vr')\av{\sum_b (w_r^b)^2\delta_D(\vr-\vr_b)}, 
\eeq
giving 
\beq\label{eq: pixel-like-config}
    \av{d^id^j} &\equiv& \C^{ij} = \Sig^{ij} + \N^{ij}\\\nonumber
    \Sig^{ij} &=& \int d\vr\,d\vr'\,n(\vr)n(\vr')\xi(\vr,\vr')\psi(\vr-\vr_i)\psi(\vr'-\vr_j), \quad \N^{ij} \equiv \int d\vr\,n'(\vr)\psi(\vr-\vr_i)\psi(\vr-\vr_j)\left[1+\alpha^2/\beta\right],
\eeq
where $\resub{\xi}(\vr,\vr')$ is the 2PCF of the underlying field $\delta$. Notably the gridded number density field $\N$ is not diagonal in pixel-space. \eqref{eq: pixel-like-config} uses the twice-weighted field $n'(\vr)$, defined by
\beq
    n'(\vr) = \lim_{\beta\rightarrow 0}\beta \sum_{b}(w_r^b)^2\delta_D(\vr-\vr_b) = \av{\sum_a (w_d^a)^2\delta_D(\vr-\vr_a)},
\eeq
assuming data and randoms to have the same weight distribution with $\beta=\sum_a (w_d^a)^2/\sum_b (w_r^b)^2$. For an alternative approximation, we can write
\beq
    \av{\sum_a (w_d^a)^2\delta_D(\vr-\vr_a)+\alpha^2 \sum_b (w_r^b)^2\delta_D(\vr-\vr_b)} \approx \frac{\av{w_d^2}+\alpha\av{w_r^2}}{\av{w_d}}n(\vr) \equiv \mathcal{S}n(\vr),
\eeq
in terms of the average squared weights. Note that the above relations depend on the \textit{ungridded} density field $n(\vr)$. In \citep{2021PhRvD.103j3504P}, it was assumed that this could be approximated by the gridded random field, however this necessarily introduces an extra factor of $\psi$ and can cause inversion errors when the number of randoms in a cell is small. Motivated by this and further testing of the power spectrum estimators, we instead use a smooth model for $n(\vr)$ computed from the survey mask and redshift distribution, as discussed in \S\ref{sec: implementation}.

% For a density field with uniform $n(\vr) = \bar{n}$ and uniform weights, this simplifies, since $n'(\vr) = w_d\bar{n}$, $\alpha = N_dw_d/N_rw_r$ (for $N_g$ galaxies and $N_r$ randoms), $\beta = N_dw_d^2/N_rw_r^2$, thus $n'(\vr)[1+\alpha^2/\beta]\rightarrow \bar{n}w_d[1+N_d/N_r]$. 

To obtain a tractable form of \eqref{eq: pixel-like-config}, we rewrite the 2PCF in Fourier-space, leading to
\beq\label{eq: sig-cov-pix}
    \Sig^{ij} &=& \int_{\vk} d\vr\,d\vr'\,n(\vr)n(\vr')\sum_\ell P_\ell(k)\mathcal{L}_\ell(\hk\cdot\hr')e^{i\vk\cdot(\vr-\vr')}\psi(\vr-\vr_i)\psi(\vr'-\vr_j),
\eeq
where we have expanded the power spectrum as a Legendre series about the first galaxy's line-of-sight using the Legendre polynomial $\mathcal{L}_\ell(\hk\cdot\hr')$, adopting the Yamamoto approximation \citep{2006PASJ...58...93Y}.\footnote{More nuanced approaches are also available, such as pairwise, or multiple, lines-of-sight \citep{2018MNRAS.476.4403C,2021PhRvD.103l3509P}.} Given that the background density $n(\vr)$ is also a pixelized (but not $\psi$-convolved) field, the $\vr$ and $\vr'$ summations can be evaluated as FFTs. Considering the action of $\Sig^{ij}$ and $\N^{ij}$ on a map $y_j$, we find
\beq\label{eq: pixel-correlator-tmp}
    [\Sig \vy]_i &=& \frac{1}{V_\mathrm{cell}}\ifft{\tilde\psi\,\fft{n\,\ifft{\sum_\ell P_\ell\,\frac{4\pi}{2\ell+1}\sum_{m=-\ell}^\ell Y_{\ell m}^{}\fft{Y_{\ell m}^*\,n\,\ifft{\tilde\psi\,\fft{y}}}}}}_i\\\nonumber
    [\N\vy]_i &=& \frac{\mathcal{S}}{V_\mathrm{cell}}\ifft{\tilde\psi\,\fft{n\,\ifft{\tilde\psi\,\fft{y}}}}_i,
\eeq
where we have additionally normalized by the pixel volume $V_{\rm cell}$, written $\psi$ in terms of its Fourier transform $\tilde\psi$, and expanded $\mathcal{L}_\ell(\hk\cdot\hr') = (4\pi)/(2\ell+1)\sum_{m=-\ell}^\ell Y_{\ell m}^{}(\hk)Y_{\ell m}^*(\hr')$ for spherical harmonics $Y^{}_{\ell m}$. Whilst somewhat imposing, \eqref{eq: pixel-correlator-tmp} is simple to implement, and requires just the gridded power spectrum, background number density field and spherical harmonics. We caution that $\tilde\psi$ is the \textit{forwards} gridding transform, which is the reciprocal of that usually applied in conventional power spectrum estimators.

A similar form may be derived for the FKP covariance \eqref{eq: fkp-weights}, including the effects of pixelation. Explicitly, we obtain:
\beq
    [\H_{\rm FKP}\vy]_i &=& \frac{1}{V_{\rm cell}}\ifft{\tilde\psi\,\fft{n(\mathcal{S}+nP_\fkp)\,\ifft{\tilde\psi\,\fft{y}}}}_i,
\eeq
for FKP power $P_{\rm FKP}\sim 10^4h^{-3}\mathrm{Mpc}^{3}$. This can be straightforwardly inverted:
\beq\label{eq: fkp-with-pix}
    [\Hi_{\rm FKP} \vy]_i &=& \ifft{\tilde\psi^{-1}\,\fft{\frac{V_{\rm cell}}{n(\mathcal{S}+P_\fkp n)}\,\ifft{\tilde\psi^{-1}\,\fft{y}}}}_i.
\eeq

% We additionally give the form of the inverse noise transform, which is analytic:
% \beq
%     [\N^{-1}y]_i &=& \frac{V_\mathrm{cell}}{1+\alpha^2/\beta}\ifft{\tilde\psi^{-1}\,\fft{\frac{1}{n'}\,\ifft{\tilde\psi^{-1}\,\fft{y}}}}_i
% \eeq
% where $\tilde\psi^{-1}(\vk)$ is the inverse mass assignment map.

Additionally of interest to this work is the covariance of uniformly distributed random particles with constant number density $\bar n$. In continuous form, this can be shown to equal
\beq\label{eq: unif-covariance-A}
    \mathsf{A}^{ij} &=& \bar{n}\int d\vr\,\psi(\vr-\vr_i)\psi(\vr-\vr_j).
\eeq
The action of $\mathsf{A}^{-1}$ on a field $\vec y$ is straightforwardly expressed in discrete form as
\beq\label{eq: unif-inv-covariance-Ainv}
    [\mathsf{A}^{-1}y]_i = \frac{V_\mathrm{cell}}{\bar{n}}\ifft{\tilde\psi^{-2}\fft{y}}_i.
\eeq
This can be straightforwardly tested by generating uniform maps $\vec a$ and verifying that $\av{\vec a^T\mathsf{A}^{-1}\vec a} = N_{\rm pix}$.

\subsection{Three-Point Statistics}\label{appen: pixelized-3pt}

We now consider the pixel-space three-point correlator, $\B^{ijk} \equiv \av{d^id^jd^k}$. Applying the same reasoning as before, this takes the full form (cf.\,\ref{eq: Bijk-def-Fourier})
\beq\label{eq: Bijk-pixel}
    \B^{ijk} &=& \int d\vr_1\,d\vr_2\,d\vr_3\,n(\vr_1)n(\vr_2)n(\vr_3)\int_{\vk_1\vk_2\vk_3}B(\vk_1,\vk_2,\vk_3; \hr_1,\hr_2,\hr_3)e^{i\vk_1\cdot\vr_1+i\vk_2\cdot\vr_2+i\vk_3\cdot\vr_3}\\\nonumber
    &&\quad\,\times\,\delD{\vk_1+\vk_2+\vk_3}\psi(\vr_1-\vr_i)\psi(\vr_2-\vr_j)\psi(\vr_3-\vr_k)\\\nonumber
    &&\,+\,\left[\int d\vr\,d\vr'\,n'(\vr)n(\vr')\int_{\vk}P(\vk; \hr,\hr')e^{i\vk\cdot(\vr-\vr')}\psi(\vr-\vr_i)\psi(\vr'-\vr_j)\psi(\vr-\vr_k) + \text{2 perms.}\right]\\\nonumber
    &&\,+\,[1+\alpha^3/\gamma]\int d\vr\,n''(\vr)\psi(\vr-\vr_i)\psi(\vr-\vr_j)\psi(\vr-\vr_k),
\eeq
where we allow for dependence of the power spectrum and bispectrum on the various lines-of-sight, with, for example, $P(\vk;\hr,\hr') = \sum_\ell P_\ell(k)\mathcal{L}_\ell(\hk\cdot\hr')$ in the Yamamoto formalism. Analogous to the two-point noise covariance, $\N$, we have introduced the doubly weighted field $n''(\vr) = \av{\sum_a (w_d^a)^3\delta_D(\vr-\vr_a)}$, with $\gamma = \sum_a (w_d^a)^3/\sum_b (w_r^b)^3$, to account for the re-weighted shot-noise contributions. These do not affect the statistics of this work.

From \eqref{eq: Bijk-pixel}, we can obtain the parameter derivatives:
\beq\label{eq: Bijk-pixel-deriv}
    \B_{,\alpha}^{ijk} &=& \int d\vr_1\,d\vr_2\,d\vr_3\,n(\vr_1)n(\vr_2)n(\vr_3)\int_{\vk_1\vk_2\vk_3}\omega_\alpha(\vk_1,\vk_2,\vk_3; \hr_1,\hr_2,\hr_3)e^{i\vk_1\cdot\vr_1+i\vk_2\cdot\vr_2+i\vk_3\cdot\vr_3}\\\nonumber
    &&\quad\,\times\,\delD{\vk_1+\vk_2+\vk_3}\psi(\vr_1-\vr_i)\psi(\vr_2-\vr_j)\psi(\vr_3-\vr_k),
\eeq
where $\omega_\alpha$ arises from the basis decomposition of \eqref{eq: Bk-decomposition}. Specializing to \resub{rotationally-averaged} binned bispectrum estimates, as in \eqref{eq: Bk-decomposition-binned}, and performing the integral over $\vr$, we find
\beq\label{eq: Bijk-pixel-deriv2}
    \B_{,\alpha}^{ijk} &=& \frac{1}{\Delta_\alpha}\int d\vr\,\left[\int_{\vk_1} d\vr_1\,e^{i\vk_1\cdot(\vr_1-\vr)}\Theta^a(k_1)n(\vr_1)\psi(\vr_1-\vr_i)\right]\left[\int_{\vk_2} d\vr_2\,e^{i\vk_2\cdot(\vr_2-\vr)}\Theta^b(k_2)n(\vr_2)\psi(\vr_2-\vr_j)\right]\\\nonumber
    &&\,\times\,\left[\int_{\vk_3}d\vr_3\,e^{i\vk_3\cdot(\vr_3-\vr)}\Theta^c(k_3)n(\vr_3)\psi(\vr_3-\vr_k)\right] + \text{5 perms.}
\eeq
Following a little algebra, we obtain the binned $\hat{q}_\alpha$ estimate of \eqref{eq: q-alpha-binned}, but with the redefinition
\beq\label{eq: g-alpha-pixel}
    g^a[\vy](\vr) &=& \int_{\vk}e^{-i\vk\cdot\vr}\Theta^a(k)\int d\vr' e^{i\vk\cdot\vr'}n(\vr')\int d\vx\,\psi(\vr'-\vx)[\Hi\vy](\vx)\\\nonumber
    g^a[\vy]_i &=& \ifft{\Theta^a\,\fft{n\,\ifft{\tilde\psi\,\fft{\Hi\vy}}}}_i
\eeq
(cf.\,\ref{eq: g-alpha-binned}), where we give the (properly normalized) gridded form involving discrete Fourier transforms in the second line. $\tilde{g}^a[\vy]_i$ follows similarly. Similarly, the Fisher matrix retains the form of \eqref{eq: F-ab-binned}, but with the redefined $\phi$ coefficients (and likewise the $\tilde\phi$ coefficients)
\beq\label{eq: phi-alpha-pixel}
    \phi_\alpha[\vy](\vx) &=&\frac{2}{\Delta_\alpha}\int d\vr\,g^{b}[\vy](\vr)g^{c}[\vy](\vr)\int_{\vk}e^{-i\vk\cdot\vr}\Theta^a(k)\int d\vr'\,n(\vr')\psi(\vr'-\vx)+\text{2 perms.}\\\nonumber
    \phi^i_\alpha[\vy] &=& \frac{2}{V_\mathrm{cell}\Delta_\alpha} \ifft{\tilde\psi\,\fft{n\,\ifft{\Theta^a\,\fft{g^b[\vy]g^c[\vy]}}}}_i + \text{2 perms.}
\eeq
(cf.\,\ref{eq: phi-alpha-binned}), which can again be implemented using FFTs. We note an additional overall normalization factor of $V_\mathrm{cell}$.

One further comment is of note. In \S\ref{sec: application}, we consider the estimators both with and without the pixelation corrections given above. Even when these effects are not included, it is important to add a factor $\tilde\psi^{-2}(\vk)$ to $\hat{q}_\alpha$ for the power spectrum estimators, or $\tilde\psi^{-1}(\vk_1)\tilde\psi^{-1}(\vk_2)\tilde\psi^{-1}(\vk_3)$ for the bispectrum estimators. This is necessary to remove the leading order pixelation effects in the statistic, and is included in all standard estimators. The treatment above, forward models this effect rather than simply dividing it out in the final step, thus should somewhat reduce the dependence on the grid-size, and slightly decrease the estimator error.

\section{Limiting Forms of the Estimators}\label{appen: limits}

It is instructive to consider the limiting forms of the ML bispectrum estimator on large- and small-scales. Assuming a uniform number density $n(\vr)\approx\bar n$ and ignoring pixelation effects and RSD, the covariance matrix given in \eqref{eq: simple-cov} becomes
\beq
    \C(\vr,\vr') &\to& \bar{n}^2 \int_{\vk}e^{i\vk\cdot(\vr-\vr')}P(\vk) + \bar{n}\,\delta_\mathrm{D}(\vr-\vr')\approx \begin{cases} \bar{n}^2 \int_{\vk}e^{i\vk\cdot(\vr-\vr')}P(\vk) & \bar{n}P\gg 1 \\ \bar{n}\,\delta_\mathrm{D}(\vr-\vr') &  \bar{n}P\ll 1,\end{cases}
\eeq
for some characteristic power spectrum amplitude $P$. The two approximations are appropriate on large ($P(\vk)$-dominated) and small (shot-noise dominated) scales respectively. Both forms are invertible:
\beq
    \Ci(\vr,\vr') &\to& \begin{cases} \frac{1}{\bar{n}^2} \int_{\vk}e^{i\vk\cdot(\vr-\vr')}\frac{1}{P(\vk)} & \bar{n}P\gg 1 \\ \frac{1}{\bar{n}}\delta_\mathrm{D}(\vr-\vr') &  \bar{n}\,P\ll 1,\end{cases}
\eeq
such that the full form of the estimators may be computed analytically. 

\subsection{Large-Scale Limit}
Following some computation, the large-scale limiting forms of $g^a$, $\tilde g^a$, $\phi_\alpha$, and $\tilde\phi_\alpha$ are found to be
\beq
    g^a[\vy](\vr) &\approx& \frac{1}{\bar n}\int_{\vk}e^{-i\vk\cdot\vr}\Theta^a(k)\frac{y(\vk)}{P(\vk)}, \qquad    \tilde g^a[\vy](\vr) \approx \frac{\bar n}{\bar n_A^2}\int_{\vk}e^{-i\vk\cdot\vr}\Theta^a(k)y(\vk)\\\nonumber
    \phi_\alpha[\va](\vr) &\approx& \frac{2}{\bar{n}\Delta_\alpha}\int_{\vk\,\vk'\,\vk''}\delD{\vk+\vk'+\vk''}e^{i\vk\cdot\vr}\Theta^a(k)\Theta^b(k')\Theta^c(k'')\frac{a(\vk')a(\vk'')}{P(\vk')P(\vk'')} + \text{2 perms.}\\\nonumber
    \tilde\phi_\alpha[\va](\vr) &=&\frac{2\bar n^3}{\bar{n}_A^4\Delta_\alpha}\int_{\vk\,\vk'\,\vk''}\delD{\vk+\vk'+\vk''}e^{i\vk\cdot\vr}\Theta^{a}(k)\Theta^b(k')\Theta^c(k'')a(\vk')a(\vk'')+\text{2 perms.}
\eeq
(from \ref{eq: g-alpha-binned}\,\&\,\ref{eq: phi-alpha-binned}), where we have assumed that the quantity of interest is the binned bispectrum monopole as in \eqref{eq: Bk-decomposition-binned}, and written the number density of uniform randoms as $\bar{n}_A$. Note that all $\Ci$-weighted density fields are multiplied $1/P(\vk)$ in this limit, as in \citep{2012PhRvD..86f3511F}. The expectations of these are also of use:
\beq
    \av{g^b[\va](\vr)\tilde g^c[\va](\vr)} &\approx& \delta^\mathrm{K}_{bc}\int_{\vk}\Theta^b(k)\frac{1}{P(\vk)}\\\nonumber
    %\av{\phi_\alpha[\vm](\vr)} &\approx& \delta^\mathrm{K}_{a0}\delta^\mathrm{K}_{bc}\times\frac{2\bar{n}}{\Delta_\alpha}\int_{\vk'}\frac{1}{P(\vk')}\Theta^b(k')+\text{2 perms.}\\\nonumber
    \av{\phi_\alpha[\va](\vr)} &\approx& \delta^{\rm K}_{a0}\delta^{\rm K}_{bc}\times\frac{2\bar n_A^2}{\bar{n}\Delta_\alpha}\int_{\vk'}\frac{1}{P^2(\vk)}\Theta^b(k) + \text{2 perms.}\\\nonumber
    %\av{\tilde\phi_\alpha[\va](\vr)} &\approx& \delta^\mathrm{K}_{a0}\delta^\mathrm{K}_{bc}\times\frac{2\bar{n}}{\Delta_\alpha}\int_{\vk'}\frac{1}{P(\vk')}\Theta^b(k')+\text{2 perms.}
    \av{\tilde\phi_\alpha[\va](\vr)} &\approx&\delta^{\rm K}_{a0}\delta^{\rm K}_{bc}\times\frac{2\bar n^3}{\bar{n}_A^2\Delta_\alpha}\int_{\vk}\Theta^b(k)+\text{2 perms.},
\eeq
where we assume $\av{a(\vk)a(\vk')} = \bar{n}^2_A\delD{\vk+\vk'}$ for Fourier-space maps $a(\vk)$. Notably $\av{g^b\tilde g^c}$ is diagonal with respect to $a,b$ and the $\av{\phi^\alpha}$ and $\av{\tilde\phi^\alpha}$ contribute only if one $k$-bin includes $\vk=\vec 0$ (designated by $\delta^\mathrm{K}_{a0}$).

These lead to the following $\hat{q}_\alpha^\mathrm{ML}$ and $F_{\alpha\beta}^\mathrm{ML}$ forms (from \ref{eq: q-alpha-binned}\,\&\,\ref{eq: F-ab-binned}):
\beq\label{eq: large-scale-tmp}
    \hat{q}^\mathrm{ML}_\alpha &\approx& \frac{1}{\bar{n}^3\Delta_\alpha}\int_{\vk\,\vk'\,\vk''}\delD{\vk+\vk'+\vk''}\Theta^a(k)\Theta^b(k')\Theta^c(k'')\frac{d(\vk)d(\vk')d(\vk'')}{P(\vk)P(\vk')P(\vk'')}\\\nonumber
    %&&\,-\frac{1}{\bar{n}\Delta_\alpha}\left(\delta^\mathrm{K}_{bc}\int_{\vk}\Theta^a(k)\frac{d(\vk)}{P(\vk)}\int_{\vp}\Theta^b(p)\frac{1}{P(\vp)}+\text{2 perms.}\right)\\\nonumber
    F_{\alpha\beta}^\mathrm{ML} &\approx& \frac{1}{12}\av{\phi_\alpha^i\Ci_{il}\phi_\alpha^l}\\\nonumber
    &\approx&  \frac{\delta^\mathrm{K}_{aa'}\left(\delta^\mathrm{K}_{bb'}\delta^\mathrm{K}_{cc'}+\delta^{\rm K}_{bc'}\delta^{\rm K}_{cb'}\right)}{3\Delta_\alpha\Delta_\beta}\int_{\vk\,\vk'\,\vk''}\delD{\vk+\vk'+\vk''}\Theta^a(k)\Theta^b(k')\Theta^c(k'')\frac{1}{P(\vk)P(\vk')P(\vk'')}+\text{8 perms.},
\eeq
dropping any zero-lag terms. Notably, neither the second term in $\hat{q}_\alpha^\mathrm{ML}$ nor the Fisher matrix terms involving $\av{\phi_\alpha^i}$ contribute in this limit. In the final line of \eqref{eq: large-scale-tmp}, we label the bins by $\alpha=\{a,b,c\}$, $\beta=\{a',b',c'\}$, and note that the Fisher matrix is non-zero only when the two triplets of indices contain the same members. Fixing $a\leq b\leq c$, the permutation symmetries allow simplification:
\beq
    F_{\alpha\beta}^\mathrm{ML} &\approx& \delta^\mathrm{K}_{\alpha\beta}\times\frac{1}{\Delta_\alpha}\int_{\vk\,\vk'\,\vk''}\delD{\vk+\vk'+\vk''}\Theta^a(k)\Theta^b(k')\Theta^c(k'')\frac{1}{P(\vk)P(\vk')P(\vk'')},
\eeq
giving a diagonal Fisher matrix. This estimator matches that of \citep{2012PhRvD..86f3511F}.

\subsection{Small-Scale Limit}
The small-scale (shot-noise dominated) form is simpler still. Via a similar calculation, we find
\beq
    g^a[\vy](\vr) &\approx& \int_{\vk}e^{-i\vk\cdot\vr}\Theta^a(k)y(\vk), \qquad \tilde g^a[\vy](\vr) \approx \frac{\bar n}{\bar n_A^2}\int_{\vk}e^{-i\vk\cdot\vr}\Theta^a(k)y(\vk)\\\nonumber
    \phi_\alpha[\va](\vr) &\approx& \frac{2\bar{n}}{\Delta_\alpha}\int_{\vk\,\vk'\,\vk''}\delD{\vk+\vk'+\vk''}e^{i\vk\cdot\vr}\Theta^a(k)\Theta^b(k')\Theta^c(k'')a(\vk')a(\vk'') + \text{2 perms.}\\\nonumber
    \tilde\phi_\alpha[\va](\vr) &\approx&\frac{2\bar n^3}{\bar{n}_A^4\Delta_\alpha}\int_{\vk\,\vk'\,\vk''}\delD{\vk+\vk'+\vk''}e^{i\vk\cdot\vr}\Theta^{a}(k)\Theta^b(k')\Theta^c(k'')a(\vk')a(\vk'')+\text{2 perms.},
\eeq
with expectations
\beq    
    \av{g^b[\va](\vr)\tilde g^c[\va](\vr)} &\approx& \delta^\mathrm{K}_{bc}\,\bar{n}\int_{\vk}\Theta^b(k)\\\nonumber
    \av{\phi_\alpha[\va](\vr)} &\approx& \delta^\mathrm{K}_{bc}\delta^\mathrm{K}_{a0}\times\frac{2\bar{n}\bar{n}_A^2}{\Delta_\alpha}\int_{\vk}\Theta^b(k)+\text{2 perms.}\\\nonumber
    \av{\tilde\phi_\alpha[\va](\vr)} &\approx&\delta^{\rm K}_{a0}\delta^{\rm K}_{bc}\times\frac{2\bar n^3}{\bar{n}_A^2\Delta_\alpha}\int_{\vk}\Theta^b(k)+\text{2 perms.}
\eeq
The $\av{\phi^\alpha}$ and $\av{\tilde\phi^\alpha}$ terms contribute only to zero-lag configurations, as before. These lead to the short-scale limits of the estimators
\beq
    \hat{q}_\alpha^\mathrm{ML} &\approx& \frac{1}{\Delta_\alpha}\int_{\vk\,\vk'\,\vk''}\delD{\vk+\vk'+\vk''}\Theta^a(k)\Theta^b(k')\Theta^c(k'')d(\vk)d(\vk')d(\vk'')\\\nonumber
    %&&\,-\,\frac{1}{\Delta_\alpha}\left(\delta^{\rm K}_{bc}\int_{\vk}\delD{\vk}\Theta^a(k)d(\vk)\right)\\\nonumber
    F_{\alpha\beta}^\mathrm{ML} &\approx& \frac{\bar{n}^3\delta^\mathrm{K}_{aa'}\left(\delta^\mathrm{K}_{bb'}\delta^\mathrm{K}_{cc'}+\delta^\mathrm{K}_{bc'}\delta^\mathrm{K}_{cb'}\right)}{3\Delta_\alpha\Delta_\beta}\int_{\vk\,\vk'\,\vk''}\delD{\vk+\vk'+\vk''}\Theta^a(k)\Theta^{b}(k')\Theta^c(k'') + \text{8 perms.},
\eeq
again ignoring zero-lag contributions. As before, incorporating the permutation symmetries leads to the simplified Fisher matrix
\beq
    F_{\alpha\beta}^\mathrm{ML} &\approx& \delta^\mathrm{K}_{\alpha\beta}\times\frac{\bar{n}^3}{\Delta_\alpha}\int_{\vk\,\vk'\,\vk''}\delD{\vk+\vk'+\vk''}\Theta^a(k)\Theta^{b}(k')\Theta^c(k'') \equiv  \delta^\mathrm{K}_{\alpha\beta}\frac{\bar{n}^3}{\Delta_\alpha}V_\alpha,
\eeq
where $V_\alpha$ is the combined bin volume. The limit of the full binned bispectrum estimator is thus:
\beq\label{eq: p-alpha-uniform-limit}
    \hat{b}_\alpha &\approx& \frac{1}{V_\alpha}\int_{\vk\,\vk'\,\vk''}\delD{\vk+\vk'+\vk''}\Theta^a(k)\Theta^b(k')\Theta^c(k'')\delta(\vk)\delta(\vk')\delta(\vk''),
\eeq
where we have written $d(\vk) = \bar{n}\,\delta(\vk)$ for fractional overdensity $\delta(\vk)$. This is the simple bispectrum estimator used in N-body simulations \citep[e.g.,][]{2017MNRAS.472.2436W}, often implemented by Fourier transforming the density field, then counting triangles.

% \begin{itemize}
%     \item List FKP and large-scale forms of $\hat{q}_\alpha$, $\phi_\alpha$ and $F_{\alpha\beta}$ and their assumptions.
% \end{itemize}

\bibliographystyle{JHEP}
\bibliography{adslib,otherlib}% Produces the bibliography via BibTeX.

\providecommand{\href}[2]{#2}\begingroup\raggedright\begin{thebibliography}{100}

\bibitem{1982PhLB..116..335L}
A.~D. {Linde}, \emph{{Scalar field fluctuations in the expanding universe and
  the new inflationary universe scenario}},
  \href{https://doi.org/10.1016/0370-2693(82)90293-3}{\emph{Physics Letters B}
  {\bfseries 116} (1982) 335}.

\bibitem{1982PhRvL..48.1220A}
A.~{Albrecht} and P.~J. {Steinhardt}, \emph{{Cosmology for Grand Unified
  Theories with Radiatively Induced Symmetry Breaking}},
  \href{https://doi.org/10.1103/PhysRevLett.48.1220}{\emph{\prl} {\bfseries 48}
  (1982) 1220}.

\bibitem{2020A&A...641A...6P}
{Planck Collaboration}, N.~{Aghanim}, Y.~{Akrami}, M.~{Ashdown}, J.~{Aumont},
  C.~{Baccigalupi} et~al., \emph{{Planck 2018 results. VI. Cosmological
  parameters}}, \href{https://doi.org/10.1051/0004-6361/201833910}{\emph{\aap}
  {\bfseries 641} (2020) A6}
  [\href{https://arxiv.org/abs/1807.06209}{{\ttfamily 1807.06209}}].

\bibitem{2015PhRvD..92l3522S}
M.~{Schmittfull}, Y.~{Feng}, F.~{Beutler}, B.~{Sherwin} and M.~Y. {Chu},
  \emph{{Eulerian BAO reconstructions and N -point statistics}},
  \href{https://doi.org/10.1103/PhysRevD.92.123522}{\emph{\prd} {\bfseries 92}
  (2015) 123522} [\href{https://arxiv.org/abs/1508.06972}{{\ttfamily
  1508.06972}}].

\bibitem{2016arXiv161100036D}
{DESI Collaboration}, A.~{Aghamousa}, J.~{Aguilar}, S.~{Ahlen}, S.~{Alam},
  L.~E. {Allen} et~al., \emph{{The DESI Experiment Part I: Science,Targeting,
  and Survey Design}}, {\emph{arXiv e-prints} (2016) arXiv:1611.00036}
  [\href{https://arxiv.org/abs/1611.00036}{{\ttfamily 1611.00036}}].

\bibitem{2011arXiv1110.3193L}
R.~{Laureijs}, J.~{Amiaux}, S.~{Arduini}, J.~L. {Augu{\`e}res},
  J.~{Brinchmann}, R.~{Cole} et~al., \emph{{Euclid Definition Study Report}},
  {\emph{arXiv e-prints} (2011) arXiv:1110.3193}
  [\href{https://arxiv.org/abs/1110.3193}{{\ttfamily 1110.3193}}].

\bibitem{2007ApJ...664..675E}
D.~J. {Eisenstein}, H.-J. {Seo}, E.~{Sirko} and D.~N. {Spergel},
  \emph{{Improving Cosmological Distance Measurements by Reconstruction of the
  Baryon Acoustic Peak}}, \href{https://doi.org/10.1086/518712}{\emph{\apj}
  {\bfseries 664} (2007) 675}
  [\href{https://arxiv.org/abs/astro-ph/0604362}{{\ttfamily
  astro-ph/0604362}}].

\bibitem{2017MNRAS.467..928G}
P.~{Gagrani} and L.~{Samushia}, \emph{{Information Content of the Angular
  Multipoles of Redshift-Space Galaxy Bispectrum}},
  \href{https://doi.org/10.1093/mnras/stx135}{\emph{\mnras} {\bfseries 467}
  (2017) 928} [\href{https://arxiv.org/abs/1610.03488}{{\ttfamily
  1610.03488}}].

\bibitem{2021JCAP...03..021A}
N.~{Agarwal}, V.~{Desjacques}, D.~{Jeong} and F.~{Schmidt}, \emph{{Information
  content in the redshift-space galaxy power spectrum and bispectrum}},
  \href{https://doi.org/10.1088/1475-7516/2021/03/021}{\emph{\jcap} {\bfseries
  2021} (2021) 021} [\href{https://arxiv.org/abs/2007.04340}{{\ttfamily
  2007.04340}}].

\bibitem{2018MNRAS.478.1341K}
D.~{Karagiannis}, A.~{Lazanu}, M.~{Liguori}, A.~{Raccanelli}, N.~{Bartolo} and
  L.~{Verde}, \emph{{Constraining primordial non-Gaussianity with bispectrum
  and power spectrum from upcoming optical and radio surveys}},
  \href{https://doi.org/10.1093/mnras/sty1029}{\emph{\mnras} {\bfseries 478}
  (2018) 1341} [\href{https://arxiv.org/abs/1801.09280}{{\ttfamily
  1801.09280}}].

\bibitem{2021JCAP...05..015M}
A.~{Moradinezhad Dizgah}, M.~{Biagetti}, E.~{Sefusatti}, V.~{Desjacques} and
  J.~{Nore{\~n}a}, \emph{{Primordial non-Gaussianity from biased tracers:
  likelihood analysis of real-space power spectrum and bispectrum}},
  \href{https://doi.org/10.1088/1475-7516/2021/05/015}{\emph{\jcap} {\bfseries
  2021} (2021) 015} [\href{https://arxiv.org/abs/2010.14523}{{\ttfamily
  2010.14523}}].

\bibitem{2019JCAP...11..034C}
A.~{Chudaykin} and M.~M. {Ivanov}, \emph{{Measuring neutrino masses with
  large-scale structure: Euclid forecast with controlled theoretical error}},
  \href{https://doi.org/10.1088/1475-7516/2019/11/034}{\emph{\jcap} {\bfseries
  2019} (2019) 034} [\href{https://arxiv.org/abs/1907.06666}{{\ttfamily
  1907.06666}}].

\bibitem{2021JCAP...04..029H}
C.~{Hahn} and F.~{Villaescusa-Navarro}, \emph{{Constraining
  M$_{{\ensuremath{\nu}}}$ with the bispectrum. Part II. The information
  content of the galaxy bispectrum monopole}},
  \href{https://doi.org/10.1088/1475-7516/2021/04/029}{\emph{\jcap} {\bfseries
  2021} (2021) 029} [\href{https://arxiv.org/abs/2012.02200}{{\ttfamily
  2012.02200}}].

\bibitem{2020arXiv201105771A}
S.~{Alam}, A.~{Aviles}, R.~{Bean}, Y.-C. {Cai}, M.~{Cautun}, J.~L.
  {Cervantes-Cota} et~al., \emph{{Testing the theory of gravity with DESI:
  estimators, predictions and simulation requirements}}, {\emph{arXiv e-prints}
  (2020) arXiv:2011.05771} [\href{https://arxiv.org/abs/2011.05771}{{\ttfamily
  2011.05771}}].

\bibitem{1998MNRAS.299..805H}
A.~F. {Heavens}, \emph{{Estimating non-Gaussianity in the microwave
  background}},
  \href{https://doi.org/10.1046/j.1365-8711.1998.01820.x}{\emph{\mnras}
  {\bfseries 299} (1998) 805}
  [\href{https://arxiv.org/abs/astro-ph/9804222}{{\ttfamily
  astro-ph/9804222}}].

\bibitem{2000MNRAS.313..141V}
L.~{Verde}, L.~{Wang}, A.~F. {Heavens} and M.~{Kamionkowski},
  \emph{{Large-scale structure, the cosmic microwave background and primordial
  non-Gaussianity}},
  \href{https://doi.org/10.1046/j.1365-8711.2000.03191.x}{\emph{\mnras}
  {\bfseries 313} (2000) 141}
  [\href{https://arxiv.org/abs/astro-ph/9906301}{{\ttfamily
  astro-ph/9906301}}].

\bibitem{2000PhRvD..62j3004G}
A.~{Gangui} and J.~{Martin}, \emph{{Best unbiased estimators for the
  three-point correlators of the cosmic microwave background radiation}},
  \href{https://doi.org/10.1103/PhysRevD.62.103004}{\emph{\prd} {\bfseries 62}
  (2000) 103004} [\href{https://arxiv.org/abs/astro-ph/0001361}{{\ttfamily
  astro-ph/0001361}}].

\bibitem{2003MNRAS.341..623S}
M.~G. {Santos}, A.~{Heavens}, A.~{Balbi}, J.~{Borrill}, P.~G. {Ferreira},
  S.~{Hanany} et~al., \emph{{Multiple methods for estimating the bispectrum of
  the cosmic microwave background with application to the MAXIMA data}},
  \href{https://doi.org/10.1046/j.1365-8711.2003.06438.x}{\emph{\mnras}
  {\bfseries 341} (2003) 623}
  [\href{https://arxiv.org/abs/astro-ph/0211123}{{\ttfamily
  astro-ph/0211123}}].

\bibitem{2005PhRvD..72d3003B}
D.~{Babich}, \emph{{Optimal estimation of non-Gaussianity}},
  \href{https://doi.org/10.1103/PhysRevD.72.043003}{\emph{\prd} {\bfseries 72}
  (2005) 043003} [\href{https://arxiv.org/abs/astro-ph/0503375}{{\ttfamily
  astro-ph/0503375}}].

\bibitem{2006JCAP...05..004C}
P.~{Creminelli}, A.~{Nicolis}, L.~{Senatore}, M.~{Tegmark} and
  M.~{Zaldarriaga}, \emph{{Limits on non-Gaussianities from WMAP data}},
  \href{https://doi.org/10.1088/1475-7516/2006/05/004}{\emph{\jcap} {\bfseries
  2006} (2006) 004} [\href{https://arxiv.org/abs/astro-ph/0509029}{{\ttfamily
  astro-ph/0509029}}].

\bibitem{2009PhRvD..80d3510F}
J.~R. {Fergusson} and E.~P.~S. {Shellard}, \emph{{Shape of primordial
  non-Gaussianity and the CMB bispectrum}},
  \href{https://doi.org/10.1103/PhysRevD.80.043510}{\emph{\prd} {\bfseries 80}
  (2009) 043510} [\href{https://arxiv.org/abs/0812.3413}{{\ttfamily
  0812.3413}}].

\bibitem{2014A&A...571A..24P}
{Planck Collaboration}, P.~A.~R. {Ade}, N.~{Aghanim}, C.~{Armitage-Caplan},
  M.~{Arnaud}, M.~{Ashdown} et~al., \emph{{Planck 2013 results. XXIV.
  Constraints on primordial non-Gaussianity}},
  \href{https://doi.org/10.1051/0004-6361/201321554}{\emph{\aap} {\bfseries
  571} (2014) A24} [\href{https://arxiv.org/abs/1303.5084}{{\ttfamily
  1303.5084}}].

\bibitem{1982ApJ...259..474F}
J.~N. {Fry} and M.~{Seldner}, \emph{{Transform analysis of the high-resolution
  Shane-Wirtanen Catalog - The power spectrum and the bispectrum}},
  \href{https://doi.org/10.1086/160184}{\emph{\apj} {\bfseries 259} (1982)
  474}.

\bibitem{2001ApJ...546..652S}
R.~{Scoccimarro}, H.~A. {Feldman}, J.~N. {Fry} and J.~A. {Frieman}, \emph{{The
  Bispectrum of IRAS Redshift Catalogs}},
  \href{https://doi.org/10.1086/318284}{\emph{\apj} {\bfseries 546} (2001) 652}
  [\href{https://arxiv.org/abs/astro-ph/0004087}{{\ttfamily
  astro-ph/0004087}}].

\bibitem{2005PhRvD..71f3001S}
E.~{Sefusatti} and R.~{Scoccimarro}, \emph{{Galaxy bias and halo-occupation
  numbers from large-scale clustering}},
  \href{https://doi.org/10.1103/PhysRevD.71.063001}{\emph{\prd} {\bfseries 71}
  (2005) 063001} [\href{https://arxiv.org/abs/astro-ph/0412626}{{\ttfamily
  astro-ph/0412626}}].

\bibitem{2015MNRAS.451..539G}
H.~{Gil-Mar{\'\i}n}, J.~{Nore{\~n}a}, L.~{Verde}, W.~J. {Percival},
  C.~{Wagner}, M.~{Manera} et~al., \emph{{The power spectrum and bispectrum of
  SDSS DR11 BOSS galaxies - I. Bias and gravity}},
  \href{https://doi.org/10.1093/mnras/stv961}{\emph{\mnras} {\bfseries 451}
  (2015) 539} [\href{https://arxiv.org/abs/1407.5668}{{\ttfamily 1407.5668}}].

\bibitem{2017MNRAS.465.1757G}
H.~{Gil-Mar{\'\i}n}, W.~J. {Percival}, L.~{Verde}, J.~R. {Brownstein}, C.-H.
  {Chuang}, F.-S. {Kitaura} et~al., \emph{{The clustering of galaxies in the
  SDSS-III Baryon Oscillation Spectroscopic Survey: RSD measurement from the
  power spectrum and bispectrum of the DR12 BOSS galaxies}},
  \href{https://doi.org/10.1093/mnras/stw2679}{\emph{\mnras} {\bfseries 465}
  (2017) 1757} [\href{https://arxiv.org/abs/1606.00439}{{\ttfamily
  1606.00439}}].

\bibitem{2017MNRAS.468.1070S}
Z.~{Slepian}, D.~J. {Eisenstein}, F.~{Beutler}, C.-H. {Chuang}, A.~J. {Cuesta},
  J.~{Ge} et~al., \emph{{The large-scale three-point correlation function of
  the SDSS BOSS DR12 CMASS galaxies}},
  \href{https://doi.org/10.1093/mnras/stw3234}{\emph{\mnras} {\bfseries 468}
  (2017) 1070} [\href{https://arxiv.org/abs/1512.02231}{{\ttfamily
  1512.02231}}].

\bibitem{2018MNRAS.478.4500P}
D.~W. {Pearson} and L.~{Samushia}, \emph{{A Detection of the Baryon Acoustic
  Oscillation features in the SDSS BOSS DR12 Galaxy Bispectrum}},
  \href{https://doi.org/10.1093/mnras/sty1266}{\emph{\mnras} {\bfseries 478}
  (2018) 4500} [\href{https://arxiv.org/abs/1712.04970}{{\ttfamily
  1712.04970}}].

\bibitem{2020JCAP...05..005D}
G.~{d'Amico}, J.~{Gleyzes}, N.~{Kokron}, K.~{Markovic}, L.~{Senatore},
  P.~{Zhang} et~al., \emph{{The cosmological analysis of the SDSS/BOSS data
  from the Effective Field Theory of Large-Scale Structure}},
  \href{https://doi.org/10.1088/1475-7516/2020/05/005}{\emph{\jcap} {\bfseries
  2020} (2020) 005} [\href{https://arxiv.org/abs/1909.05271}{{\ttfamily
  1909.05271}}].

\bibitem{1975ApJ...196....1P}
P.~J.~E. {Peebles} and E.~J. {Groth}, \emph{{Statistical analysis of catalogs
  of extragalactic objects. V. Three-point correlation function for the galaxy
  distribution in the Zwicky catalog.}},
  \href{https://doi.org/10.1086/153390}{\emph{\apj} {\bfseries 196} (1975) 1}.

\bibitem{2001ASPC..252..201P}
P.~J.~E. {Peebles}, \emph{{The Galaxy and Mass N-Point Correlation Functions: a
  Blast from the Past}},  in \emph{Historical Development of Modern Cosmology},
  V.~J. {Mart{\'\i}nez}, V.~{Trimble} and M.~J. {Pons-Border{\'\i}a}, eds.,
  vol.~252 of \emph{Astronomical Society of the Pacific Conference Series},
  p.~201, Jan., 2001, \href{https://arxiv.org/abs/astro-ph/0103040}{{\ttfamily
  astro-ph/0103040}}.

\bibitem{1998ApJ...503...37J}
Y.~P. {Jing} and G.~{B{\"o}rner}, \emph{{The Three-Point Correlation Function
  of Galaxies Determined from the Las Campanas Redshift Survey}},
  \href{https://doi.org/10.1086/305997}{\emph{\apj} {\bfseries 503} (1998) 37}
  [\href{https://arxiv.org/abs/astro-ph/9802011}{{\ttfamily
  astro-ph/9802011}}].

\bibitem{2004PASJ...56..415K}
I.~{Kayo}, Y.~{Suto}, R.~C. {Nichol}, J.~{Pan}, I.~{Szapudi}, A.~J. {Connolly}
  et~al., \emph{{Three-Point Correlation Functions of SDSS Galaxies in Redshift
  Space: Morphology, Color, and Luminosity Dependence}},
  \href{https://doi.org/10.1093/pasj/56.3.415}{\emph{\pasj} {\bfseries 56}
  (2004) 415} [\href{https://arxiv.org/abs/astro-ph/0403638}{{\ttfamily
  astro-ph/0403638}}].

\bibitem{2006MNRAS.368.1507N}
R.~C. {Nichol}, R.~K. {Sheth}, Y.~{Suto}, A.~J. {Gray}, I.~{Kayo}, R.~H.
  {Wechsler} et~al., \emph{{The effect of large-scale structure on the SDSS
  galaxy three-point correlation function}},
  \href{https://doi.org/10.1111/j.1365-2966.2006.10239.x}{\emph{\mnras}
  {\bfseries 368} (2006) 1507}
  [\href{https://arxiv.org/abs/astro-ph/0602548}{{\ttfamily
  astro-ph/0602548}}].

\bibitem{2011ApJ...737...97M}
F.~{Mar{\'\i}n}, \emph{{The Large-scale Three-point Correlation Function of
  Sloan Digital Sky Survey Luminous Red Galaxies}},
  \href{https://doi.org/10.1088/0004-637X/737/2/97}{\emph{\apj} {\bfseries 737}
  (2011) 97} [\href{https://arxiv.org/abs/1011.4530}{{\ttfamily 1011.4530}}].

\bibitem{2015MNRAS.449L..95G}
H.~{Guo}, Z.~{Zheng}, Y.~P. {Jing}, I.~{Zehavi}, C.~{Li}, D.~H. {Weinberg}
  et~al., \emph{{Modelling the redshift-space three-point correlation function
  in SDSS-III.}}, \href{https://doi.org/10.1093/mnrasl/slv020}{\emph{\mnras}
  {\bfseries 449} (2015) L95}
  [\href{https://arxiv.org/abs/1409.7389}{{\ttfamily 1409.7389}}].

\bibitem{2017MNRAS.469.1738S}
Z.~{Slepian}, D.~J. {Eisenstein}, J.~R. {Brownstein}, C.-H. {Chuang},
  H.~{Gil-Mar{\'\i}n}, S.~{Ho} et~al., \emph{{Detection of baryon acoustic
  oscillation features in the large-scale three-point correlation function of
  SDSS BOSS DR12 CMASS galaxies}},
  \href{https://doi.org/10.1093/mnras/stx488}{\emph{\mnras} {\bfseries 469}
  (2017) 1738} [\href{https://arxiv.org/abs/1607.06097}{{\ttfamily
  1607.06097}}].

\bibitem{2018MNRAS.474.2109S}
Z.~{Slepian}, D.~J. {Eisenstein}, J.~A. {Blazek}, J.~R. {Brownstein}, C.-H.
  {Chuang}, H.~{Gil-Mar{\'\i}n} et~al., \emph{{Constraining the baryon-dark
  matter relative velocity with the large-scale three-point correlation
  function of the SDSS BOSS DR12 CMASS galaxies}},
  \href{https://doi.org/10.1093/mnras/stx2723}{\emph{\mnras} {\bfseries 474}
  (2018) 2109} [\href{https://arxiv.org/abs/1607.06098}{{\ttfamily
  1607.06098}}].

\bibitem{npcf_algo}
O.~H.~E. {Philcox}, Z.~{Slepian}, J.~{Hou}, C.~{Warner}, R.~N. {Cahn} and D.~J.
  {Eisenstein}, \emph{{ENCORE: Estimating Galaxy $N$-point Correlation
  Functions in $\mathcal{O}(N_{\rm g}^2)$ Time}}, {\emph{arXiv e-prints} (2021)
  arXiv:2105.08722} [\href{https://arxiv.org/abs/2105.08722}{{\ttfamily
  2105.08722}}].

\bibitem{4pcf_boss}
O.~H.~E. {Philcox}, J.~{Hou} and Z.~{Slepian}, ``{A First Measurement of the
  BOSS Non-Gaussian Four-Point Function}.'' in prep.

\bibitem{2017MNRAS.472.2436W}
C.~A. {Watkinson}, S.~{Majumdar}, J.~R. {Pritchard} and R.~{Mondal}, \emph{{A
  fast estimator for the bispectrum and beyond - a practical method for
  measuring non-Gaussianity in 21-cm maps}},
  \href{https://doi.org/10.1093/mnras/stx2130}{\emph{\mnras} {\bfseries 472}
  (2017) 2436} [\href{https://arxiv.org/abs/1705.06284}{{\ttfamily
  1705.06284}}].

\bibitem{2015PhRvD..92h3532S}
R.~{Scoccimarro}, \emph{{Fast estimators for redshift-space clustering}},
  \href{https://doi.org/10.1103/PhysRevD.92.083532}{\emph{\prd} {\bfseries 92}
  (2015) 083532} [\href{https://arxiv.org/abs/1506.02729}{{\ttfamily
  1506.02729}}].

\bibitem{2017JCAP...12..020R}
D.~{Regan}, \emph{{An inventory of bispectrum estimators for redshift space
  distortions}},
  \href{https://doi.org/10.1088/1475-7516/2017/12/020}{\emph{\jcap} {\bfseries
  2017} (2017) 020} [\href{https://arxiv.org/abs/1708.05303}{{\ttfamily
  1708.05303}}].

\bibitem{2020MNRAS.492.1214P}
O.~H.~E. {Philcox} and D.~J. {Eisenstein}, \emph{{Computing the small-scale
  galaxy power spectrum and bispectrum in configuration space}},
  \href{https://doi.org/10.1093/mnras/stz3335}{\emph{\mnras} {\bfseries 492}
  (2020) 1214} [\href{https://arxiv.org/abs/1912.01010}{{\ttfamily
  1912.01010}}].

\bibitem{philcox_fastfft}
O.~H.~E. {Philcox}, \emph{{A faster Fourier transform? Computing small-scale
  power spectra and bispectra for cosmological simulations in
  $\mathcal{O}(N^{2})$ time}},
  \href{https://doi.org/10.1093/mnras/staa3882}{\emph{\mnras} {\bfseries 501}
  (2021) 4004} [\href{https://arxiv.org/abs/2005.01739}{{\ttfamily
  2005.01739}}].

\bibitem{1998ApJ...496..586S}
R.~{Scoccimarro}, S.~{Colombi}, J.~N. {Fry}, J.~A. {Frieman}, E.~{Hivon} and
  A.~{Melott}, \emph{{Nonlinear Evolution of the Bispectrum of Cosmological
  Perturbations}}, \href{https://doi.org/10.1086/305399}{\emph{\apj} {\bfseries
  496} (1998) 586} [\href{https://arxiv.org/abs/astro-ph/9704075}{{\ttfamily
  astro-ph/9704075}}].

\bibitem{1997MNRAS.290..651M}
S.~{Matarrese}, L.~{Verde} and A.~F. {Heavens}, \emph{{Large-scale bias in the
  Universe: bispectrum method}},
  \href{https://doi.org/10.1093/mnras/290.4.651}{\emph{\mnras} {\bfseries 290}
  (1997) 651} [\href{https://arxiv.org/abs/astro-ph/9706059}{{\ttfamily
  astro-ph/9706059}}].

\bibitem{1998MNRAS.300..747V}
L.~{Verde}, A.~F. {Heavens}, S.~{Matarrese} and L.~{Moscardini},
  \emph{{Large-scale bias in the Universe - II. Redshift-space bispectrum}},
  \href{https://doi.org/10.1046/j.1365-8711.1998.01937.x}{\emph{\mnras}
  {\bfseries 300} (1998) 747}
  [\href{https://arxiv.org/abs/astro-ph/9806028}{{\ttfamily
  astro-ph/9806028}}].

\bibitem{1999ApJ...517..531S}
R.~{Scoccimarro}, H.~M.~P. {Couchman} and J.~A. {Frieman}, \emph{{The
  Bispectrum as a Signature of Gravitational Instability in Redshift Space}},
  \href{https://doi.org/10.1086/307220}{\emph{\apj} {\bfseries 517} (1999) 531}
  [\href{https://arxiv.org/abs/astro-ph/9808305}{{\ttfamily
  astro-ph/9808305}}].

\bibitem{2001MNRAS.325.1312S}
R.~{Scoccimarro} and H.~M.~P. {Couchman}, \emph{{A fitting formula for the
  non-linear evolution of the bispectrum}},
  \href{https://doi.org/10.1046/j.1365-8711.2001.04281.x}{\emph{\mnras}
  {\bfseries 325} (2001) 1312}
  [\href{https://arxiv.org/abs/astro-ph/0009427}{{\ttfamily
  astro-ph/0009427}}].

\bibitem{2000ApJ...544..597S}
R.~{Scoccimarro}, \emph{{The Bispectrum: From Theory to Observations}},
  \href{https://doi.org/10.1086/317248}{\emph{\apj} {\bfseries 544} (2000) 597}
  [\href{https://arxiv.org/abs/astro-ph/0004086}{{\ttfamily
  astro-ph/0004086}}].

\bibitem{2015JCAP...05..007B}
T.~{Baldauf}, L.~{Mercolli}, M.~{Mirbabayi} and E.~{Pajer}, \emph{{The
  bispectrum in the Effective Field Theory of Large Scale Structure}},
  \href{https://doi.org/10.1088/1475-7516/2015/05/007}{\emph{\jcap} {\bfseries
  2015} (2015) 007} [\href{https://arxiv.org/abs/1406.4135}{{\ttfamily
  1406.4135}}].

\bibitem{2015JCAP...10..039A}
R.~E. {Angulo}, S.~{Foreman}, M.~{Schmittfull} and L.~{Senatore}, \emph{{The
  one-loop matter bispectrum in the Effective Field Theory of Large Scale
  Structures}},
  \href{https://doi.org/10.1088/1475-7516/2015/10/039}{\emph{\jcap} {\bfseries
  2015} (2015) 039} [\href{https://arxiv.org/abs/1406.4143}{{\ttfamily
  1406.4143}}].

\bibitem{2020PhRvD.102l3541N}
T.~{Nishimichi}, G.~{D'Amico}, M.~M. {Ivanov}, L.~{Senatore},
  M.~{Simonovi{\'c}}, M.~{Takada} et~al., \emph{{Blinded challenge for
  precision cosmology with large-scale structure: Results from effective field
  theory for the redshift-space galaxy power spectrum}},
  \href{https://doi.org/10.1103/PhysRevD.102.123541}{\emph{\prd} {\bfseries
  102} (2020) 123541} [\href{https://arxiv.org/abs/2003.08277}{{\ttfamily
  2003.08277}}].

\bibitem{2001PhRvD..64d3516C}
A.~{Cooray}, \emph{{Squared temperature-temperature power spectrum as a probe
  of the CMB bispectrum}},
  \href{https://doi.org/10.1103/PhysRevD.64.043516}{\emph{\prd} {\bfseries 64}
  (2001) 043516} [\href{https://arxiv.org/abs/astro-ph/0105415}{{\ttfamily
  astro-ph/0105415}}].

\bibitem{1998astro.ph.12271M}
D.~{Munshi}, A.~L. {Melott} and P.~{Coles}, \emph{{Generalised Cumulant
  Correlators and Hierarchical Clustering}}, {\emph{arXiv e-prints} (1998)
  astro} [\href{https://arxiv.org/abs/astro-ph/9812271}{{\ttfamily
  astro-ph/9812271}}].

\bibitem{2015PhRvD..91d3530S}
M.~{Schmittfull}, T.~{Baldauf} and U.~{Seljak}, \emph{{Near optimal bispectrum
  estimators for large-scale structure}},
  \href{https://doi.org/10.1103/PhysRevD.91.043530}{\emph{\prd} {\bfseries 91}
  (2015) 043530} [\href{https://arxiv.org/abs/1411.6595}{{\ttfamily
  1411.6595}}].

\bibitem{2013ApJ...762..115O}
D.~{Obreschkow}, C.~{Power}, M.~{Bruderer} and C.~{Bonvin}, \emph{{A Robust
  Measure of Cosmic Structure beyond the Power Spectrum: Cosmic Filaments and
  the Temperature of Dark Matter}},
  \href{https://doi.org/10.1088/0004-637X/762/2/115}{\emph{\apj} {\bfseries
  762} (2013) 115} [\href{https://arxiv.org/abs/1211.5213}{{\ttfamily
  1211.5213}}].

\bibitem{2015MNRAS.453..797E}
A.~{Eggemeier}, T.~{Battefeld}, R.~E. {Smith} and J.~{Niemeyer}, \emph{{The
  anisotropic line correlation function as a probe of anisotropies in galaxy
  surveys}}, \href{https://doi.org/10.1093/mnras/stv1602}{\emph{\mnras}
  {\bfseries 453} (2015) 797}
  [\href{https://arxiv.org/abs/1504.04036}{{\ttfamily 1504.04036}}].

\bibitem{2014JCAP...05..048C}
C.-T. {Chiang}, C.~{Wagner}, F.~{Schmidt} and E.~{Komatsu},
  \emph{{Position-dependent power spectrum of the large-scale structure: a
  novel method to measure the squeezed-limit bispectrum}},
  \href{https://doi.org/10.1088/1475-7516/2014/05/048}{\emph{\jcap} {\bfseries
  2014} (2014) 048} [\href{https://arxiv.org/abs/1403.3411}{{\ttfamily
  1403.3411}}].

\bibitem{2015JCAP...09..028C}
C.-T. {Chiang}, C.~{Wagner}, A.~G. {S{\'a}nchez}, F.~{Schmidt} and
  E.~{Komatsu}, \emph{{Position-dependent correlation function from the
  SDSS-III Baryon Oscillation Spectroscopic Survey Data Release 10 CMASS
  sample}}, \href{https://doi.org/10.1088/1475-7516/2015/09/028}{\emph{\jcap}
  {\bfseries 2015} (2015) 028}
  [\href{https://arxiv.org/abs/1504.03322}{{\ttfamily 1504.03322}}].

\bibitem{2012PhRvD..86f3511F}
J.~R. {Fergusson}, D.~M. {Regan} and E.~P.~S. {Shellard}, \emph{{Rapid
  separable analysis of higher order correlators in large-scale structure}},
  \href{https://doi.org/10.1103/PhysRevD.86.063511}{\emph{\prd} {\bfseries 86}
  (2012) 063511} [\href{https://arxiv.org/abs/1008.1730}{{\ttfamily
  1008.1730}}].

\bibitem{2012PhRvD..86l3524R}
D.~M. {Regan}, M.~M. {Schmittfull}, E.~P.~S. {Shellard} and J.~R. {Fergusson},
  \emph{{Universal non-Gaussian initial conditions for N-body simulations}},
  \href{https://doi.org/10.1103/PhysRevD.86.123524}{\emph{\prd} {\bfseries 86}
  (2012) 123524} [\href{https://arxiv.org/abs/1108.3813}{{\ttfamily
  1108.3813}}].

\bibitem{2013PhRvD..88f3512S}
M.~M. {Schmittfull}, D.~M. {Regan} and E.~P.~S. {Shellard}, \emph{{Fast
  estimation of gravitational and primordial bispectra in large scale
  structures}}, \href{https://doi.org/10.1103/PhysRevD.88.063512}{\emph{\prd}
  {\bfseries 88} (2013) 063512}
  [\href{https://arxiv.org/abs/1207.5678}{{\ttfamily 1207.5678}}].

\bibitem{2021JCAP...03..105B}
J.~{Byun}, A.~{Oddo}, C.~{Porciani} and E.~{Sefusatti}, \emph{{Towards
  cosmological constraints from the compressed modal bispectrum: a robust
  comparison of real-space bispectrum estimators}},
  \href{https://doi.org/10.1088/1475-7516/2021/03/105}{\emph{\jcap} {\bfseries
  2021} (2021) 105} [\href{https://arxiv.org/abs/2010.09579}{{\ttfamily
  2010.09579}}].

\bibitem{2000MNRAS.317..965H}
A.~F. {Heavens}, R.~{Jimenez} and O.~{Lahav}, \emph{{Massive lossless data
  compression and multiple parameter estimation from galaxy spectra}},
  \href{https://doi.org/10.1046/j.1365-8711.2000.03692.x}{\emph{\mnras}
  {\bfseries 317} (2000) 965}
  [\href{https://arxiv.org/abs/astro-ph/9911102}{{\ttfamily
  astro-ph/9911102}}].

\bibitem{2018MNRAS.476.4045G}
D.~{Gualdi}, M.~{Manera}, B.~{Joachimi} and O.~{Lahav}, \emph{{Maximal
  compression of the redshift-space galaxy power spectrum and bispectrum}},
  \href{https://doi.org/10.1093/mnras/sty261}{\emph{\mnras} {\bfseries 476}
  (2018) 4045} [\href{https://arxiv.org/abs/1709.03600}{{\ttfamily
  1709.03600}}].

\bibitem{2019MNRAS.484.3713G}
D.~{Gualdi}, H.~{Gil-Mar{\'\i}n}, R.~L. {Schuhmann}, M.~{Manera}, B.~{Joachimi}
  and O.~{Lahav}, \emph{{Enhancing BOSS bispectrum cosmological constraints
  with maximal compression}},
  \href{https://doi.org/10.1093/mnras/stz051}{\emph{\mnras} {\bfseries 484}
  (2019) 3713} [\href{https://arxiv.org/abs/1806.02853}{{\ttfamily
  1806.02853}}].

\bibitem{2019MNRAS.484L..29G}
D.~{Gualdi}, H.~{Gil-Mar{\'\i}n}, M.~{Manera}, B.~{Joachimi} and O.~{Lahav},
  \emph{{Geometrical compression: a new method to enhance the BOSS galaxy
  bispectrum monopole constraints}},
  \href{https://doi.org/10.1093/mnrasl/sly242}{\emph{\mnras} {\bfseries 484}
  (2019) L29} [\href{https://arxiv.org/abs/1901.00987}{{\ttfamily
  1901.00987}}].

\bibitem{2021PhRvD.103d3508P}
O.~H.~E. {Philcox}, M.~M. {Ivanov}, M.~{Zaldarriaga}, M.~{Simonovi{\'c}} and
  M.~{Schmittfull}, \emph{{Fewer mocks and less noise: Reducing the
  dimensionality of cosmological observables with subspace projections}},
  \href{https://doi.org/10.1103/PhysRevD.103.043508}{\emph{\prd} {\bfseries
  103} (2021) 043508} [\href{https://arxiv.org/abs/2009.03311}{{\ttfamily
  2009.03311}}].

\bibitem{2018MNRAS.476L..60A}
J.~{Alsing} and B.~{Wandelt}, \emph{{Generalized massive optimal data
  compression}}, \href{https://doi.org/10.1093/mnrasl/sly029}{\emph{\mnras}
  {\bfseries 476} (2018) L60}
  [\href{https://arxiv.org/abs/1712.00012}{{\ttfamily 1712.00012}}].

\bibitem{2019MNRAS.484..364S}
N.~S. {Sugiyama}, S.~{Saito}, F.~{Beutler} and H.-J. {Seo}, \emph{{A complete
  FFT-based decomposition formalism for the redshift-space bispectrum}},
  \href{https://doi.org/10.1093/mnras/sty3249}{\emph{\mnras} {\bfseries 484}
  (2019) 364} [\href{https://arxiv.org/abs/1803.02132}{{\ttfamily
  1803.02132}}].

\bibitem{2021PhRvD.103j3504P}
O.~H.~E. {Philcox}, \emph{{Cosmology without window functions: Quadratic
  estimators for the galaxy power spectrum}},
  \href{https://doi.org/10.1103/PhysRevD.103.103504}{\emph{\prd} {\bfseries
  103} (2021) 103504} [\href{https://arxiv.org/abs/2012.09389}{{\ttfamily
  2012.09389}}].

\bibitem{2021arXiv210606324B}
F.~{Beutler} and P.~{McDonald}, \emph{{Unified galaxy power spectrum
  measurements from 6dFGS, BOSS, and eBOSS}}, {\emph{arXiv e-prints} (2021)
  arXiv:2106.06324} [\href{https://arxiv.org/abs/2106.06324}{{\ttfamily
  2106.06324}}].

\bibitem{1997PhRvD..55.5895T}
M.~{Tegmark}, \emph{{How to measure CMB power spectra without losing
  information}}, \href{https://doi.org/10.1103/PhysRevD.55.5895}{\emph{\prd}
  {\bfseries 55} (1997) 5895}
  [\href{https://arxiv.org/abs/astro-ph/9611174}{{\ttfamily
  astro-ph/9611174}}].

\bibitem{1998ApJ...499..555T}
M.~{Tegmark}, A.~J.~S. {Hamilton}, M.~A. {Strauss}, M.~S. {Vogeley} and A.~S.
  {Szalay}, \emph{{Measuring the Galaxy Power Spectrum with Future Redshift
  Surveys}}, \href{https://doi.org/10.1086/305663}{\emph{\apj} {\bfseries 499}
  (1998) 555} [\href{https://arxiv.org/abs/astro-ph/9708020}{{\ttfamily
  astro-ph/9708020}}].

\bibitem{1997ApJ...480...22T}
M.~{Tegmark}, A.~N. {Taylor} and A.~F. {Heavens}, \emph{{Karhunen-Lo{\`e}ve
  Eigenvalue Problems in Cosmology: How Should We Tackle Large Data Sets?}},
  \href{https://doi.org/10.1086/303939}{\emph{\apj} {\bfseries 480} (1997) 22}
  [\href{https://arxiv.org/abs/astro-ph/9603021}{{\ttfamily
  astro-ph/9603021}}].

\bibitem{1998PhRvD..57.2117B}
J.~R. {Bond}, A.~H. {Jaffe} and L.~{Knox}, \emph{{Estimating the power spectrum
  of the cosmic microwave background}},
  \href{https://doi.org/10.1103/PhysRevD.57.2117}{\emph{\prd} {\bfseries 57}
  (1998) 2117} [\href{https://arxiv.org/abs/astro-ph/9708203}{{\ttfamily
  astro-ph/9708203}}].

\bibitem{1999ApJ...510..551O}
S.~P. {Oh}, D.~N. {Spergel} and G.~{Hinshaw}, \emph{{An Efficient Technique to
  Determine the Power Spectrum from Cosmic Microwave Background Sky Maps}},
  \href{https://doi.org/10.1086/306629}{\emph{\apj} {\bfseries 510} (1999) 551}
  [\href{https://arxiv.org/abs/astro-ph/9805339}{{\ttfamily
  astro-ph/9805339}}].

\bibitem{2002MNRAS.335..887T}
M.~{Tegmark}, A.~J.~S. {Hamilton} and Y.~{Xu}, \emph{{The power spectrum of
  galaxies in the 2dF 100k redshift survey}},
  \href{https://doi.org/10.1046/j.1365-8711.2002.05622.x}{\emph{\mnras}
  {\bfseries 335} (2002) 887}
  [\href{https://arxiv.org/abs/astro-ph/0111575}{{\ttfamily
  astro-ph/0111575}}].

\bibitem{2002ApJ...571..191T}
M.~{Tegmark}, S.~{Dodelson}, D.~J. {Eisenstein}, V.~{Narayanan},
  R.~{Scoccimarro}, R.~{Scranton} et~al., \emph{{The Angular Power Spectrum of
  Galaxies from Early Sloan Digital Sky Survey Data}},
  \href{https://doi.org/10.1086/339894}{\emph{\apj} {\bfseries 571} (2002) 191}
  [\href{https://arxiv.org/abs/astro-ph/0107418}{{\ttfamily
  astro-ph/0107418}}].

\bibitem{2005astro.ph..3603H}
A.~J.~S. {Hamilton}, \emph{{Power Spectrum Estimation I. Basics}}, {\emph{arXiv
  e-prints} (2005) astro}
  [\href{https://arxiv.org/abs/astro-ph/0503603}{{\ttfamily
  astro-ph/0503603}}].

\bibitem{2005astro.ph..3604H}
A.~J.~S. {Hamilton}, \emph{{Power Spectrum Estimation II. Linear Maximum
  Likelihood}}, {\emph{arXiv e-prints} (2005) astro}
  [\href{https://arxiv.org/abs/astro-ph/0503604}{{\ttfamily
  astro-ph/0503604}}].

\bibitem{2004ApJ...606..702T}
M.~{Tegmark}, M.~R. {Blanton}, M.~A. {Strauss}, F.~{Hoyle}, D.~{Schlegel},
  R.~{Scoccimarro} et~al., \emph{{The Three-Dimensional Power Spectrum of
  Galaxies from the Sloan Digital Sky Survey}},
  \href{https://doi.org/10.1086/382125}{\emph{\apj} {\bfseries 606} (2004) 702}
  [\href{https://arxiv.org/abs/astro-ph/0310725}{{\ttfamily
  astro-ph/0310725}}].

\bibitem{2011MNRAS.417....2S}
K.~M. {Smith} and M.~{Zaldarriaga}, \emph{{Algorithms for bispectra:
  forecasting, optimal analysis and simulation}},
  \href{https://doi.org/10.1111/j.1365-2966.2010.18175.x}{\emph{\mnras}
  {\bfseries 417} (2011) 2}
  [\href{https://arxiv.org/abs/astro-ph/0612571}{{\ttfamily
  astro-ph/0612571}}].

\bibitem{2020JCAP...05..042I}
M.~M. {Ivanov}, M.~{Simonovi{\'c}} and M.~{Zaldarriaga}, \emph{{Cosmological
  parameters from the BOSS galaxy power spectrum}},
  \href{https://doi.org/10.1088/1475-7516/2020/05/042}{\emph{\jcap} {\bfseries
  2020} (2020) 042} [\href{https://arxiv.org/abs/1909.05277}{{\ttfamily
  1909.05277}}].

\bibitem{2020MNRAS.497.1684S}
N.~S. {Sugiyama}, S.~{Saito}, F.~{Beutler} and H.-J. {Seo}, \emph{{Perturbation
  theory approach to predict the covariance matrices of the galaxy power
  spectrum and bispectrum in redshift space}},
  \href{https://doi.org/10.1093/mnras/staa1940}{\emph{\mnras} {\bfseries 497}
  (2020) 1684} [\href{https://arxiv.org/abs/1908.06234}{{\ttfamily
  1908.06234}}].

\bibitem{2020JCAP...06..041G}
D.~{Gualdi} and L.~{Verde}, \emph{{Galaxy redshift-space bispectrum: the
  importance of being anisotropic}},
  \href{https://doi.org/10.1088/1475-7516/2020/06/041}{\emph{\jcap} {\bfseries
  2020} (2020) 041} [\href{https://arxiv.org/abs/2003.12075}{{\ttfamily
  2003.12075}}].

\bibitem{2021arXiv210403976G}
D.~{Gualdi}, H.~{Gil-Marin} and L.~{Verde}, \emph{{Joint analysis of
  anisotropic power spectrum, bispectrum and trispectrum: application to N-body
  simulations}}, {\emph{arXiv e-prints} (2021) arXiv:2104.03976}
  [\href{https://arxiv.org/abs/2104.03976}{{\ttfamily 2104.03976}}].

\bibitem{1994ApJ...426...23F}
H.~A. {Feldman}, N.~{Kaiser} and J.~A. {Peacock}, \emph{{Power-Spectrum
  Analysis of Three-dimensional Redshift Surveys}},
  \href{https://doi.org/10.1086/174036}{\emph{\apj} {\bfseries 426} (1994) 23}
  [\href{https://arxiv.org/abs/astro-ph/9304022}{{\ttfamily
  astro-ph/9304022}}].

\bibitem{2006PASJ...58...93Y}
K.~{Yamamoto}, M.~{Nakamichi}, A.~{Kamino}, B.~A. {Bassett} and H.~{Nishioka},
  \emph{{A Measurement of the Quadrupole Power Spectrum in the Clustering of
  the 2dF QSO Survey}},
  \href{https://doi.org/10.1093/pasj/58.1.93}{\emph{\pasj} {\bfseries 58}
  (2006) 93} [\href{https://arxiv.org/abs/astro-ph/0505115}{{\ttfamily
  astro-ph/0505115}}].

\bibitem{2017JCAP...07..002H}
N.~{Hand}, Y.~{Li}, Z.~{Slepian} and U.~{Seljak}, \emph{{An optimal FFT-based
  anisotropic power spectrum estimator}},
  \href{https://doi.org/10.1088/1475-7516/2017/07/002}{\emph{\jcap} {\bfseries
  2017} (2017) 002} [\href{https://arxiv.org/abs/1704.02357}{{\ttfamily
  1704.02357}}].

\bibitem{2019JCAP...09..010C}
E.~{Castorina}, N.~{Hand}, U.~{Seljak}, F.~{Beutler}, C.-H. {Chuang}, C.~{Zhao}
  et~al., \emph{{Redshift-weighted constraints on primordial non-Gaussianity
  from the clustering of the eBOSS DR14 quasars in Fourier space}},
  \href{https://doi.org/10.1088/1475-7516/2019/09/010}{\emph{\jcap} {\bfseries
  2019} (2019) 010} [\href{https://arxiv.org/abs/1904.08859}{{\ttfamily
  1904.08859}}].

\bibitem{2017arXiv170903452S}
E.~{Sellentin}, A.~H. {Jaffe} and A.~F. {Heavens}, \emph{{On the use of the
  Edgeworth expansion in cosmology I: how to foresee and evade its pitfalls}},
  {\emph{arXiv e-prints} (2017) arXiv:1709.03452}
  [\href{https://arxiv.org/abs/1709.03452}{{\ttfamily 1709.03452}}].

\bibitem{2009arXiv0906.0664H}
A.~{Heavens}, \emph{{Statistical techniques in cosmology}}, {\emph{arXiv
  e-prints} (2009) arXiv:0906.0664}
  [\href{https://arxiv.org/abs/0906.0664}{{\ttfamily 0906.0664}}].

\bibitem{2017MNRAS.470.2617A}
S.~{Alam}, M.~{Ata}, S.~{Bailey}, F.~{Beutler}, D.~{Bizyaev}, J.~A. {Blazek}
  et~al., \emph{{The clustering of galaxies in the completed SDSS-III Baryon
  Oscillation Spectroscopic Survey: cosmological analysis of the DR12 galaxy
  sample}}, \href{https://doi.org/10.1093/mnras/stx721}{\emph{\mnras}
  {\bfseries 470} (2017) 2617}
  [\href{https://arxiv.org/abs/1607.03155}{{\ttfamily 1607.03155}}].

\bibitem{2016MNRAS.460.1173R}
S.~A. {Rodr{\'\i}guez-Torres}, C.-H. {Chuang}, F.~{Prada}, H.~{Guo},
  A.~{Klypin}, P.~{Behroozi} et~al., \emph{{The clustering of galaxies in the
  SDSS-III Baryon Oscillation Spectroscopic Survey: modelling the clustering
  and halo occupation distribution of BOSS CMASS galaxies in the Final Data
  Release}}, \href{https://doi.org/10.1093/mnras/stw1014}{\emph{\mnras}
  {\bfseries 460} (2016) 1173}
  [\href{https://arxiv.org/abs/1509.06404}{{\ttfamily 1509.06404}}].

\bibitem{2016MNRAS.456.4156K}
F.-S. {Kitaura}, S.~{Rodr{\'\i}guez-Torres}, C.-H. {Chuang}, C.~{Zhao},
  F.~{Prada}, H.~{Gil-Mar{\'\i}n} et~al., \emph{{The clustering of galaxies in
  the SDSS-III Baryon Oscillation Spectroscopic Survey: mock galaxy catalogues
  for the BOSS Final Data Release}},
  \href{https://doi.org/10.1093/mnras/stv2826}{\emph{\mnras} {\bfseries 456}
  (2016) 4156} [\href{https://arxiv.org/abs/1509.06400}{{\ttfamily
  1509.06400}}].

\bibitem{2017MNRAS.466.2242B}
F.~{Beutler}, H.-J. {Seo}, S.~{Saito}, C.-H. {Chuang}, A.~J. {Cuesta}, D.~J.
  {Eisenstein} et~al., \emph{{The clustering of galaxies in the completed
  SDSS-III Baryon Oscillation Spectroscopic Survey: anisotropic galaxy
  clustering in Fourier space}},
  \href{https://doi.org/10.1093/mnras/stw3298}{\emph{\mnras} {\bfseries 466}
  (2017) 2242} [\href{https://arxiv.org/abs/1607.03150}{{\ttfamily
  1607.03150}}].

\bibitem{2018AJ....156..160H}
N.~{Hand}, Y.~{Feng}, F.~{Beutler}, Y.~{Li}, C.~{Modi}, U.~{Seljak} et~al.,
  \emph{{nbodykit: An Open-source, Massively Parallel Toolkit for Large-scale
  Structure}}, \href{https://doi.org/10.3847/1538-3881/aadae0}{\emph{\aj}
  {\bfseries 156} (2018) 160}
  [\href{https://arxiv.org/abs/1712.05834}{{\ttfamily 1712.05834}}].

\bibitem{2020PhRvD.102f3533C}
A.~{Chudaykin}, M.~M. {Ivanov}, O.~H.~E. {Philcox} and M.~{Simonovi{\'c}},
  \emph{{Nonlinear perturbation theory extension of the Boltzmann code CLASS}},
  \href{https://doi.org/10.1103/PhysRevD.102.063533}{\emph{\prd} {\bfseries
  102} (2020) 063533} [\href{https://arxiv.org/abs/2004.10607}{{\ttfamily
  2004.10607}}].

\bibitem{2021JCAP...03..020S}
M.~{Schmittfull} and A.~M. {Dizgah}, \emph{{Galaxy skew-spectra in
  redshift-space}},
  \href{https://doi.org/10.1088/1475-7516/2021/03/020}{\emph{\jcap} {\bfseries
  2021} (2021) 020} [\href{https://arxiv.org/abs/2010.14267}{{\ttfamily
  2010.14267}}].

\bibitem{2020JCAP...04..011M}
A.~{Moradinezhad Dizgah}, H.~{Lee}, M.~{Schmittfull} and C.~{Dvorkin},
  \emph{{Capturing non-Gaussianity of the large-scale structure with weighted
  skew-spectra}},
  \href{https://doi.org/10.1088/1475-7516/2020/04/011}{\emph{\jcap} {\bfseries
  2020} (2020) 011} [\href{https://arxiv.org/abs/1911.05763}{{\ttfamily
  1911.05763}}].

\bibitem{2021PhRvD.103b3538P}
O.~H.~E. {Philcox}, B.~D. {Sherwin}, G.~S. {Farren} and E.~J. {Baxter},
  \emph{{Determining the Hubble constant without the sound horizon:
  Measurements from galaxy surveys}},
  \href{https://doi.org/10.1103/PhysRevD.103.023538}{\emph{\prd} {\bfseries
  103} (2021) 023538} [\href{https://arxiv.org/abs/2008.08084}{{\ttfamily
  2008.08084}}].

\bibitem{2005ApJ...620..559J}
Y.~P. {Jing}, \emph{{Correcting for the Alias Effect When Measuring the Power
  Spectrum Using a Fast Fourier Transform}},
  \href{https://doi.org/10.1086/427087}{\emph{\apj} {\bfseries 620} (2005) 559}
  [\href{https://arxiv.org/abs/astro-ph/0409240}{{\ttfamily
  astro-ph/0409240}}].

\bibitem{2018MNRAS.476.4403C}
E.~{Castorina} and M.~{White}, \emph{{Beyond the plane-parallel approximation
  for redshift surveys}},
  \href{https://doi.org/10.1093/mnras/sty410}{\emph{\mnras} {\bfseries 476}
  (2018) 4403} [\href{https://arxiv.org/abs/1709.09730}{{\ttfamily
  1709.09730}}].

\bibitem{2021PhRvD.103l3509P}
O.~H.~E. {Philcox} and Z.~{Slepian}, \emph{{Beyond the Yamamoto approximation:
  Anisotropic power spectra and correlation functions with pairwise lines of
  sight}}, \href{https://doi.org/10.1103/PhysRevD.103.123509}{\emph{\prd}
  {\bfseries 103} (2021) 123509}
  [\href{https://arxiv.org/abs/2102.08384}{{\ttfamily 2102.08384}}].

\end{thebibliography}\endgroup

\end{document}